%% file: main.tex
\documentclass[reprint,showpacs,superscriptaddress,english,twocolumn,notitlepage,longbibliography]{revtex4-2}
\usepackage[utf8]{inputenc}
\pdfoutput=1
\usepackage{times}
\usepackage{color}
\usepackage{array}
\usepackage{verbatim}
\usepackage{multirow}
\usepackage{amsmath}
\usepackage{extarrows}
\usepackage{amssymb}
\usepackage{dsfont}
\usepackage{graphicx}
\usepackage{placeins}
\usepackage{tikz}
\usepackage{esint}
\usepackage{makecell}
\usepackage{longtable}
\usepackage{changepage}
\PassOptionsToPackage{obeyspaces}{url}
\usepackage{xurl}
\usepackage[unicode=true,
 bookmarks=true,bookmarksnumbered=true,bookmarksopen=false,
 breaklinks=true,pdfborder={0 0 0},backref=false,colorlinks=true]
 {hyperref}
\usepackage{enumitem}
\usepackage[normalem]{ulem}

\usepackage{rotating}

\hypersetup{
    linkcolor=purple,          
    citecolor=blue,        
    filecolor=magenta,      
    urlcolor=magenta           
}

\renewcommand{\paragraph}[1]{\vspace{0.2cm}{\textbf{\textit{#1}}}}

\usepackage{simpler-wick}
\allowdisplaybreaks[1] 

\newcommand{\ovl}{\overline}

\def\ket#1{\vert #1 \rangle}

\def\inn#1{\langle #1 \rangle}

\def\tr{\mathrm{tr}}
\def\Tr{\mathrm{Tr}}
\def\bf#1{\mathbf{#1}}

\begin{document}

\title{A Fractional Chern Semimetal: Exact Wavefunction and Conformal Criticality}

\author{Fa-Jie Wang}
\affiliation{International Center for Quantum Materials, School of Physics, Peking University, Beijing 100871, China}

\author{Yang Qi}
\affiliation{ State Key Laboratory of Surface Physics, Fudan University, 200433 Shanghai, China}
\affiliation{Center for Field Theory and Particle Physics,
Department of Physics, Fudan University, 200433 Shanghai, China}
\affiliation{Collaborative Innovation Center of Advanced Microstructures, 210093 Nanjing, China}

\author{Zhi-Da Song}
\email{songzd@pku.edu.cn}
\affiliation{International Center for Quantum Materials, School of Physics, Peking University, Beijing 100871, China}
\affiliation{Hefei National Laboratory, Hefei 230088, China}

\begin{abstract}
Unlike fractional quantum Hall states in uniform Landau levels, fractional Chern insulators can host dispersive anyon bands, raising the question of what happens when the anyon gap closes.
We construct an exact finite-bond-dimension tensor-network state for a fractional Chern semimetal by Gutzwiller projecting a critical parton state. 
This state exhibits divergent correlation lengths in charged sectors but a fractionally quantized many-body Chern number of $1/3$.
Although it has one electron per unit cell, its modular $S$ and $T$ matrices and entanglement spectrum match those of the $\nu=1/3$ Laughlin state.
We identify a $(1+1)$--dimensional $U(3)_1\times SU(3)_1$ conformal field theory (CFT) governing the wavefunction norm in the infrared, an unexpected result for a state with $U(1)_3$ topology.
The $SU(3)_1$ sector is a nonlinear sigma model whose Wess--Zumino level and kinetic stiffness are determined by the parton Chern number and Fubini--Study metric, respectively.
Under renormalization, it flows to the conformal fixed point, with an  initial distance to this fixed point controlled by the idealness of parton quantum geometry.
Favorable geometry enables precise agreement with CFT predictions even on narrow cylinders.
\end{abstract}

\date{\today}
\maketitle

\paragraph{Introduction.}
The observation of the fractional quantum anomalous Hall effect in moir\'e materials \cite{cai_signatures_2023, xu_observation_2023, Park2023FractionalQAH, zeng_thermodynamic_2023, lu_fractional_2024, li_fractional_2026} has renewed interest in fractional Chern insulators (FCIs) \cite{tang_high-temperature_2011, Neupert2011Fractional, sun_nearly_2011, sheng2011fractional, Regnault2011FCI}.
Their conventional description starts from a partially filled, nearly flat Chern band, which, under the ideal quantum geometry condition \cite{Wang2021IdealFlatband, Roy2014Geometry, claassen_position_2015, wang_origin_2023, Ledwith2023Vortexability, Estienne2023Ideal}, maps to Landau levels (LLs) \cite{qi_generic_2011, wu_gauge-fixed_2012, ledwith_fractional_2020, zhao2026ideal, li2026quantum}. 
Despite this correspondence, lattice periodicity allows charged anyons in FCIs to acquire dispersion \cite{Goncalves2026ChargeExcitations, Schleith2026AnyonSphere, Liu2025FCIQuasiparticles, Yan2026AnyonDispersion, Iyer2026AnyonBloch, Xiao2026AnyonChern, Sarkar2026AnyonBands}, unlike their counterparts in uniform LLs. 
An extreme possibility is that the anyon excitation gap closes, motivating the study of interplay between fractional topology and bulk criticality.

Tensor-network states (TNSs) offer a systematic framework for constructing trial wavefunctions \cite{cirac_matrix_2021}, including fractional quantum Hall states \cite{Iblisdir2007Entropy, Zaletel2012Exact, Estienne2013Matrix}. 
However, exact finite-bond-dimension representations of gapped chiral topological states face fundamental obstructions in the thermodynamic limit \cite{chen_impossibility_2014, wahl_projected_2013, DubailRead2015ChiralTNS, Lemm2019Spectral}.
In particular, for Gaussian TNSs in two and higher dimensions, nontrivial band topology necessarily entails algebraic correlations \cite{read_compactly-supported_2017}.
Local projection of multiple critical Gaussian copies nevertheless offers a route to fractionalization, as proposed in Ref.~\cite{DubailRead2015ChiralTNS}.
Subsequent numerical constructions revealed chiral topological orders alongside bulk criticality in spin systems \cite{yang_chiral_2015, Poilblanc2015Chiral, Poilblanc2016SU2}.
These examples motivate a fundamental question: how can fermionic Laughlin-type topology coexist with bulk criticality, and what field theory governs this criticality?

\nocite{yang_fractional_2025}
Critical topological flat bands (CTFBs) provide a systematic symmetry-based recipe for constructing exact Gaussian TNSs with prescribed topology \cite{Li2026StableTopology}, offering starting points for studying these questions.
These flat bands touch dispersive bands at isolated momenta, where their projectors remain continuous but become nonanalytic \cite{DubailRead2015ChiralTNS, yang_fractional_2025, Li2026StableTopology, liu2026general}, allowing well-defined band topology together with algebraic correlations.
The corresponding TNSs are built from compactly supported orbitals and hence have finite bond dimensions. 

Here, we Gutzwiller project a CTFB-based parton state to construct an exact fermionic TNS whose topological data match those of the $\nu=1/3$ Laughlin state.
Yet it preserves translation symmetry at one electron per unit cell, realizing fractional topology at integer filling, a possibility also explored in fractional excitonic insulators \cite{hu_fractional_2018, gassner_flux_2025}.
More importantly, we identify a $U(3)_1\times SU(3)_1$ conformal field theory (CFT) of decoupled Dirac fermion and Wess--Zumino--Witten (WZW) sectors governing the wavefunction norm in the infrared, and numerically confirm this description through exact finite-size transfer spectra.
The resulting charged correlations exhibit critical scaling consistent with isolated Fermi points; we therefore call this state a fractional Chern semimetal (FCSM).
We further show that the deviation of the bare stiffness from its conformal fixed-point value is controlled by the idealness of the parton quantum geometry.
Related interest has emerged in critical Laughlin wavefunctions \cite{Moitra2026Instability}, topological order coexisting with Goldstone modes \cite{Han2026Anyon, Wang2026Anyon, jian2026theory}, and FCIs in unconventional band settings \cite{lin2026fractional,Liu2026Topological}.

\paragraph{Construction of the TNS.}
We consider a honeycomb lattice with one physical $s$ orbital per site.
We label each site by $(\mathbf{r},a)$, where $\mathbf{r}$ labels the unit cell and $a=1,2$ labels the honeycomb sublattice.
At each site, we represent the electron operator as $c_{\mathbf{r}a}=\prod_{\alpha=1}^{3}f_{\mathbf{r}a\alpha}$, where $\alpha$ labels the parton color.
This decomposition has a local $SU(3)$ gauge redundancy because an independent color rotation $g$ at each site only multiplies the electron operator by $\det g=1$. 
The occupied bands of a bilinear parton Hamiltonian define a Slater determinant $\ket{\Phi}$.
We Gutzwiller-project $\ket{\Phi}$ onto the physical Hilbert space by retaining only configurations with  $\sum_{\alpha}f_{\mathbf{r}a\alpha}^{\dagger}f_{\mathbf{r}a\alpha}=0$ or 3 at every site.
The resulting state $\ket{\Psi}=P_{\rm G}\ket{\Phi}$ can exhibit topological orders if the parton bands are topological \cite{LuRan2012PartonFCI}. 

$\vert\Psi\rangle$ can be expressed as a TNS with a finite bond dimension if $\ket{\Phi}$ can be built from compact Bloch wavefunctions 
$S_{(a,\alpha),n}(\bf{k})= \sum_{\Delta\mathbf{r}} \mathrm{e}^{-\mathrm{i}\mathbf{k}\cdot(\Delta\mathbf{r}+\mathbf d_a)}S_{(a,\alpha),n}(\Delta\mathbf{r})$ for a finite support $(\Delta\mathbf{r},a)$, where $\mathbf d_a$ is the position of sublattice $a$ within the cell. 
Here the columns need not be orthonormal. 
By defining the local operator $\mathcal{S}_{n}(\bf{r})=\sum_{a,\alpha, \Delta\mathbf{r} } S^*_{(a,\alpha),n}(\Delta\mathbf{r}) f_{\mathbf{r}+\Delta\mathbf{r},a\alpha}$, one can verify
\begin{equation}
        \vert\Psi\rangle = P_{\rm G} \prod_{\mathbf{r}} \prod_{n\in \mathrm{occ}} \mathcal{S}^\dagger_n(\mathbf{r})\ket{0} \ .
    \label{eq:Gutzwiller_real}
\end{equation}
Here $\ket{0}$ is the vacuum state and the product over $n$ runs over the occupied states. 
Compact and non-vanishing $S_n(\bf{k})$ usually contradicts with a nonzero Chern number.
However, by allowing finite-ranged $S_n(\bf{k})$ to vanish while the projector $P=S (S^\dagger S)^{-1} S^\dagger$ remains continuous at high-symmetry points, a CTFB with nonzero Chern number can still be realized. 
The resulted $\ket{\Phi}$ admits an exact TNS representation with algebraic correlations.


\begin{figure}[!t]
    \centering
    \includegraphics[width=\columnwidth]{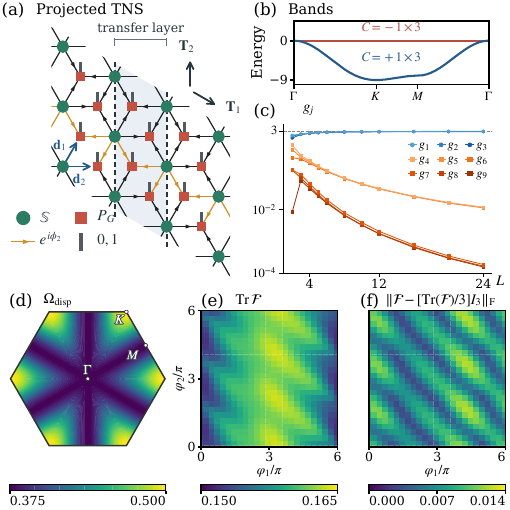}
    \caption{The tensor-network state.
    (a) Green circles denote the $\mathbb{S}$ tensors, red squares Gutzwiller projectors, and gray legs are physical states.
    Orange bonds indicate a Peierls phase string. 
    (b) The parton band structure. 
    (c) Eigenvalues of the Gram matrix of states corresponding to the nine parton boundary conditions compatible with the physical boundary condition $(\pi,\pi)$, as functions of system size. 
    (d) Single-particle Berry curvature $\Omega(\mathbf{k})$ of the occupied parton band. 
    (e) Trace $\operatorname{Tr}\mathcal F$ of the many-body Berry curvature of the ground-state manifold, orthonormalized from states with parton twists $(\phi_1,\phi_2)=(\varphi_1/3,\varphi_2/3+2\pi a/3)$, $a=0,1,2$.
    The system size is $L=12$. 
    (f) Deviation from scalar curvature, measured by the Frobenius norm $\|\mathcal F-\frac13 \operatorname{Tr} [\mathcal F] \mathbb{I}_3\|_{\rm F}$.
    }
    \label{fig:mean-field}
\end{figure}

We introduce three local operators $\mathcal{S}_n(\mathbf{r})$ ($n=1,2,3$) at each hexagon center $\mathbf{r}$. 
Each is a linear combination of parton operators on the six nearest honeycomb sites, numbered $j=1,\ldots,6$ counterclockwise starting from $\mathbf{d}_1$ in Fig.~\ref{fig:mean-field}(a).
We require each parton orbital to transforms as $p_x-\mathrm{i}p_y$ under $C_{6z}$ and each  $\mathcal{S}_n(\mathbf{r})$ to transform as an $s$ orbital.
This fixes the six coefficients $S_{(a,\alpha),n}(\Delta\mathbf{r})$ to $s_{\alpha,n}\mathrm{e}^{\mathrm{i}(j-1)\pi/3}$, where $s_{\alpha,n}$ is arbitrary. 
Since each $\mathcal{S}_n^\dagger(\mathbf{r})$ has a finite support, we can represent $\prod_{n\in\mathrm{occ}}\mathcal{S}_n^\dagger(\mathbf{r})$ by a tensor $\mathbb{S}(\mathbf{r})$ centered at $\mathbf{r}$, whose legs contract with the Gutzwiller projectors $P_{\rm G}$ on neighboring parton sites [Fig.~\ref{fig:mean-field}(a)].

Due to the gauge redundancy, any full-rank $s$ gives the same normalized projected state $\ket{\bar\Psi}$ up to phase, hence we take $s_{\alpha,n}=\delta_{\alpha, n}$ hereafter.
With this choice, the three colors have identical Slater wavefunctions.
Let $\Phi_1[\sigma]=\inn{\sigma|\Phi_1}$ be the amplitude of a real-space occupation configuration $\sigma$ in the single-color Slater state $\ket{\Phi_1}$.
Gutzwiller projection selects the same configuration $\sigma$ for all three colors, so the physical wavefunction has amplitudes
\begin{equation} 
\label{eq:projected-amplitude}
    \inn{\sigma|\Psi}= (\Phi_1[\sigma])^3 \ . 
\end{equation}

Topology of the unprojected state $\ket{\Phi}$ is most easily seen from its parent bilinear parton Hamiltonian $ H(\mathbf{k}) = - S(\mathbf{k}) S^\dagger (\mathbf{k})$, which is also studied in Ref.~\cite{yang_fractional_2025}. 
As shown in Fig.~\ref{fig:mean-field}(b), the band structure contains three  degenerate CTFBs touching three degenerate lower dispersive bands at the $\Gamma$.
Apart from $\Gamma$, the flat bands and dispersive bands respectively span $\operatorname{Ker} S^\dagger(\bf{k})$ and $\operatorname{Im}S(\bf{k})$ since $S(\bf{k})$ has full column rank there.
Thus, for boundary conditions that exclude $\Gamma$ from the allowed momenta, $\ket{\Phi}$ constructed from $\mathcal{S}_n^\dagger(\mathbf{r})$ fills all three dispersive bands.

Near $\Gamma$, rotation symmetry fixes the linear term in $S$ to be
$S(\mathbf{k})=k_+ S_0 \otimes \mathbb{I}_3 +\mathcal{O}(|\mathbf{k}|^2)$, where $k_+=k_x+\mathrm{i}k_y$ and $S_0$ is a two-by-one matrix. 
The occupied subspace $\operatorname{Im}S(\mathbf{k})$ and the flat-band subspace $\operatorname{Ker} S^\dagger(\mathbf{k})$ thus have direction-independent limits at $\Gamma$.
The orthonormal frame $S(S^\dagger S)^{-1/2}$ defines a Wannier gauge for the occupied bands that is singular only at $\Gamma$.
For each color, the phase $k_+/|k_+|$ winds once as $\mathbf{k}$ encircles $\Gamma$.
Applying Stokes' theorem to the punctured Brillouin zone gives a Chern number of $+1$ per color and hence $C=3$ for the occupied bands \cite{Li2026StableTopology}.
Fig.~\ref{fig:mean-field}(d) shows the corresponding total Berry curvature, which remains finite as $\mathbf{k}\to\Gamma$.
Assigning charge $e/3$ to each parton, the normalized unprojected state $\ket{\bar\Phi}$ describes a parton semimetal at filling $\nu_f=3$ with mean-field Hall conductance $C(e/3)^2/h=e^2/(3h)$.

\paragraph{Topological order.}
If the parton bands were gapped, the projected state $\ket{\bar\Psi}$ would be a gapped FCI with $U(1)_{3}$ topological order~\cite{McGreevy2012PartonFCI,LuRan2012PartonFCI}.
However, the absence of a parton band gap requires further investigations. 
We will show that its ground state degeneracy, many-body Chern number, and Wilson loop operators are all consistent with the $U(1)_3$ topological order.

Since $P_{\rm G}$ is local, $\vert\bar\Psi\rangle$ preserves the crystalline symmetry of $\ket{\bar\Phi}$ and has an integer physical filling $\nu=\frac{\nu_f}{3}=1$, which contrasts with existing FCIs with partial fillings. 

On a torus, a topological order in $\ket{\bar\Psi}$ should induce other locally indistinguishable ground states.
Given the physical boundary condition where an electron acquires phase $\mathrm{e}^{\mathrm{i}\varphi_i}$ after traveling around the torus along $\bf{T}_i$ ($i=1,2$), the corresponding parton phase angle can be $\phi_{i}=\frac{\varphi_{i}}{3} +\frac{2\pi n_i}{3}$ with $n_i=0$, 1, 2.
Therefore, one can consider nine ansatz projected states on a torus with parton twisted boundary conditions given by $n_1,n_2=0$, 1, 2.
These twists can be implemented by attaching Peierls phases $(\phi_1,\phi_2)$ to $\mathcal{S}_n(\mathbf{r})$ operators that cross some noncontractible cuts, as illustrated in the TNS representation in Fig.~\ref{fig:mean-field}(a). 
Since local $SU(3)$ gauge transformations can move the cuts away from any contractible region, all nine states share identical tensors within that region.
Numerical overlaps suggest that the nine states span a three-dimensional space in the thermodynamic limit, as illustrated in Fig.~\ref{fig:mean-field}(c); and their reduced density matrices provide evidence for local indistinguishability, as detailed in Sec.~\ref{app:twist-rank}. 

To determine the many-body Chern number, we first follow a single branch at fixed $(n_1,n_2)$.
As the electron twists vary over $[0,6\pi)^2$, this branch covers the parton twist torus exactly once.
Although the parton Slater determinant $\ket{\Phi}$ is critical, the occupied Bloch subspace has a direction-independent limit at $\Gamma$, as shown above.
The normalized $\ket{\bar\Phi}$ therefore has a well-defined limit up to an overall phase as $\boldsymbol{\phi}\to0$, where the momentum mesh reaches $\Gamma$.
Over the full parton twist torus, the unprojected branch has Chern number $C=3$, the total Chern number of the occupied parton bands.
Gutzwiller projection preserves this integer as long as it does not annihilate $\ket{\bar\Phi}$ at any twist.
Indeed, at each twist the orthonormal occupied Bloch states define a normalized single-color Slater state $\ket{\bar\Phi_1}$, so at least one of its real-space amplitudes $\bar\Phi_1[\sigma]$ is nonzero.
By Eq.~\ref{eq:projected-amplitude}, the projected amplitude $\inn{\sigma|P_{\rm G}|\bar\Phi}=\bar\Phi_1[\sigma]^3$ is also nonzero.

Assuming three linearly independent branches throughout the extended twist torus, 
their Chern numbers add to $3C=9$.
A $2\pi$ flux insertion permutes the twist states, so their span is periodic in each physical twist.
Since the extended torus covers the physical twist torus nine times, the total Chern number of the ground-state space is $C_{\rm g} = 3C/9=1$, giving an average many-body Chern number per ground state of $C_{\mathrm{MB}}=1/3$.

Despite the gapless bulk, the many-body Berry curvature behaves as expected for a gapped state.
Its trace integrates to $2\pi C_{\rm g}=2\pi$ over the physical twist torus, giving $C_{\mathrm{MB}}=C_{\rm g}/3=1/3$, and varies very little at large system sizes [Fig.~\ref{fig:mean-field}(e)].
The full curvature matrix is also nearly proportional to $\mathbb{I}_3$ [Fig.~\ref{fig:mean-field}(f)], suggesting that all three ground states experience the same, twist-independent Berry curvature in the thermodynamic limit.

The Wilson loop operators $W_x$ and $W_y$ describe parallel transport within the ground-state space as the physical twists $\varphi_x$ and $\varphi_y$, respectively, increase by $2\pi$.
If the many-body Berry curvature is proportional to $\mathbb{I}_3$ throughout the twist torus, the two orders of flux insertion differ only by a Berry phase, giving $W_xW_yW_x^{-1}W_y^{-1} = \mathrm{e}^{2\pi\mathrm{i}/3}\mathbb{I}_3$. 
Our numerical Wilson loops approach this relation with increasing system size, and $W_y$ cycles through the three eigenstates of $W_x$ [Sec.~\ref{app:wilson-numerics}].
This cycling through all three sectors supports a $\mathbb{Z}_3$ anyon fusion group, with $\mathrm{e}^{2\pi\mathrm{i}/3}$ the mutual braiding phase of two copies of its generator.

We obtain the modular matrix $S_{ab}=\langle x_a|y_b\rangle$ from the overlaps between the eigenstates $\ket{x_a}$ of $W_x$ and $\ket{y_b}$ of $W_y$~\cite{Zhang2012MES,LiMong2022Modular}. 
The modular $T$ matrix acts on the ground-state space as $C_{6z}S$ up to an overall phase, allowing us to extract the relative phases $T_{aa}/T_{00}$.
We find $S_{ab}\simeq\mathrm{e}^{2\pi\mathrm{i}ab/3}/\sqrt{3}$ and $T_{aa}/T_{00}\simeq\mathrm{e}^{4\pi\mathrm{i}/3}$ for $a=1,2$, in agreement with fermionic $U(1)_3$ topological order~\cite{Okuda2021SpinCS}.
[See Sec.~\ref{app:modular-extraction} for more details.]

\begin{figure}[t]
    \centering
    \includegraphics[width=\columnwidth]{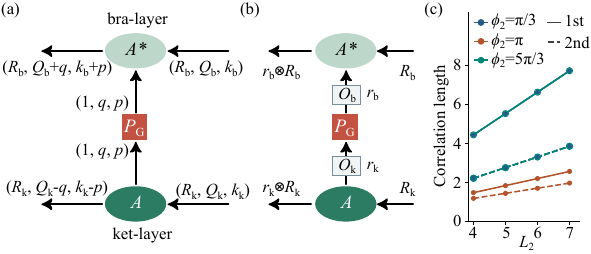}
    \caption{
The transfer matrix. 
(a,b) $A$ ($A^*$) denotes a tensor column in the ket (bra) layer, with each external-leg dimension $2^{3L_2}$. $R_{\tau}$, $Q_{\tau}$, and $k_{\tau}$ ($\tau=\mathrm{k,b}$) label the $SU(3)$ representation, parton charge, and momentum, while $r_{\tau}$ label the irreps of inserted operators. 
The vertical arrows follow virtual partons from ket to bra. In (a), $q$ and $p$ give the net charge and momentum transferred between layers relative to the filled-cut background.
(c) Leading correlation lengths as functions of $L_2$. 
    }
    \label{fig:tns-transfer}
\end{figure}
\begin{table}[t]
    \caption{Transfer gaps $\Delta$ in different quantum number channels, compared with CFT predictions. Here $\phi_2=\pi$. Fit windows are listed in Table~\ref{tab:transfer-branches}.}
    \label{tab:transfer-dimensions}
    \centering
    \setlength{\tabcolsep}{4pt}
    \renewcommand{\arraystretch}{1.12}
    \begin{ruledtabular}
    \setlength{\tabcolsep}{2pt}
    \begin{tabular}{@{$($}c@{$\,$}c@{$\,$}c@{$\,$}c@{$)$\hspace{5pt}}ccc}
    $R_{\mathrm{k}}$ & $R_{\mathrm{b}}$ & $Q$ & $K_y$
    & Interpretation
    & \makecell{Extrapolated\\$\Delta$}
    & \makecell{CFT\\$\Delta$}\\
    \hline
    $\mathbf{3}$ & $\mathbf{1}$ & $1$ & $1/2$
        & $\psi_{\rm k}^\dagger$ & $0.50277(59)$ & $1/2$\\
    $\mathbf{8}$ & $\mathbf{1}$ & $0$ & $1$
        & $\psi_{\rm k1}^\dagger \psi_{\rm k2}$ & $0.99576(84)$ & $1$\\
    $\mathbf{1}$ & $\mathbf{1}$ & $3$ & $3/2$
        & $\psi_{\rm k1}^\dagger \psi_{\rm k2}^\dagger \psi_{\rm k3}^\dagger$ & $1.5322(59)$ & $3/2$\\
    $\mathbf{1}$ & $\mathbf{3}$ & $1$ & $-1/2$
        & $\psi_{\rm b}^\dagger$ & $0.50277(59)$ & $1/2$\\
    $\mathbf{3}$ & $\ovl{\mathbf{3}}$ & $0$ & $0$
        & $\psi_{\rm k}^\dagger \psi_{\rm b}$ & $0.99578(97)$ & $1$\\
    $\mathbf{3}$ & $\ovl{\mathbf{3}}$ & $0$ & $0$
        & $g$ & $0.66576(31)$ & $2/3$\\
    $\mathbf{8}$ & $\mathbf{1}$ & 0 & $-1$ & $g^{-1} \partial_{\ovl z} g$ & $0.9946(18)$ & $1$ \\
    $\mathbf{1}$ & $\mathbf{8}$ & 0 & $1$ & $(\partial_{z} g) g^{-1}$ & $0.9946(18)$ & $1$ \\
    $\ovl{\mathbf{3}}$ & $\ovl{\mathbf{3}}$ & $1$ & $1/2$
        & $\psi_{\rm k}^\dagger g$ & $1.17253(54)$ & $7/6$\\
    $\mathbf{6}$ & $\ovl{\mathbf{6}}$ & 0 & 0 & $\psi_{\rm k}^\dagger \psi_{\rm b} g$ & $1.6887(29)$ & $5/3$ 
    \end{tabular}
    \end{ruledtabular}
\end{table}

\paragraph{Transfer matrix.}
To test whether the projected state $\ket{\bar\Psi}$ remains critical, we examine its bulk correlations, which must decay exponentially in a gapped local system \cite{HastingsKoma2006Clustering}.
We can extract the correlations by decomposing the norm $\inn{\Psi|\Psi}$ into layers according to the dashed lines in Fig.~\ref{fig:mean-field}(a), where each cut virtual bond has dimension  $2^3$. 
The resulting transfer matrix is illustrated in Fig.~\ref{fig:tns-transfer}.
The virtual bonds connecting to the right (left) neighboring layer form the column (row) index of dimension $2^{6L_2}$, where $L_2$ is the circumference along $\mathbf{T}_2$.
The eigenvalues $\lambda$ of the transfer matrix determine the correlation lengths $\xi=\frac{{\sqrt3}/{2}}{\ln|\lambda_0/\lambda|}$, where $\lambda_0$ is the largest eigenvalue \cite{Zauner2015Transfer}, and ${\sqrt3}/{2}$ is the perpendicular distance advanced by one layer.
For all three states with parton twists $\phi_2=\pi/3,\pi,5\pi/3$, the leading correlation lengths grow approximately linearly with $L_2$ [Fig.~\ref{fig:tns-transfer}(c)].
This supports their bulk criticality and the absence of additional topological degeneracy within each twist sector.
We also define the scaled transfer gaps
\begin{equation}    \label{eq:bulk-scaled-transfer-gap}
    \Delta = \frac{L_2}{2\pi \xi} \ .
\end{equation}

The transfer eigenmodes can be labeled by $(R_{\mathrm{k}}$, $R_{\mathrm{b}}$, $Q$, $K_y)$, where $R_{\mathrm{k}}$ and $R_{\mathrm{b}}$ are the $SU(3)$ representations \footnote{For $SU(3)$, $\mathbf{1}$ denotes the trivial representation, while $\mathbf{3}=[1]$, $\ovl{\mathbf{3}}=[1,1]$, $\mathbf{6}=[2]$, $\ovl{\mathbf{6}}=[2,2]$, and $\mathbf{8}=[2,1]$ in Young-diagram notation, with brackets specifying row lengths.} in the ket and bra layers, $Q$ is the total parton charge relative to the dominant mode ($3L_2$), and $K_y=-L_2(k_2-k_{2,0})/(2\pi)\pmod{L_2}$ labels transverse momentum, with $k_{2,0}=3\phi_2+3\pi(L_2-1)\pmod{2\pi}$. An active translation by $+\mathbf T_2$ gives the Bloch phase $\mathrm{e}^{-\mathrm{i}k_2}$.
As illustrated in Fig.~\ref{fig:tns-transfer}(a), $P_{\rm G}$ selects only $SU(3)$ singlets, so the transfer matrix is invariant under independent color rotations in the ket and bra layers, forming an $SU(3)^{\mathrm{k}}\times SU(3)^{\mathrm{b}}$ symmetry.
Consequently, $R_{\mathrm{k}}$ and $R_{\mathrm{b}}$ are separately conserved.
Parton charge and momentum can be exchanged between the two layers through $P_{\rm G}$, so only their totals are conserved.
Block diagonalization using these quantum numbers allows us to reach $L_2=7$.

We focus on $\phi_2=\pi$, corresponding to antiperiodic parton boundary conditions.
In the CFT description, $\lambda_0$ then represents the identity vacuum, so $\Delta$ gives scaling dimensions~\cite{Cardy1984FiniteSize}.
The dominant eigenvalue $\lambda_0$ lies in the $(\mathbf{1},\mathbf{1},0,0)$ sector.
Local correlations are contributed by subleading eigenvalues in sectors selected by the operator insertions $O_{\tau}$ ($\tau=\mathrm{k,b}$) [Fig.~\ref{fig:tns-transfer}(b)].
The creation operators $\psi_{\tau\alpha}^\dagger$ form the fundamental representation of $SU(3)_\tau$ and transform as
\begin{equation}
    \psi_{\tau\alpha}^\dagger
    \mapsto
    \sum_{\beta=1}^{3}\psi_{\tau\beta}^\dagger U_{\beta\alpha},
    \qquad U\in SU(3)^\tau,
    \quad \tau=\mathrm{k},\mathrm{b}.
\end{equation}
For example, particle $\psi_{\mathrm{k}}^\dagger$, particle--hole pair $\psi_{\mathrm{k}}^\dagger \psi_{\mathrm{b}}$, and three-parton singlet $\psi_{\mathrm{k}1}^\dagger \psi_{\mathrm{k}2}^\dagger \psi_{\mathrm{k}3}^\dagger$ insertions access the $(\mathbf{3},\mathbf{1},1,1/2)$, $({\mathbf{3}}, \ovl{\mathbf{3}},0,0)$, and $(\mathbf{1},\mathbf{1},3,3/2)$ sectors, respectively.
The particle--hole field has components $\psi_{\mathrm{k}\alpha}^\dagger\psi_{\mathrm{b}\beta}$ with independent ket and bra color indices $\alpha,\beta$.

The first five rows of Table~\ref{tab:transfer-dimensions} are consistent with auxiliary Dirac fields $\psi$ of scaling dimension $1/2$. 
The ket (bra) excitations have $K_y>0$ ($K_y<0$), indicating chiral (anti-chiral) modes.
This assignment is also suggested by $S(\mathbf{k})$: in a Bloch gauge continuous at $\Gamma$, the leading nonanalytic ket Bloch amplitude is proportional to $k_-^2/k_+$, while the bra amplitude is its complex conjugate \cite{Li2026StableTopology,liu2026general}.
Thus, the auxiliary propagators should have poles at $k_\pm=0$. 
Taking the transfer direction as imaginary time, Wick rotation $\mathrm{i}k_x\to2\pi\Delta/L_2$ gives $\Delta=\pm K_y$, consistent with the observed chiralities.
Together, the three colors of $\psi$ form a $U(3)_1$ Dirac fermion sector.

\paragraph{Emergent WZW sector.}
The spectrum also contains modes beyond $U(3)_1$.
In particular, the sixth row of Table~\ref{tab:transfer-dimensions} suggests a neutral field $g$ in the $(\mathbf{3},\ovl{\mathbf{3}})$ representation with scaling dimension $2/3$.
A natural interpretation is an additional $SU(3)_1$ WZW sector \cite{Witten1984Bosonization}. 
Its fixed point action is 
\begin{equation} \label{eq:bulk-color-action}
\mathcal{S}_G[g] = 
    \kappa_\star \int  \mathrm{d}^2\mathbf r\,\Tr(\partial_\mu g^{-1}\partial_\mu g)
    +   \mathrm{i} \, \eta \,  S_{\mathrm{WZ}}[g] \ ,
\end{equation}
where $g\in SU(3)$, $\kappa_\star=1/(8\pi)$, $S_{\mathrm{WZ}}[g]$ is the WZ integral defined by extending $g$ into an auxiliary third dimension, and $\eta=\pm1$ will be fixed later. 
This theory has two independent $SU(3)$ symmetries acting from the left and right.
Identifying them with the bra and ket color symmetries gives
\begin{equation} \label{eq:color-symmetry}
    g\mapsto V^\dagger gU,
    \qquad
    U\in SU(3)^{\mathrm{k}},\quad
    V\in SU(3)^{\mathrm{b}},
\end{equation}
so $g$ carries precisely the observed $(\mathbf{3},\ovl{\mathbf{3}})$ representation.

For $\eta=-1$, the ket $SU(3)$ Noether current $g^{-1}\partial_{\bar z}g$ is antiholomorphic, while the bra current $(\partial_zg)g^{-1}$ is holomorphic; $\eta=+1$ reverses their chiralities.
The antiholomorphic ket current carries quantum numbers $(\mathbf{8},\mathbf{1},0,-1)$ and has $\Delta=1$, matching the observed mode and thereby fixing $\eta=-1$.
Its negative $K_y$ also distinguishes it from the free-parton $(\mathbf{8},\mathbf{1})$ current, given by the traceless part of $\psi^\dagger_{\mathrm{k}\alpha} \psi_{\mathrm{k}\beta}$, which has $K_y=+1$.

This interpretation also accounts for the two mixed modes:
the product $\psi_{\mathrm{k}}^\dagger g$ contains a $(\ovl{\mathbf{3}},\ovl{\mathbf{3}})$ component with additive dimension $\Delta=1/2+2/3=7/6$, while $\psi_{\mathrm{k}}^\dagger\psi_{\mathrm{b}}g$ contains a $(\mathbf{6},\ovl{\mathbf{6}})$ component with $\Delta=1+2/3=5/3$.
These mixed modes further support a decoupled $U(3)_1\times SU(3)_1$ description.

A Dirac mode contributes central charge $c=1$, while the $SU(3)_1$ WZW sector contributes $c=2$; the suggested $U(3)_1\times SU(3)_1$ CFT therefore has a total central charge $c=5$~\cite{Witten1984Bosonization,Knizhnik1984WZW}.
As an independent check, a linear fit of the cylinder free-energy scaling~\cite{Blote1986CentralCharge,Affleck1986ConformalAnomaly}, $\frac{\ln\lambda_0}{L_2}=\mathrm{A}+\frac{\pi\sqrt{3}}{12L_2^2}c_{\rm fit}$, over $L_2=4$--$7$ gives $c_{\rm fit}\approx5.1$.

\paragraph{Field-theory derivation.}
We now provide an analytical argument for the emergent $U(3)_1\times SU(3)_1$ field theory. 
Since the local projector $P_{\rm G}$ is equivalent to an $SU(3)$ Haar average retaining the empty and three-parton singlets, we have
\begin{equation}
        \inn{\Psi|\Psi}
        =\det (S^\dagger S) \int\prod_{\mathbf r,a} \mathrm{d} g_{\mathbf r a}\,
        \frac{\det(S^\dagger g S)}{\det(S^\dagger S)} \ . 
    \label{eq:bulk-norm-factorization}
\end{equation}
The factor $\det(S^\dagger S)=\inn{\Phi|\Phi}$ contributes the $U(3)_1$ Dirac spectrum, as follows from the unprojected transfer matrix in Sec.~\ref{app:unprojected-transfer}.

The remaining factor $\mathrm{e}^{-\mathcal{S}_G[g]}=\frac{\det S^\dagger g S}{\det S^\dagger S}$ describes the collective color fluctuations.
This action is invariant under the independent $SU(3)^{\mathrm{k}}$ and $SU(3)^{\mathrm{b}}$ transformations defined in Eq.~\ref{eq:color-symmetry}.
A gradient expansion about a uniform $g$ yields the WZ action in Eq.~\ref{eq:bulk-color-action}, with the stiffness 
$\kappa = \frac1{48 \pi^2} \int \mathrm{d}^2\mathbf{k} \ \mathrm{tr}[\mathfrak{g}]$ and the coefficient $\eta = - C/3$. 
Here $\mathfrak {g}_{\mu\nu}=\frac12\Tr[(\partial_{k_\mu}P)(\partial_{k_\nu}P)]$ is the quantum metric of the occupied parton bands, and $C=3$ is the parton Chern number. 
In this expansion, each spatial derivative acting on $g$ is accompanied by a momentum derivative acting on the occupied-band projector $P(\mathbf k)$.
The symmetric and antisymmetric parts of the resulting quantum geometry tensor determine the stiffness and WZ term, respectively [see Sec.~\ref{app:bulk-cft}].
The leading omitted terms are $\mathcal{O}(|\mathbf q|^4\ln|\mathbf q|)$, where $\mathbf q$ is the fluctuation momentum.
The logarithmic correction arises from the gapless partons but remains subleading in the long-wave limit.

\begin{figure}[t]
    \centering
    \includegraphics[width=\linewidth]{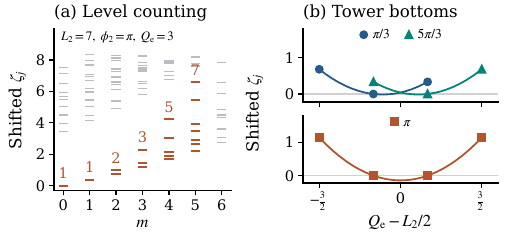}
    \caption{Half-cylinder entanglement spectrum at $L_2=7$.
    (a) Levels at $\phi_2=\pi$ and $Q_{\rm e}=3$. Energies are measured from the tower bottom, whose momentum is $K_{y,0}=5/2$. The horizontal axis shows $m=(K_{y,0}-K_y)\bmod L_2$. Numbers give the highlighted level counts.
    (b) Lowest level in each fitted charge sector, measured from the minimum at each twist. Points show data and curves are quadratic fits. At $\phi_2=\pi/3$ and $5\pi/3$ (top), both fits give $K=3.075$, with charge centers shifted by $-0.330$ and $+0.330$, respectively, from the center at $\phi_2=\pi$ (bottom, $K=3.057$).}
    \label{fig:transfer-entanglement}
\end{figure}

Remarkably, the ideal quantum geometry condition $\mathrm{tr}[\mathfrak{g}]=|\Omega(\mathbf{k})|$ would give precisely the WZW fixed-point stiffness $\kappa_\star=1/(8\pi)$.
The inequality $\mathrm{tr}[\mathfrak{g}] \ge |\Omega(\mathbf{k})|$ ensures $\kappa\geq\kappa_\star$, placing the theory on the side expected to flow to the $SU(3)_1$ WZW fixed point~\cite{Witten1984Bosonization,Demulder2018WZ}.
Thus, proximity to ideal quantum geometry places the bare theory close to this fixed point.

\paragraph{Entanglement spectrum.}
Having established charged topological order and bulk criticality, we probe edge properties through the entanglement spectrum.
The eigenstates of transfer matrix can be reshaped into matrices according to the tensor product structure of bra and ket [Fig.~\ref{fig:tns-transfer}(a)].
Denoting the reshaped dominant left and right eigenstates with eigenvalue $\lambda_0$ by $L_0$ and $R_0$, the reduced density matrix of a half-infinite cylinder can be written as $ \rho_0 =\frac{\sqrt{L_0} R_0 \sqrt{L_0} }{\mathrm{Tr}(L_0 R_0)}$ \cite{Cirac2011Boundary}. 
Since $\lambda_0$ lies in the singlet sector of both $SU(3)$ symmetries, $L_0$ and $R_0$ have support only on auxiliary states with integer electron numbers.
The eigenvalues $p_j$ of $\rho_0$ define the entanglement levels $\zeta_j=-\ln p_j$, labeled by electron charge $Q_{\rm e}$ and the dimensionless momentum $K_y$. 
Figure~\ref{fig:transfer-entanglement}(a) shows the $Q_{\rm e}=3$ spectrum at $L_2=7$ as a function of the momentum $m$ relative to the tower bottom. 
The lowest tower has $1,1,2,3,5,7$ levels at $m=0,\ldots,5$, respectively, as expected for the single chiral boson of a $U(1)_3$ edge. 

Following Refs.~\cite{Wen1990ChiralLuttinger,Qi2012EntanglementEdge}, we extract the Chern-Simons level $K$ from the parabolic dependence of the tower-bottom levels on $Q_{\rm e}$.
For all three boundary conditions, the fits yield $K\approx3$ [Fig.~\ref{fig:transfer-entanglement}(b)], consistent with the $U(1)_3$ topological order. 

\paragraph{Discussions.}
Unlike spin-chain realizations of WZW criticality \cite{affleck_critical_1988,itoi_extended_1997,Tu2014SUNWZW}, where independent left and right $SU(3)$ symmetries emerge only in the infrared, our transfer matrix has independent onsite ket and bra $SU(3)$ symmetries.
Their WZW anomalies are canceled by the opposite-chirality level-one Dirac currents \cite{Witten1984Bosonization}, allowing symmetry-preserving gapping in principle.
This is consistent with the parton gap opening without changing the Chern number \cite{Li2026StableTopology}.
However, the widely held belief that exact TNSs cannot represent gapped chiral states would, if correct, pin the wavefunction at criticality, manifested here in decoupled Dirac and WZW sectors.

Several open questions deserve further exploration:
(i) Although the many-body Chern number fixes the adiabatic limit of Hall response $\lim_{L\to\infty}\lim_{\omega\to0}\sigma_H=e^2/(3h)$ \cite{Niu1985ManyBodyChern}, whether this quantization survives in the dc transport limit $\lim_{\omega\to0}\lim_{L\to\infty}\sigma_H$ needs further investigation.
The robustness of the parton Chern number under generic perturbative gaps \cite{Li2026StableTopology} motivates the conjecture that it does.
(ii) The Dirac--WZW theory describes the wavefunction at this gap-closing point, raising further questions about the full $(2+1)$--dimensional criticality, including its dynamics. 
(iii) While the standard TNS construction yields a frustration-free parent Hamiltonian, its local terms can be complex, motivating the search for a simpler parent Hamiltonian for the FCSM.
(iv) More broadly, the variety of CTFBs offers a systematic route to exploring interplay between fractional topology and bulk criticality.

\paragraph{Acknowledgments.}
We thank Jiabin Yu for discussions on analytical lattice parent Hamiltonians for FCIs \cite{Zhao2026Analytical}.
We are also grateful to helpful discussions with Ady Stern, Yu-An Chen, Wei Zhu, and Nicolas Regnault. 
We used large language models to assist with coding, manuscript editing, and analytical derivations.
Z.-D.~S. and F.-J.~W. were supported by National Natural Science Foundation of China (No.~12625410, No.~12521006, No.~12274005), National Key Research and Development Program of China (No.~2021YFA1401900), and Innovation Program for Quantum Science and Technology (No.~2021ZD0302403).  

\bibliography{ref}

\clearpage

\onecolumngrid

\setcounter{section}{0}
\setcounter{subsection}{0}
\setcounter{subsubsection}{0}
\setcounter{figure}{0}
\setcounter{table}{0}
\setcounter{equation}{0}
\renewcommand{\thesection}{S\arabic{section}}
\renewcommand{\thesubsection}{\thesection.\arabic{subsection}}
\renewcommand{\thesubsubsection}{\thesubsection.\arabic{subsubsection}}
\renewcommand{\thefigure}{S\arabic{figure}}
\renewcommand{\thetable}{S\arabic{table}}
\renewcommand{\theequation}{S\arabic{equation}}
\renewcommand{\theHsection}{supp.\arabic{section}}
\renewcommand{\theHsubsection}{\theHsection.\arabic{subsection}}
\renewcommand{\theHsubsubsection}{\theHsubsection.\arabic{subsubsection}}
\renewcommand{\theHfigure}{supp.\arabic{figure}}
\renewcommand{\theHtable}{supp.\arabic{table}}
\renewcommand{\theHequation}{supp.\arabic{equation}}
\renewcommand{\figurename}{Fig.}
\renewcommand{\tablename}{Table}
\makeatletter
\def\p@subsection{}
\def\p@subsubsection{}
\def\@sectioncntformat#1{Sec.~\thesection.\quad}
\def\@hangfrom@section#1#2#3{\@hangfrom{#1#2}#3}
\makeatother
\pdfbookmark[0]{Supplementary Materials}{supplementary-materials}
\begin{center}
    \large\bfseries Supplementary Materials
\end{center}

\input{appendix}

\end{document}

%% file: appendix.tex
\FloatBarrier
\begin{center}
\begin{minipage}{0.94\textwidth}
\centering\textbf{Contents}\par\vspace{0.6em}
\begingroup
\small
\setlength{\tabcolsep}{0.25em}
\renewcommand{\arraystretch}{1.12}
\begin{tabular}{llr}
 & & \textbf{Page}\\[2pt]
\hline\\[-0.5em]
\hyperref[app:twist-rank]{\textbf{\ref*{app:twist-rank}}} & \parbox[t]{\dimexpr\linewidth-9em\relax}{\hyperref[app:twist-rank]{\textbf{Twist states and local indistinguishability}}\nobreak\hspace{0.5em}\dotfill} & \pageref{app:twist-rank}\\
\hspace*{0.9em}\hyperref[app:gram-spectrum-data]{\ref*{app:gram-spectrum-data}} & \parbox[t]{\dimexpr\linewidth-9em\relax}{\hspace*{0.7em}\hyperref[app:gram-spectrum-data]{Gram spectrum of the nine twist states}\nobreak\hspace{0.5em}\dotfill} & \pageref{app:gram-spectrum-data}\\
\hspace*{0.9em}\hyperref[app:local-indistinguishability-data]{\ref*{app:local-indistinguishability-data}} & \parbox[t]{\dimexpr\linewidth-9em\relax}{\hspace*{0.7em}\hyperref[app:local-indistinguishability-data]{Local indistinguishability}\nobreak\hspace{0.5em}\dotfill} & \pageref{app:local-indistinguishability-data}\\
\noalign{\vskip 0.5em}
\hyperref[app:tns-construction]{\textbf{\ref*{app:tns-construction}}} & \parbox[t]{\dimexpr\linewidth-9em\relax}{\hyperref[app:tns-construction]{\textbf{Transfer matrix}}\nobreak\hspace{0.5em}\dotfill} & \pageref{app:tns-construction}\\
\hspace*{0.9em}\hyperref[app:tns-column-transfer]{\ref*{app:tns-column-transfer}} & \parbox[t]{\dimexpr\linewidth-9em\relax}{\hspace*{0.7em}\hyperref[app:tns-column-transfer]{Transfer matrix from column contractions}\nobreak\hspace{0.5em}\dotfill} & \pageref{app:tns-column-transfer}\\
\hspace*{0.9em}\hyperref[app:transfer-quantum-numbers]{\ref*{app:transfer-quantum-numbers}} & \parbox[t]{\dimexpr\linewidth-9em\relax}{\hspace*{0.7em}\hyperref[app:transfer-quantum-numbers]{Cut quantum numbers and local insertions}\nobreak\hspace{0.5em}\dotfill} & \pageref{app:transfer-quantum-numbers}\\
\hspace*{0.9em}\hyperref[app:unprojected-transfer]{\ref*{app:unprojected-transfer}} & \parbox[t]{\dimexpr\linewidth-9em\relax}{\hspace*{0.7em}\hyperref[app:unprojected-transfer]{Transfer spectrum of the unprojected state}\nobreak\hspace{0.5em}\dotfill} & \pageref{app:unprojected-transfer}\\
\noalign{\vskip 0.5em}
\hyperref[app:wilson-modular-data]{\textbf{\ref*{app:wilson-modular-data}}} & \parbox[t]{\dimexpr\linewidth-9em\relax}{\hyperref[app:wilson-modular-data]{\textbf{Wilson loops and modular matrices}}\nobreak\hspace{0.5em}\dotfill} & \pageref{app:wilson-modular-data}\\
\hspace*{0.9em}\hyperref[app:wilson-numerics]{\ref*{app:wilson-numerics}} & \parbox[t]{\dimexpr\linewidth-9em\relax}{\hspace*{0.7em}\hyperref[app:wilson-numerics]{Finite-size convergence of the Wilson loops}\nobreak\hspace{0.5em}\dotfill} & \pageref{app:wilson-numerics}\\
\hspace*{0.9em}\hyperref[app:modular-extraction]{\ref*{app:modular-extraction}} & \parbox[t]{\dimexpr\linewidth-9em\relax}{\hspace*{0.7em}\hyperref[app:modular-extraction]{Modular matrices from Wilson loops and sixfold rotation}\nobreak\hspace{0.5em}\dotfill} & \pageref{app:modular-extraction}\\
\noalign{\vskip 0.5em}
\hyperref[app:rotation-wilson-proof]{\textbf{\ref*{app:rotation-wilson-proof}}} & \parbox[t]{\dimexpr\linewidth-9em\relax}{\hyperref[app:rotation-wilson-proof]{\textbf{Twofold rotation and Wilson-loop constraints}}\nobreak\hspace{0.5em}\dotfill} & \pageref{app:rotation-wilson-proof}\\
\hspace*{0.9em}\hyperref[app:c2-wilson-loops]{\ref*{app:c2-wilson-loops}} & \parbox[t]{\dimexpr\linewidth-9em\relax}{\hspace*{0.7em}\hyperref[app:c2-wilson-loops]{Wilson-loop reversal under twofold rotation}\nobreak\hspace{0.5em}\dotfill} & \pageref{app:c2-wilson-loops}\\
\hspace*{0.9em}\hyperref[app:c2-wilson-determinants]{\ref*{app:c2-wilson-determinants}} & \parbox[t]{\dimexpr\linewidth-9em\relax}{\hspace*{0.7em}\hyperref[app:c2-wilson-determinants]{Wilson-loop determinants from rotation eigenvalues}\nobreak\hspace{0.5em}\dotfill} & \pageref{app:c2-wilson-determinants}\\
\noalign{\vskip 0.5em}
\hyperref[app:bulk-cft]{\textbf{\ref*{app:bulk-cft}}} & \parbox[t]{\dimexpr\linewidth-9em\relax}{\hyperref[app:bulk-cft]{\textbf{Bulk field theory of the projected norm}}\nobreak\hspace{0.5em}\dotfill} & \pageref{app:bulk-cft}\\
\hspace*{0.9em}\hyperref[app:bulk-norm-construction]{\ref*{app:bulk-norm-construction}} & \parbox[t]{\dimexpr\linewidth-9em\relax}{\hspace*{0.7em}\hyperref[app:bulk-norm-construction]{Fermion and color sectors of the projected norm}\nobreak\hspace{0.5em}\dotfill} & \pageref{app:bulk-norm-construction}\\
\hspace*{0.9em}\hyperref[app:bulk-gradient-action]{\ref*{app:bulk-gradient-action}} & \parbox[t]{\dimexpr\linewidth-9em\relax}{\hspace*{0.7em}\hyperref[app:bulk-gradient-action]{Color action and the WZW fixed point}\nobreak\hspace{0.5em}\dotfill} & \pageref{app:bulk-gradient-action}\\
\hspace*{0.9em}\hyperref[app:bulk-transfer-energy-momentum]{\ref*{app:bulk-transfer-energy-momentum}} & \parbox[t]{\dimexpr\linewidth-9em\relax}{\hspace*{0.7em}\hyperref[app:bulk-transfer-energy-momentum]{CFT levels and their transfer eigenvalues}\nobreak\hspace{0.5em}\dotfill} & \pageref{app:bulk-transfer-energy-momentum}\\
\hspace*{0.9em}\hyperref[app:bulk-gradient-convergence]{\ref*{app:bulk-gradient-convergence}} & \parbox[t]{\dimexpr\linewidth-9em\relax}{\hspace*{0.7em}\hyperref[app:bulk-gradient-convergence]{Error bounds for the gradient expansion}\nobreak\hspace{0.5em}\dotfill} & \pageref{app:bulk-gradient-convergence}\\
\noalign{\vskip 0.5em}
\hyperref[app:transfer-spectrum-data]{\textbf{\ref*{app:transfer-spectrum-data}}} & \parbox[t]{\dimexpr\linewidth-9em\relax}{\hyperref[app:transfer-spectrum-data]{\textbf{Finite-size scaling of the transfer spectrum}}\nobreak\hspace{0.5em}\dotfill} & \pageref{app:transfer-spectrum-data}\\
\noalign{\vskip 0.5em}
\hyperref[app:entanglement-spectrum]{\textbf{\ref*{app:entanglement-spectrum}}} & \parbox[t]{\dimexpr\linewidth-9em\relax}{\hyperref[app:entanglement-spectrum]{\textbf{Chiral towers in the entanglement spectrum}}\nobreak\hspace{0.5em}\dotfill} & \pageref{app:entanglement-spectrum}\\
\end{tabular}
\endgroup
\end{minipage}
\end{center}
\clearpage

\section{Twist states and local indistinguishability}
\label{app:twist-rank}

\subsection{Gram spectrum of the nine twist states}
\label{app:gram-spectrum-data}

We evaluate the full Gram matrix on $L\times L$ tori with antiperiodic physical boundary conditions along both cycles,
using all nine parton twists $(\phi_1,\phi_2)=(\pi/3+2\pi m/3,\pi/3+2\pi n/3)$ with $m,n=0,1,2$.
All nine momentum meshes avoid $\Gamma$, so all the projected states acquire TNS representations with the same occupation orbitals, and the twisting phases are placed on the boundary bonds [Fig.~\ref{fig:mean-field}(a)].
The rank of the Gram matrix $G_{st}=\langle\bar\Psi_s|\bar\Psi_t\rangle$ of the normalized twist states counts the linearly independent states, and $\Tr G=9$.

We compute the overlaps by exact enumeration for $L=2,3$ and by variational Monte Carlo at larger sizes.
Let $\mathcal X$ denote the set of $L^2$ sites occupied by electrons on the $2L^2$-site torus.
Write $\Psi_s(\mathcal X)$ for the unnormalized projected amplitude at twist $s$. It is the cube of its single-color Slater determinant.
A local Metropolis proposal moves one electron to an empty site. We sample with weight $q(\mathcal X)=\sum_s c_s|\Psi_s(\mathcal X)|^2$, where the sum runs over the states sampled together.
Choosing $c_s\propto\langle\Psi_s|\Psi_s\rangle^{-1}$ would give every twist state equal weight in the mixture, regardless of its normalization.
To estimate the weights, we first set all $c_s=1$ and average $|\Psi_s(\mathcal X)|^2/q(\mathcal X)$ after equilibration.
The resulting $\hat f_s$ estimates the norm fraction $\langle\Psi_s|\Psi_s\rangle/\sum_t\langle\Psi_t|\Psi_t\rangle$. We use the inverse estimated fractions to set the weights for the measurement runs.
Averaging with probability proportional to $q(\mathcal X)$ gives the normalized overlap
\begin{equation}
 G_{st}=
 \frac{
   \left\langle\dfrac{\sqrt{c_s c_t}\,\Psi_s^*(\mathcal X)\Psi_t(\mathcal X)}{q(\mathcal X)}\right\rangle
 }{
   \sqrt{\left\langle\dfrac{c_s|\Psi_s(\mathcal X)|^2}{q(\mathcal X)}\right\rangle
         \left\langle\dfrac{c_t|\Psi_t(\mathcal X)|^2}{q(\mathcal X)}\right\rangle}
 }
 =\frac{\langle\Psi_s|\Psi_t\rangle}
        {\sqrt{\langle\Psi_s|\Psi_s\rangle\langle\Psi_t|\Psi_t\rangle}} ,
 \label{eq:sm-vmc-normalized-overlap}
\end{equation}
where $\langle\cdot\rangle$ denotes the average over the sampled configurations.

\begin{figure}[htbp]
    \centering
    \includegraphics[width=0.84\textwidth]{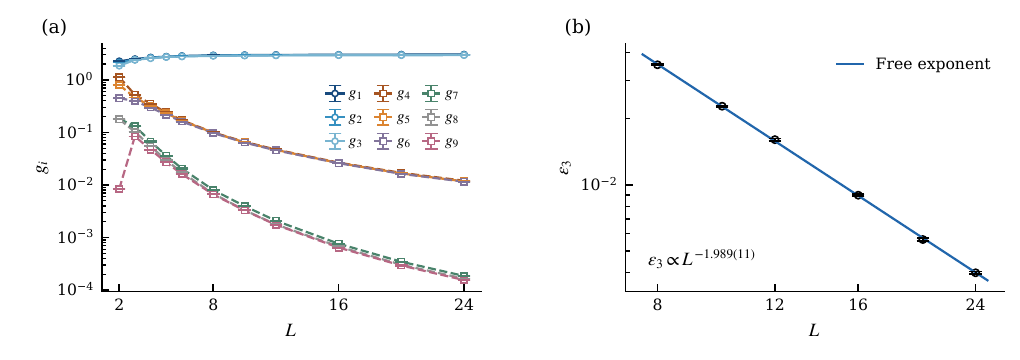}
    \caption{Gram spectrum of nine center-twist states with antiperiodic physical boundaries along both cycles.
    (a) Three leading eigenvalues (circles) and six subleading eigenvalues (squares).
    (b) Discarded weight $\varepsilon_3$, with a power-law fit over $L=8$--$24$ (solid line).
    Error bars are one jackknife standard error and can be smaller than the markers; $L=2,3$ use exact enumeration. The sampling series is independent of the Wilson-loop runs.}
    \label{fig:modular-scaling}
\end{figure}

Denote the nine eigenvalues of $G$ as $g_1\geq\cdots\geq g_9$.
Fig.~\ref{fig:modular-scaling}(a) shows that $g_{1,2,3}$ are nearly degenerate and constant as $L$ increases, while the other six decrease.
To quantify this separation, we calculate the normalized weight $\varepsilon_3=\frac{1}{9}\sum_{j=4}^{9}g_j$ outside the leading three-dimensional subspace.
At $L=24$, the three largest eigenvalues are $3.025(12)$, $2.989(12)$, and $2.951(13)$, while $\varepsilon_3=0.003977(58)$, i.e.  $g_{1,2,3}$ retain $99.602(6)\%$ of the total Gram weight.
Over $L=8$--$24$, fitting $\varepsilon_3\propto L^{-p}$ gives $p=1.989(11)$, with $\chi^2/\mathrm{dof}=4.22/4$.
The data therefore support a three-dimensional limiting space, with an approximately $L^{-2}$ finite-size correction.
For each $L>3$, we use four independent Monte Carlo chains of 8000 measured sweeps each, with $L^2$ proposals per sweep.
We repeat normalization and diagonalization after deleting each block of 100, 200, or 500 consecutive sweeps within a chain, and after deleting each whole chain. For each observable, we report the largest of the four jackknife standard errors.

\subsection{Local indistinguishability}
\label{app:local-indistinguishability-data}

Let $|\chi_a\rangle$, $a=0,1,2$, be the orthonormal states obtained from the three largest Gram eigenvalues. For a fixed local region $A$, the local indistinguishability requires
\begin{equation}
    \rho_A^{ab}\equiv\operatorname{Tr}_{\bar A}|\chi_a\rangle\langle\chi_b|
    \longrightarrow\delta_{ab}\rho_A,
    \qquad L\longrightarrow\infty,
    \label{eq:app-local-indistinguishability}
\end{equation}
where $\bar A$ is the complement.
The diagonal condition requires local operators in region $A$ to have equal expectation values in the three states; the off-diagonal condition requires the local operators to induce vanishing tunneling between them. 
We measure deviations from these conditions by
\begin{equation}
    \delta_A^2=\frac{1}{3}\sum_{a,b=0}^2
    \left\|\rho_A^{ab}-\delta_{ab}\left(\frac{1}{3}\sum_{a=0}^2\rho_A^{aa}\right)\right\|_{\rm F}^2
    \label{eq:app-local-deviation}
\end{equation}
where $\|M\|_{\rm F}^2=\sum_{i,j}|M_{ij}|^2$ is the Frobenius norm.

\begin{figure}[htbp]
    \centering
    \includegraphics[width=0.88\textwidth]{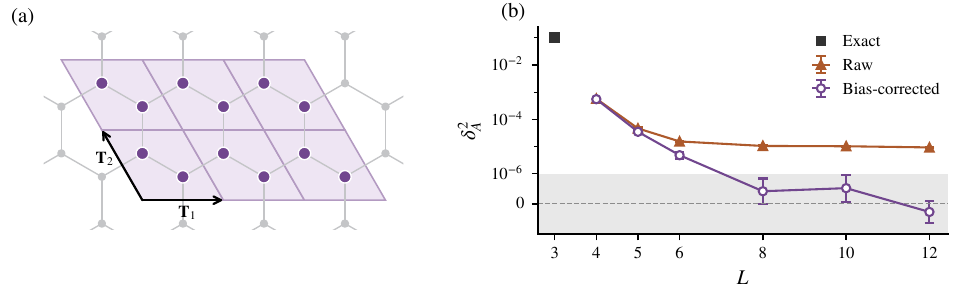}
    \caption{(a) The twelve highlighted sites form $A$. Each shaded cell contains both honeycomb sublattices.
    (b) Raw and jackknife-corrected estimates of $\delta_A^2$ for the same samples. The square at $L=3$ is the exact value. Each curve shows one jackknife standard error at the block length selected for that size.
    The gray band marks the linear range of the vertical axis, $|\delta_A^2|\leq10^{-6}$. Outside the band the scale is logarithmic.}
    \label{fig:app-local-indistinguishability}
\end{figure}

Region $A$ contains the twelve honeycomb sites in a $3\times2$ block of unit cells [Fig.~\ref{fig:app-local-indistinguishability}(a)]. The cells lie at $m\mathbf T_1+n\mathbf T_2$ with $m=0,1,2$ and $n=0,1$.
We use the nine parton twists and antiperiodic physical boundary conditions of Sec.~\ref{app:gram-spectrum-data}. For each of the $\sum_{p=0}^{12}\binom{12}{p}^2=2\,704\,156$ local matrix elements that conserve the particle number $p$, we compute the $9\times9$ matrix between twist states and transform it to the three-state basis $|\chi_a\rangle$.

At $L=3$, the block wraps around the torus along $\mathbf T_1$, and exact enumeration gives $\delta_A^2=0.10207$.
For $L\geq4$, we use the variational Monte Carlo method of Sec.~\ref{app:gram-spectrum-data}, keeping $c_s=1$ for all nine twists throughout sampling.
For each sample, we hold the occupations outside $A$ fixed and enumerate all configurations inside $A$ with the required particle number. We average the resulting matrices over six placements of $A$ at $L=4$ and nine selected placements at each larger size.

Figure~\ref{fig:app-local-indistinguishability}(b) shows two estimates of $\delta_A^2$ from four chains of 8000 measured sweeps each. The raw estimate uses the sample-averaged matrices in Eq.~\ref{eq:app-local-deviation}. The corrected estimate subtracts the finite-sample bias estimated by a jackknife with 200-sweep blocks. We compute the two standard errors separately, repeating the bias correction in each resample for the corrected curve. At each $L$, both error bars use the block length that gives the largest standard error for the corrected estimate among 200, 400, 800, 1600, and 8000 sweeps.

\FloatBarrier

\section{Transfer matrix}
\label{app:tns-construction}
We express the projected norm as a matrix element in an auxiliary Fock space and organize it by physical columns to obtain the transfer matrix $\mathcal T$. Throughout Sec.~\ref{app:tns-construction}, both $|\Phi\rangle$ and $|\Psi\rangle$ are unnormalized.

\subsection{Transfer matrix from column contractions}
\label{app:tns-column-transfer}

We introduce two copies of the auxiliary fermions, with creation operators $d_{\mathbf r,\alpha,\tau}^\dagger$ at each orbital center $\mathbf r$, where $\tau=\mathrm{k},\mathrm{b}$ labels the ket and bra layers, respectively.
For each layer, define the linear combinations
\begin{equation}
 K_{\mathbf R,\alpha,\mathrm k}=\sum_{\mathbf r}s_{\mathbf R,\mathbf r}\,d_{\mathbf r,\alpha,\mathrm k},
 \qquad K_{\mathbf R,\alpha,\mathrm b}^\dagger=\sum_{\mathbf r}s_{\mathbf R,\mathbf r}^*\,d_{\mathbf r,\alpha,\mathrm b}^\dagger .
\end{equation}
The physical sites $\mathbf R$ form a honeycomb lattice, with two sites $\mathbf R=\mathbf r+\mathbf d_a$ in each unit cell centered at $\mathbf r$, where $a=1,2$ labels the sublattice [Fig.~\ref{fig:mean-field}(a)].
The coefficients are those of the compact orbitals,
\begin{equation}
 \mathcal S_n^\dagger(\mathbf r)=\sum_{\mathbf R,\alpha} S_{(\mathbf R,\alpha),(\mathbf r,n)} f_{\mathbf R,\alpha}^\dagger,
 \qquad S_{(\mathbf R,\alpha),(\mathbf r,n)}=\delta_{\alpha n}s_{\mathbf R,\mathbf r} .
 \label{eq:app-tns-orbital}
\end{equation}
On the six neighboring honeycomb sites $\mathbf R\in\mathcal N(\mathbf r)$, $s_{\mathbf R,\mathbf r}=\mathrm{e}^{\mathrm{i}(j-1)\pi/3}$ with $j=1,\ldots,6$ increasing counterclockwise from $\mathbf r+\mathbf d_1$, and the amplitude vanishes elsewhere [Fig.~\ref{fig:mean-field}(a)].
A parton twist $\boldsymbol\phi$ multiplies this amplitude by $\mathrm{e}^{-\mathrm{i}\sum_\mu n_{\mathbf T_\mu}\phi_\mu}$, where $n_{\mathbf T_\mu}$ counts signed boundary crossings from $\mathbf r$ to $\mathbf R$ and the physical twist is $\varphi_\mu=3\phi_\mu$ modulo $2\pi$.

Let $\sigma$ be a set of $N=L_1L_2$ occupied physical sites.
We define $S[\sigma]$ as the submatrix of $S$ formed by the rows $(\mathbf R,\alpha)$ with $\mathbf R\in\sigma$ and $\alpha=1,2,3$.
Since each site contributes three rows, $S[\sigma]$ is a $3N\times3N$ square matrix.
The Slater-determinant identity then gives
\begin{equation}
 \begin{aligned}
 \prod_{\mathbf R\in\sigma}^{\rightarrow}\prod_\alpha^{\rightarrow} K_{\mathbf R,\alpha,\mathrm k}
 &=\det S[\sigma]\prod_{\mathbf r,\alpha}^{\rightarrow} d_{\mathbf r,\alpha,\mathrm k},\\
 \prod_{\mathbf R\in\sigma}^{\leftarrow}\prod_\alpha^{\leftarrow} K^\dagger_{\mathbf R,\alpha,\mathrm b}
 &=(\det S[\sigma])^*\prod_{\mathbf r,\alpha}^{\leftarrow} d^\dagger_{\mathbf r,\alpha,\mathrm b}.
 \end{aligned}
\end{equation}
Here $\prod^{\leftarrow}$ places smaller labels farther to the right and $\prod^{\rightarrow}$ places them farther to the left.
We order the $K$ labels lexicographically by $(\ell_1,\ell_2,a,\alpha,\tau)$ and the $d$ labels by $(\ell_1,\ell_2,\alpha,\tau)$, with $\mathrm k$ before $\mathrm b$, $\mathbf r_{\ell_1\ell_2}=\ell_1\mathbf T_1+\ell_2\mathbf T_2$, and $\mathbf R_{\ell_1\ell_2 a}=\mathbf r_{\ell_1\ell_2}+\mathbf d_a$.

The norm can equivalently be written as the following matrix element in the auxiliary Fock space,
\begin{equation}
 \langle\Psi|\Psi\rangle=\langle \mathbf1_{\mathrm b}|\prod_{\mathbf R}^{\leftarrow}
 \left(1+\prod_\alpha^{\leftarrow}(K_{\mathbf R,\alpha,\mathrm b}^\dagger K_{\mathbf R,\alpha,\mathrm k})\right)|\mathbf1_{\mathrm k}\rangle .
 \label{eq:app-tns-network}
\end{equation}
Here $|\mathbf1_\tau\rangle$ has all $3N$ auxiliary modes in layer $\tau$ occupied and the other layer empty,
\begin{equation}
 |\mathbf1_\tau\rangle=\prod_{\mathbf r,\alpha}^{\leftarrow}d_{\mathbf r,\alpha,\tau}^\dagger|\mathbf0\rangle ,
 \label{eq:app-tns-center}
\end{equation}
where $|\mathbf0\rangle$ has both layers empty.
To obtain a nonzero overlap with $\langle \mathbf1_{\mathrm b}|$, the operator product must transfer all $3N$ particles from ket to bra.
Each site factor transfers either zero or three particles, so exactly $N$ sites must contribute their six-operator term.
Expanding the product gives
\begin{equation}
 \sum_{|\sigma|=N}\langle \mathbf1_{\mathrm b}|
 \left(\prod_{\mathbf R\in\sigma}^{\leftarrow}\prod_\alpha^{\leftarrow}K_{\mathbf R,\alpha,\mathrm b}^\dagger\right)
 \left(\prod_{\mathbf R\in\sigma}^{\rightarrow}\prod_\alpha^{\rightarrow}K_{\mathbf R,\alpha,\mathrm k}\right)|\mathbf1_{\mathrm k}\rangle .
\end{equation}
Moving the ket operators to the right and reversing their order give the same permutation sign, so the two signs cancel.
For a fixed $\sigma$, the two layer contractions give $\det S[\sigma]$ and its complex conjugate.
Summing over all sets $\sigma$ reproduces $\langle\Psi|\Psi\rangle=\sum_{|\sigma|=N}|\det S[\sigma]|^2$.

To proceed, we define the column operator
\begin{equation}
 \mathbb B_{\ell_1}=\prod_{\mathbf R\in\mathcal C_{\ell_1}}^{\leftarrow}
 \left(1+\prod_\alpha^{\leftarrow}(K_{\mathbf R,\alpha,\mathrm b}^\dagger K_{\mathbf R,\alpha,\mathrm k})\right),
 \qquad \mathcal C_{\ell_1}=\{\mathbf R_{\ell_1\ell_2 a}\mid 0\leq\ell_2<L_2,\ a=1,2\} .
 \label{eq:app-column-operator}
\end{equation}
It contains auxiliary operators only on cuts $\ell_1$ and $\ell_1+1$.
To evaluate $\langle \mathbf1_{\mathrm b}|\cdots\mathbb B_{\ell_1+1}\mathbb B_{\ell_1}\mathbb B_{\ell_1-1}\cdots|\mathbf1_{\mathrm k}\rangle$ on a cylinder open along $\mathbf T_1$, we insert a complete Fock basis between each pair of neighboring column operators.
Consider the cut between $\mathbb B_{\ell_1}$ and $\mathbb B_{\ell_1-1}$.
Every cut $\ell'>\ell_1$ is still in $|\mathbf1_{\mathrm k,\ell'}\rangle$ because the columns already contracted do not act there.
Likewise, every cut $\ell'<\ell_1$ must already be in $|\mathbf1_{\mathrm b,\ell'}\rangle$ because the remaining columns cannot act there.
On one or more cuts, fix the phases of the occupation states by
$|\mathbf n\rangle=\prod_i^{\leftarrow}(d^\dagger_{i,\mathrm b})^{n_{i,\mathrm b}}(d_{i,\mathrm k})^{1-n_{i,\mathrm k}}|\mathbf1_{\mathrm k}\rangle$, with the ordered product over all orbital-color labels $i$ on those cuts.
Let $<$ denote the input cut $\ell_1$ with occupations $\mathbf n$, and $>$ the output cut $\ell_1+1$ with occupations $\mathbf m$. Then
\begin{equation}
\begin{aligned}
 \mathcal T_{\ell_1}[\mathbf m,\mathbf n]
 &\equiv\langle\cdots\mathbf1_{\mathrm k,\ell_1+2},\mathbf m_{\ell_1+1},\mathbf1_{\mathrm b,\ell_1},\cdots|
 \mathbb B_{\ell_1}|\cdots\mathbf1_{\mathrm k,\ell_1+1},\mathbf n_{\ell_1},\mathbf1_{\mathrm b,\ell_1-1},\cdots\rangle\\
 &=\langle\mathbf m_{>},\mathbf1_{\mathrm b,<}|\mathbb B_{\ell_1}|\mathbf1_{\mathrm k,>},\mathbf n_{<}\rangle .
\end{aligned}
 \label{eq:app-tns-orbital-column-split}
\end{equation}
In $|\mathbf1_{\mathrm k,>},\mathbf n_<\rangle$ the input operators cross $3L_2$ ket creators on cut $>$, giving $(-1)^{3L_2(|\mathbf n_{\mathrm k}^c|+|\mathbf n_{\mathrm b}|)}$ relative to $|\mathbf1_{\mathrm k,>}\rangle\otimes|\mathbf n\rangle_<$, where $\mathbf n_\tau^c$ denotes the unoccupied modes. The omitted modes have identical input and output occupations, and $\mathbb B_{\ell_1}$ does not act on them.
Taking $\mathbf m$ as the output label and $\mathbf n$ as the input label gives the $2^{6L_2}\times2^{6L_2}$ transfer matrix $\mathcal T_{\ell_1}$ on the cut Fock space $\mathcal H_{\rm cut}=\mathcal F_{\mathrm k}\otimes\mathcal F_{\mathrm b}$, with $3L_2$ modes in each layer.
For open boundary conditions the norm takes the form
\begin{equation}
 \langle\Psi|\Psi\rangle=\boldsymbol B_{\infty} \cdot \mathcal T \cdot \mathcal T  \cdots\mathcal T \cdot \boldsymbol B_{0} ,
 \label{eq:app-tns-contracted-column-product}
\end{equation}
where $\boldsymbol B_0$ and $\boldsymbol B_\infty$ define the input and output boundary conditions along $\mathbf T_1$ in the sector $Q=0$ [Eq.~\ref{eq:app-cut-charge}]. For identical columns, $\mathcal T_{\ell_1}=\mathcal T$.

Within each cut and layer $\tau$, the modes are labelled by $i=(\ell_2,\alpha)$, and the input and output occupations are denoted by $\mathbf n_\tau$ and $\mathbf m_\tau$.
To separate the bra and ket layers, we move the fermion-operator blocks into left-to-right order $\mathrm k,\mathrm b$, then $\ell_1+1,\ell_1$, keeping descending lexicographic order of $i$ within each block.
Within one cut, a ket hole $i$ crosses an occupied bra mode $j$ when $i\leq j$, including equal labels because the bra operator originally precedes the ket operator at each mode.
This gives $\zeta(\mathbf n)=(-1)^{\sum_{i\leq j}(1-n_{i,\mathrm k})n_{j,\mathrm b}}$.
We denote the occupation counts by $|\mathbf n_\tau|=\sum_i n_{i,\tau}$ and $|\mathbf m_\tau|=\sum_i m_{i,\tau}$.
The input state therefore satisfies
\begin{equation}
 |\mathbf1_{\mathrm k,>},\mathbf n_<\rangle=\zeta(\mathbf n)
 |\mathbf1_{\mathrm k,>},\mathbf n_{\mathrm k,<},\mathbf0_{\mathrm b,>},\mathbf n_{\mathrm b,<}\rangle .
\end{equation}
The cross-cut exchange gives no sign because $\mathbf0_{\mathrm b}$ is empty.
For the output, grouping the completed cut gives $\zeta(\mathbf1_{\mathrm b})=(-1)^{\binom{3L_2+1}{2}}$, and exchanging $\mathbf m_{\mathrm b}$ with the $3L_2$ ket holes gives $(-1)^{3L_2|\mathbf m_{\mathrm b}|}$, so
\begin{equation}
 \langle\mathbf m_>,\mathbf1_{\mathrm b,<}|=\zeta(\mathbf m)\zeta(\mathbf1_{\mathrm b})(-1)^{3L_2|\mathbf m_{\mathrm b}|}
 \langle\mathbf m_{\mathrm k,>},\mathbf0_{\mathrm k,<},\mathbf m_{\mathrm b,>},\mathbf1_{\mathrm b,<}| .
\end{equation}

With the original interleaved ordering of the $K_{\mathrm b}^\dagger K_{\mathrm k}$ pairs, expanding the column operator gives
\begin{equation}
 \mathbb B_{\ell_1}=\sum_{\sigma_{\ell_1}\subseteq\mathcal C_{\ell_1}}
 \prod_{\mathbf R\in\sigma_{\ell_1}}^{\leftarrow}\prod_{\alpha}^{\leftarrow}
 (K^\dagger_{\mathbf R,\alpha,\mathrm b}K_{\mathbf R,\alpha,\mathrm k}) .
\end{equation}
For a fixed $\sigma_{\ell_1}$, there are $3|\sigma_{\ell_1}|$ operators in each layer.
Moving all ket operators to the left requires $\binom{3|\sigma_{\ell_1}|+1}{2}$ exchanges.
To separate the two layer contractions, the bra operator block must also pass the $|\mathbf n_{\mathrm k}^c|$ ket annihilation operators in the input state, giving $(-1)^{3|\sigma_{\ell_1}||\mathbf n_{\mathrm k}^c|}$.
Combining the boundary and operator signs, each term in $\mathcal T_{\ell_1}$ factorizes into a ket matrix element and a bra matrix element,
\begin{align}
 \mathcal T_{\ell_1}[\mathbf m,\mathbf n]
 =\zeta(\mathbf m)\zeta(\mathbf n)\zeta(\mathbf1_{\mathrm b})(-1)^{3L_2|\mathbf m_{\mathrm b}|}
 \sum_{\sigma_{\ell_1}}(-1)^{\binom{3|\sigma_{\ell_1}|+1}{2}+3|\sigma_{\ell_1}||\mathbf n_{\mathrm k}^c|} 
 \times A^{\sigma_{\ell_1}}[\mathbf m_{\mathrm k}^c,\mathbf n_{\mathrm k}^c]
 A^{\sigma_{\ell_1}}[\mathbf m_{\mathrm b},\mathbf n_{\mathrm b}]^* .
\end{align}
The single-layer contractions are indexed by ket holes and bra particles, respectively,
\begin{equation}
 \begin{aligned}
 A^{\sigma_{\ell_1}}_{\mathbf m_{\mathrm k}^c,\mathbf n_{\mathrm k}^c}
 &=\langle\mathbf m_{\mathrm k,>},\mathbf0_{\mathrm k,<}|
 \prod_{\mathbf R\in\sigma_{\ell_1}}^{\leftarrow}\prod_\alpha^{\leftarrow}K_{\mathbf R,\alpha,\mathrm k}
 |\mathbf1_{\mathrm k,>},\mathbf n_{\mathrm k,<}\rangle,\\
 \bigl(A^{\sigma_{\ell_1}}_{\mathbf m_{\mathrm b},\mathbf n_{\mathrm b}}\bigr)^*
 &=\langle\mathbf m_{\mathrm b,>},\mathbf1_{\mathrm b,<}|
 \prod_{\mathbf R\in\sigma_{\ell_1}}^{\leftarrow}\prod_\alpha^{\leftarrow}K^\dagger_{\mathbf R,\alpha,\mathrm b}
 |\mathbf0_{\mathrm b,>},\mathbf n_{\mathrm b,<}\rangle .
 \end{aligned}
\end{equation}
The ket transfers $3|\sigma_{\ell_1}|$ particles to the bra, while the new and completed cuts carry $3L_2$ ket and bra particles, respectively, so a nonzero contraction requires
\begin{equation}
 |\mathbf n_{\mathrm k}|-|\mathbf m_{\mathrm k}|=|\mathbf m_{\mathrm b}|-|\mathbf n_{\mathrm b}|=3|\sigma_{\ell_1}|-3L_2 .
 \label{eq:app-column-charge-flow}
\end{equation}
Using this relation for the bra layer and the cut charge $Q=|\mathbf n_{\mathrm k}|+|\mathbf n_{\mathrm b}|-3L_2$ of Eq.~\ref{eq:app-cut-charge}, the operator and filled-cut signs satisfy
\begin{equation}
 \binom{3L_2+1}{2}+3L_2|\mathbf m_{\mathrm b}|+\binom{3|\sigma_{\ell_1}|+1}{2}+3|\sigma_{\ell_1}||\mathbf n_{\mathrm k}^c|
 \equiv \binom{|\mathbf m_{\mathrm b}|+1}{2}+\binom{|\mathbf n_{\mathrm b}|+1}{2}+3|\sigma_{\ell_1}|Q\pmod 2 .
\end{equation}
Combining the first two signs on the right with $\zeta(\mathbf m)\zeta(\mathbf n)$ gives the boundary phase $\xi(\mathbf n)=\zeta(\mathbf n)(-1)^{\binom{|\mathbf n_{\mathrm b}|+1}{2}}$ on each cut.
For a boundary vector $v[\mathbf n]$ in the occupation basis, define its rephased coefficient matrix by $X_Q[\mathbf n_{\mathrm k}^c,\mathbf n_{\mathrm b}]=\xi(\mathbf n)v[\mathbf n]$.
One application of $\mathcal T_{\ell_1}$ is then evaluated as
\begin{equation}
 X'_Q=\sum_{\sigma_{\ell_1}}(-1)^{|\sigma_{\ell_1}|Q}A^{\sigma_{\ell_1}}X_Q \bigl(A^{\sigma_{\ell_1}}\bigr)^\dagger,
 \qquad (\mathcal T_{\ell_1}v)[\mathbf m]=\xi(\mathbf m)X'_Q[\mathbf m_{\mathrm k}^c,\mathbf m_{\mathrm b}] .
 \label{eq:app-cut-eigenmatrix}
\end{equation}
Taking the Hermitian conjugate exchanges the row and column occupations, so $X_Q^\dagger$ belongs to the $-Q$ sector.
The transfer sign has the same parity in both sectors, so an eigenmatrix $X_Q$ with eigenvalue $\lambda$ gives an eigenmatrix $X_Q^\dagger$ with eigenvalue $\lambda^*$.

We apply $\mathcal T$ in each symmetry block of Sec.~\ref{app:transfer-quantum-numbers} and compute its leading eigenpairs by matrix-free Arnoldi iteration (KrylovKit).
At our largest circumference $L_2=7$ and twist $\phi_2=\pi$, the blocks $(R_{\rm k},R_{\rm b},Q,K_y)=(\mathbf3,\overline{\mathbf3},0,0)$ and $(\overline{\mathbf3},\mathbf3,0,0)$ each have dimension $15\,591\,359$, the largest we diagonalized.
With a Krylov space of dimension 24 and a limit of 100 cycles, both leading eigenpairs converged in the first cycle. The two runs used 18 and 26 applications of $\mathcal T$, including residual checks, and both gave $\|\mathcal T v-\lambda v\|/(|\lambda|\,\|v\|)<6.3\times10^{-12}$.

\subsection{Cut quantum numbers and local insertions}
\label{app:transfer-quantum-numbers}

Let $\mathbf u_{\mathrm k}=\{(\mathbf R_1,\alpha_1),(\mathbf R_2,\alpha_2),\ldots\}$ be an arbitrary parton configuration in a column.
The ket-layer tensor has matrix elements
\begin{equation}
    A^{\mathbf u_{\mathrm k}}_{\mathbf m_{\mathrm k}^c,\mathbf n_{\mathrm k}^c}
    =
    \langle\mathbf m_{\mathrm k},\mathbf 0_{\mathrm k}|
    \prod_{(\mathbf R,\alpha)\in\mathbf u_{\mathrm k}}^{\leftarrow}
    K_{\mathbf R,\alpha,\mathrm k}
    |\mathbf 1_{\mathrm k},\mathbf n_{\mathrm k}\rangle .
    \label{eq:A-tensor-matrix}
\end{equation}
Let $\mathcal H_{\mathrm k}^{\mathrm p}$ be the parton Fock space of the column, with basis states $|\mathbf u_{\mathrm k}\rangle$.
With the parton leg as an output, the tensor acts as
\begin{equation}
    \begin{aligned}
        A\colon\quad
        \mathcal H_{\mathrm k}^{<}
        &\longrightarrow \mathcal H_{\mathrm k}^{>}\otimes\mathcal H_{\mathrm k}^{\mathrm p},\\
        |\mathbf n_{\mathrm k}\rangle
        &\longmapsto
        \sum_{\mathbf m_{\mathrm k},\mathbf u_{\mathrm k}}
        |\mathbf m_{\mathrm k},\mathbf u_{\mathrm k}\rangle\,
        A^{\mathbf u_{\mathrm k}}_{\mathbf m_{\mathrm k}^c,\mathbf n_{\mathrm k}^c} .
    \end{aligned}
    \label{eq:A-tensor-mapping}
\end{equation}
Here $<$ and $>$ denote the cuts before and after the column, respectively.
In the one-parton basis, $(U_\tau)_{\beta\alpha}=\langle\beta|\widehat U_\tau|\alpha\rangle$. Under a ket rotation $f^\dagger_\alpha\mapsto\sum_\beta f^\dagger_\beta(U_{\mathrm k})_{\beta\alpha}$, the auxiliary creators transform as
\begin{equation}
    d^\dagger_{\mathbf r,\alpha,\mathrm k}
    \longmapsto
    \sum_\beta d^\dagger_{\mathbf r,\beta,\mathrm k}(U_{\mathrm k})_{\beta\alpha},
    \qquad U_{\mathrm k}\in SU(3)_{\mathrm k}.
    \label{eq:app-ket-cut-rotation}
\end{equation}
Each occupied index of $|\mathbf n_{\mathrm k}\rangle$ or $|\mathbf u_{\mathrm k}\rangle$ carries $\mathbf3$. The ket holes indexing $A$ in Eq.~\ref{eq:A-tensor-matrix} carry $\overline{\mathbf3}$.
We denote the irreducible components in $\mathcal H_{\mathrm k}^{<}$, $\mathcal H_{\mathrm k}^{\mathrm p}$, and $\mathcal H_{\mathrm k}^{>}$ by $R_{\mathrm k}^{<}$, $R_{\mathrm k}^{\mathrm p}$, and $R_{\mathrm k}^{>}$, respectively.
The selection rule for Eq.~\ref{eq:A-tensor-mapping} is
\begin{equation}
    R_{\mathrm k}^{<}\subset R_{\mathrm k}^{>}\otimes R_{\mathrm k}^{\mathrm p}
    \qquad\Longleftrightarrow\qquad
    R_{\mathrm k}^{>}\subset R_{\mathrm k}^{<}\otimes (R_{\mathrm k}^{\mathrm p})^*.
\end{equation}

For a parton configuration $\mathbf u_{\mathrm b}$ in the bra layer, the complex-conjugate tensor has matrix elements
\begin{equation}
    \bigl(A^{\mathbf u_{\mathrm b}}_{\mathbf m_{\mathrm b},\mathbf n_{\mathrm b}}\bigr)^*
    =
    \langle\mathbf m_{\mathrm b},\mathbf 1_{\mathrm b}|
    \prod_{(\mathbf R,\alpha)\in\mathbf u_{\mathrm b}}^{\leftarrow}
    K^\dagger_{\mathbf R,\alpha,\mathrm b}
    |\mathbf 0_{\mathrm b},\mathbf n_{\mathrm b}\rangle .
    \label{eq:A-bra-tensor-matrix}
\end{equation}
With the parton leg as an input, the tensor acts as
\begin{equation}
    \begin{aligned}
        A^*\colon\quad
        \mathcal H_{\mathrm b}^{\mathrm p}\otimes\mathcal H_{\mathrm b}^{<}
        &\longrightarrow \mathcal H_{\mathrm b}^{>},\\
        |\mathbf u_{\mathrm b},\mathbf n_{\mathrm b}\rangle
        &\longmapsto
        \sum_{\mathbf m_{\mathrm b}}
        |\mathbf m_{\mathrm b}\rangle\,
        \bigl(A^{\mathbf u_{\mathrm b}}_{\mathbf m_{\mathrm b},\mathbf n_{\mathrm b}}\bigr)^* .
    \end{aligned}
    \label{eq:A-bra-tensor-mapping}
\end{equation}
Under an independent $SU(3)_{\mathrm b}$ rotation, the auxiliary creators transform as
\begin{equation}
    d^\dagger_{\mathbf r,\alpha,\mathrm b}
    \longmapsto
    \sum_\beta d^\dagger_{\mathbf r,\beta,\mathrm b}(U_{\mathrm b})_{\beta\alpha},
    \qquad U_{\mathrm b}\in SU(3)_{\mathrm b}.
    \label{eq:app-bra-cut-rotation}
\end{equation}
Both $|\mathbf n_{\mathrm b}\rangle$ and $|\mathbf u_{\mathrm b}\rangle$ carry one fundamental $\mathbf3$ per occupied index.
We denote the irreducible components in $\mathcal H_{\mathrm b}^{<}$, $\mathcal H_{\mathrm b}^{\mathrm p}$, and $\mathcal H_{\mathrm b}^{>}$ by $R_{\mathrm b}^{<}$, $R_{\mathrm b}^{\mathrm p}$, and $R_{\mathrm b}^{>}$, respectively.
The selection rule for Eq.~\ref{eq:A-bra-tensor-mapping} is
\begin{equation}
    R_{\mathrm b}^{>}\subset R_{\mathrm b}^{\mathrm p}\otimes R_{\mathrm b}^{<}
    \qquad\Longleftrightarrow\qquad
    R_{\mathrm b}^{<}\subset (R_{\mathrm b}^{\mathrm p})^*\otimes R_{\mathrm b}^{>}.
\end{equation}

The Gutzwiller projector maps the parton output of $A$ to the parton input of $A^*$ as
\begin{equation}
    \begin{aligned}
        P_{\rm G}\colon\quad
        \mathcal H_{\mathrm k}^{\mathrm p}
        &\longrightarrow \mathcal H_{\mathrm b}^{\mathrm p},\\
        {}_{\mathrm b}\langle\mathbf u_{\mathrm b}|P_{\rm G}|\mathbf u_{\mathrm k}\rangle_{\mathrm k}
        &=
        \begin{cases}
            1, & \mathbf u_{\mathrm b}=\mathbf u_{\mathrm k}
                 =\sigma\times\{1,2,3\}\ \text{for some }\sigma,\\
            0, & \text{otherwise}.
        \end{cases}
    \end{aligned}
    \label{eq:app-projector-mapping}
\end{equation}
Here $\sigma$ is the set of sites occupied by all three colors in both layers.
At each site, the empty and triply occupied states are both $SU(3)$ singlets.
Thus every configuration retained by $P_{\rm G}$ is a singlet in each layer, and the fusion rules give
\begin{equation}
    R_{\mathrm k}^{\mathrm p}=R_{\mathrm b}^{\mathrm p}=\mathbf 1,
    \qquad
    R_{\mathrm k}^{>}=R_{\mathrm k}^{<},
    \qquad
    R_{\mathrm b}^{>}=R_{\mathrm b}^{<}.
\end{equation}

The $U(1)$ symmetry enables a charge quantum number on the cut space. 
Each auxiliary creator transforms as $d^\dagger_\tau\mapsto \mathrm{e}^{\mathrm{i}\theta}d^\dagger_\tau$ under a common charge rotation.
In the auxiliary occupation basis, write $n_\alpha^\tau=\sum_{\ell_2}n_{\ell_2,\alpha,\tau}$ and analogously $m_\alpha^\tau$, so $|\mathbf n_\tau|=\sum_\alpha n_\alpha^\tau$. The charge $\widehat Q|\mathbf n\rangle=Q|\mathbf n\rangle$ relative to the filled ket cut $|\mathbf1_{\mathrm k}\rangle$:
\begin{equation}
 Q=|\mathbf n_{\mathrm k}|+|\mathbf n_{\mathrm b}|-3L_2.
 \label{eq:app-cut-charge}
\end{equation}
Also, the the $SU(3)$ weights are
\begin{equation}
     h_i^{\tau}=n_i^{\tau}-n_{i+1}^{\tau},\quad \tau=\mathrm{k,b},\;i=1,2,
\end{equation}
which together with the total spin $j_{12}^\tau$ of the subgroup rotating colors 1 and 2 resolves the multiplet $R^\tau$.

To define the quantum number $K_y$ of the translation symmetry, let $|v\rangle$ be a cut state of charge $Q$ and transverse momentum $k_2=k_{\mathrm k}+k_{\mathrm b}\pmod{2\pi}$. 
The active translation $\mathsf t_d$ by $+\mathbf T_2$ fixes the empty cut. On $|\mathbf1_{\mathrm k}\rangle$ it cycles the $L_2$ occupied modes of each color, taking one particle per color across the seam. Thus $k_{2,0}=3\phi_2+3\pi(L_2-1)\pmod{2\pi}$. In $L_2$ steps each of the $Q+3L_2$ occupied modes crosses the seam once and acquires $\mathrm{e}^{-\mathrm{i}\phi_2}$ [Eq.~\ref{eq:app-tns-orbital}]. With $K_y=-L_2(k_2-k_{2,0})/(2\pi)$ modulo $L_2$, this gives
\begin{equation}
 \begin{aligned}
 \mathsf t_d|v\rangle&=\mathrm{e}^{-\mathrm{i}k_2}|v\rangle,\qquad
 \mathsf t_d^{L_2}|v\rangle=\mathrm{e}^{-\mathrm{i}\phi_2(Q+3L_2)}|v\rangle,\\
 \mathrm{e}^{2\pi\mathrm{i}K_y}
 &=\mathrm{e}^{-\mathrm{i}L_2(k_2-k_{2,0})}
 =\frac{\mathrm{e}^{-\mathrm{i}\phi_2(Q+3L_2)}}{\mathrm{e}^{-\mathrm{i}\phi_2(3L_2)}}
 =\mathrm{e}^{-\mathrm{i}\phi_2 Q}.
 \end{aligned}
 \label{eq:app-twisted-transverse-momentum}
\end{equation}
At $\phi_2=\pi$, $K_y$ is integer for even $Q$ and half-integer for odd $Q$.

On the parton spaces $\mathcal H_\tau^{\mathrm p}$, the ket and bra copies of $\sigma$ carry the same parton charge $3|\sigma|$.
Their phases therefore cancel in the projector matrix element.
Likewise, a simultaneous transverse translation preserves the matching condition in Eq.~\ref{eq:app-projector-mapping}, whereas translating one layer alone generally violates it.
In a charge and momentum basis, a nonzero matrix element of $P_{\rm G}$ therefore requires
\begin{equation}
    Q_{\mathrm k}^{\mathrm p}=Q_{\mathrm b}^{\mathrm p}=3|\sigma|,
    \qquad
    k_{\mathrm k}^{\mathrm p}=k_{\mathrm b}^{\mathrm p}\pmod{2\pi}.
    \label{eq:app-projected-parton-charge-momentum}
\end{equation}
Thus $Q^{>}-Q^{<}=Q_{\mathrm b}^{\mathrm p}-Q_{\mathrm k}^{\mathrm p}=0$ and $k_2^{>}-k_2^{<}=k_{\mathrm b}^{\mathrm p}-k_{\mathrm k}^{\mathrm p}\pmod{2\pi}=0$.
Charge and momentum can thus be exchanged between layers, while the total cut labels $Q$ and $K_y$ are conserved. 
The transfer eigenmodes are labeled by $(R_{\mathrm k},R_{\mathrm b},Q,K_y)$, and the above discussion of its conservation is illustrated in Fig.~\ref{fig:tns-transfer}(a).

To identify the transfer sectors contributing to a local correlation function, we insert $O_\tau$ between each layer $\tau=\mathrm k,\mathrm b$ and the Gutzwiller projector. A local insertion can transfer several transverse momenta. Let $\widehat O$ denote a component of the resulting column operator with definite charge $\delta Q$ and momentum $\delta K_y$, so that
\begin{equation}
 \begin{aligned}
 [\widehat Q,\widehat O]&=\delta Q\,\widehat O,\qquad
 \mathsf t_d\widehat O\mathsf t_d^\dagger
 =\mathrm{e}^{2\pi\mathrm{i}\delta K_y/L_2}\widehat O,\\
 Q^{>}-Q^{<}&=\delta Q,\qquad
 K_y^{>}-K_y^{<}=\delta K_y\pmod{L_2},\qquad
 \mathrm{e}^{2\pi\mathrm{i}\delta K_y}=\mathrm{e}^{-\mathrm{i}\phi_2\delta Q}.
 \end{aligned}
 \label{eq:app-insertion-charge-momentum}
\end{equation}
The last equality follows from Eq.~\ref{eq:app-twisted-transverse-momentum} and fixes only the fractional part of $\delta K_y$.
For an input with $(Q^{<},K_y^{<})=(0,0)$, the output has $(Q^{>},K_y^{>})=(\delta Q,\delta K_y)$.
Writing $\mathrm p'$ for the parton leg next to the projector, we have
\begin{equation}
    \mathcal H_{\mathrm k}^{\mathrm p}
    \xrightarrow{\;O_{\mathrm k}\;}\mathcal H_{\mathrm k}^{\mathrm p'}
    \xrightarrow{\;P_{\rm G}\;}\mathcal H_{\mathrm b}^{\mathrm p'}
    \xrightarrow{\;O_{\mathrm b}\;}\mathcal H_{\mathrm b}^{\mathrm p}.
    \label{eq:app-insertion-mapping}
\end{equation}
Let $r_\tau$ denote the $SU(3)_\tau$ irrep of the inserted operator $O_\tau$. The matrix elements in Eq.~\ref{eq:app-insertion-mapping} can be nonzero only if
\begin{equation}
    R_{\mathrm k}^{\mathrm p'}\subset r_{\mathrm k}\otimes R_{\mathrm k}^{\mathrm p},\qquad
    R_{\mathrm b}^{\mathrm p}\subset r_{\mathrm b}\otimes R_{\mathrm b}^{\mathrm p'}.
    \label{eq:app-insertion-selection-rule}
\end{equation}
Since $P_{\rm G}$ selects $R_\tau^{\mathrm p'}=\mathbf 1$ in each layer, only $R_{\mathrm k}^{\mathrm p}=r_{\mathrm k}^*$ and $R_{\mathrm b}^{\mathrm p}=r_{\mathrm b}$ contribute.
The configurations $\mathbf u_{\mathrm b}$ and $\mathbf u_{\mathrm k}$ can differ, while $P_{\rm G}$ requires both intermediate configurations to equal $\sigma\times\{1,2,3\}$,
\begin{equation}
 {}_{\mathrm b}\langle\mathbf u_{\mathrm b}|O_{\mathrm b}P_{\rm G}O_{\mathrm k}|\mathbf u_{\mathrm k}\rangle_{\mathrm k}
 =\sum_\sigma(O_{\mathrm b})_{\mathbf u_{\mathrm b},\sigma}(O_{\mathrm k})_{\sigma,\mathbf u_{\mathrm k}} .
\end{equation}

Combining the tensor and insertion selection rules gives $R_\tau^{>}\subseteq r_\tau\otimes R_\tau^{<}$, so for a singlet input $R_\tau^{>}=r_\tau$. The occupation changes satisfy
\begin{equation}
 m_\alpha^{\mathrm k}-n_\alpha^{\mathrm k}=L_2-|\mathbf u_{\mathrm k,\alpha}|,\qquad
 m_\alpha^{\mathrm b}-n_\alpha^{\mathrm b}=|\mathbf u_{\mathrm b,\alpha}|-L_2,\qquad
 \delta Q=|\mathbf u_{\mathrm b}|-|\mathbf u_{\mathrm k}| ,
 \label{eq:app-local-insertion-charge-rule}
\end{equation}
where $\mathbf u_{\tau,\alpha}$ denotes the sites occupied by color $\alpha$ in the configuration $\mathbf u_\tau$. For a ket creator, $|\mathbf u_{\mathrm k}|=3|\sigma|-1$ before projection. A bra creator acts after projection, giving $|\mathbf u_{\mathrm b}|=3|\sigma|+1$. With the other layer at $3|\sigma|$, either insertion gives $\delta Q=+1$. The conjugate insertion $f_\alpha$ gives $\delta Q=-1$. Fourier summation selects the transverse momentum transferred to the cut.

\FloatBarrier

\subsection{Transfer spectrum of the unprojected state}
\label{app:unprojected-transfer}

In $\langle\Phi|\Phi\rangle=\det(S^\dagger S)$, bra and ket occupations match separately for each color, the column operator in Eq.~\ref{eq:app-column-operator} for column $\mathcal C=\mathcal C_{\ell_1}$ becomes
\begin{equation}
 \mathbb B_{\ell_1}^{\Phi}
 =\prod_{\mathbf R\in\mathcal C}\prod_{\alpha=1}^3
   \bigl(1+K_{\mathbf R,\alpha,\mathrm b}^\dagger K_{\mathbf R,\alpha,\mathrm k}\bigr)
 =\exp\!\left(\sum_{\mathbf R\in\mathcal C,\,\alpha}
       K_{\mathbf R,\alpha,\mathrm b}^\dagger K_{\mathbf R,\alpha,\mathrm k}\right).
 \label{eq:app-unprojected-column}
\end{equation}
Inside the exponent is a quadratic form with kernel $D^{(\ell_1)}=S[\mathcal C]^\dagger S[\mathcal C]$, where $S[\mathcal C]$ takes the rows $(\mathbf R,\alpha)$ of $S$ in Eq.~\ref{eq:app-tns-orbital} for all sites $\mathbf R\in\mathcal C$ and all three colors. 
We label the cuts $\ell_1,\ell_1+1$ by $<,>$ and take the Fourier transform $d^\dagger_{\ell,\alpha,\tau}(-q)=L_2^{-1/2}\sum_{\ell_2}\mathrm{e}^{-\mathrm{i} q\ell_2}d^\dagger_{\mathbf r_{\ell\ell_2},\alpha,\tau}$ with $q=(2\pi m-\phi_2)/L_2\in(-\pi,\pi]$, $m\in\mathbb Z$.
Relabeling the summation momentum by $q\mapsto -q$ gives
\begin{equation}
\begin{aligned}
    \sum_{\mathbf R\in\mathcal C,\,\alpha}
    K^\dagger_{\mathbf R,\alpha,\mathrm b}K_{\mathbf R,\alpha,\mathrm k}
    =\sum_{\substack{\alpha,q\\L_2q=\,\phi_2\;(\mathrm{mod}\,2\pi)}}
    \begin{pmatrix}d^\dagger_{<,\alpha,\mathrm b}(q)&d^\dagger_{>,\alpha,\mathrm b}(q)\end{pmatrix}
    D^{(\ell_1)}(-q)
    \begin{pmatrix}
    d_{<,\alpha,\mathrm k}(q)\\d_{>,\alpha,\mathrm k}(q)\end{pmatrix},\\
    D^{(\ell_1)}(q)=
    \begin{pmatrix}
    3-\cos q-\sqrt3\sin q & -(1+\mathrm{e}^{-\mathrm{i} q})\\
    -(1+\mathrm{e}^{\mathrm{i} q}) & 3-\cos q+\sqrt3\sin q
    \end{pmatrix}=:\begin{pmatrix}D_{<<}&D_{<>}\\D_{><}&D_{>>}\end{pmatrix}.
\end{aligned}
\label{eq:app-unprojected-column-exponent}
\end{equation}
The columns of $\mathcal T_\Phi(q)$ index input states $|n_{\mathrm k}n_{\mathrm b}\rangle_{<}$ and its rows index output states $|m_{\mathrm k}m_{\mathrm b}\rangle_{>}$. 
Both use the order $10,01,00,11$ and $|11\rangle=d^\dagger_{\ell,\alpha,\mathrm b}(-q)\,
  d^\dagger_{\ell,\alpha,\mathrm k}(-q)\,|00\rangle$. 
Cut $>$ starts in $|10\rangle$ and cut $<$ ends in $|01\rangle$ in Eq.~\ref{eq:app-tns-orbital-column-split}, giving
\begin{equation}
 \begin{aligned}
 \mathcal T_\Phi(q)&=
 \begin{pmatrix}
 D_{<<}&1&0&0\\
 D_{<<}D_{>>}-D_{<>}D_{><}&D_{>>}&0&0\\
 0&0&-D_{<>}&0\\
 0&0&0&-D_{><}
 \end{pmatrix}\\
 &=\begin{pmatrix}
 3-\cos q-\sqrt3\sin q&1&0&0\\
 4(1-\cos q)^2&3-\cos q+\sqrt3\sin q&0&0\\
 0&0&1+\mathrm{e}^{-\mathrm{i} q}&0\\
 0&0&0&1+\mathrm{e}^{\mathrm{i} q}
 \end{pmatrix},\qquad
 \mathcal T_\Phi=\bigotimes_{\alpha,q}\mathcal T_\Phi(q).
 \end{aligned}
 \label{eq:app-unprojected-transfer-block}
\end{equation}
The one-particle block ($\ket{10},\,\ket{01}$) has eigenvalues $\lambda_\pm$, and $|00\rangle,|11\rangle$ have eigenvalues $\lambda_{00},\lambda_{11}$:
\begin{equation}
 \lambda_\pm(q)=3-\cos q\ \pm\sqrt{(1-\cos q)(7-\cos q)},\qquad
 \lambda_{00}(q)=1+\mathrm{e}^{-\mathrm{i} q},\qquad \lambda_{11}(q)=1+\mathrm{e}^{\mathrm{i} q}.
 \label{eq:app-unprojected-transfer-eigenvalues}
\end{equation}
At $\phi_2=\pi$, the dominant eigenvalue of each $(\alpha,q)$ block is $\lambda_+(q)$. 
The dominant state has one particle per block, hence $Q=0$. Replacing one block by $|00\rangle$ or $|11\rangle$ gives $(Q,K_y)=(-1,-L_2q/(2\pi))$ or $(1,L_2q/(2\pi))$, respectively. 
The remaining $\lambda_-$ mode can therefore be viewed as making both single excitations, because it satisfies $\lambda_-/\lambda_+=(\lambda_{00}/\lambda_+)(\lambda_{11}/\lambda_+)$ and $(\delta Q,K_y)=(0,0)$.

Equation~\ref{eq:app-unprojected-transfer-eigenvalues} gives $|\lambda_{00}|=|\lambda_{11}|=2|\cos(q/2)|$. Thus, at fixed half-integer $K_y$, Eq.~\ref{eq:bulk-scaled-transfer-gap} gives
\begin{equation}
 \Delta_{00}=\Delta_{11}=\frac{L_2}{\pi\sqrt3}\ln\frac{\lambda_+}{2|\cos(q/2)|}
 \xrightarrow{L_2\to\infty}\frac{L_2|q|}{2\pi}=|K_y|.
 \label{eq:app-unprojected-scaled-gap}
\end{equation}
Gathering the three colors together, we obtain the antiperiodic spectrum of three massless Dirac fermions, $U(3)_1$~\cite{Ginsparg1988CFT,Witten1984Bosonization}. 
Also, the total dominant eigenvalue $\lambda_0^\Phi=\prod_{q}\lambda_+(q)^3$ with $\ln\lambda_+(q)=\ln2+\sqrt3|q|/2+O(q^2)$ gives 
\begin{equation}
 \frac{\ln\lambda_0^\Phi}{L_2}
 =\frac{3}{L_2}\sum_{q}\ln\lambda_+(q)
 =A_\Phi+\frac{c\,\pi\sqrt3}{12L_2^2}+O(L_2^{-4}),\qquad c=3,
 \label{eq:app-unprojected-casimir}
\end{equation}
where $A_\Phi=3\int_{-\pi}^{\pi}\frac{dq}{2\pi}\ln\lambda_+(q)$ is a nonuniversal bulk value.
The $L_2^{-2}$ coefficient gives $c=3$, the expected central charge of three massless Dirac fermions.

To assign the low-energy modes to the two layers, consider their occupations relative to the dominant state.
In the rescaled basis $\bigl(|10\rangle,\,2(1-\cos q)|01\rangle\bigr)$, the one-particle block of Eq.~\ref{eq:app-unprojected-transfer-block} is symmetric and has the small-$q$ expansion
\begin{equation}
 \left.\mathcal T_\Phi(q)\right|_{10,01}
 =\begin{pmatrix}2-\sqrt3\,q&0\\0&2+\sqrt3\,q\end{pmatrix}
 +O(q^2).
\end{equation}
In this basis the dominant state approaches $\ket{01}$ as $q\to0^+$ and $\ket{10}$ as $q\to0^-$.
A change $(\delta n_{\rm k},\delta n_{\rm b})$ relative to this state carries
\begin{equation}
 \delta Q=\delta n_{\rm k}+\delta n_{\rm b},\qquad
 \delta k_2=-q\,(\delta n_{\rm k}+\delta n_{\rm b}),\qquad
 K_y=-\frac{L_2}{2\pi}\delta k_2.
\end{equation}
In the order $(\delta n_{\rm k},\delta n_{\rm b},\delta Q,2\pi K_y/L_2)$ the single excitations are
\begin{equation}
\begin{aligned}
 q\to0^+ \quad &\left\{
 \begin{alignedat}{2}
 |11\rangle &\quad \text{ket particle} &\qquad& (+1,0,+1,+|q|)\\
 |00\rangle &\quad \text{bra hole}     &      & (0,-1,-1,-|q|)
 \end{alignedat}\right. \\[5pt]
 q\to0^- \quad &\left\{
 \begin{alignedat}{2}
 |00\rangle &\quad \text{ket hole}     &\qquad& (-1,0,-1,+|q|)\\
 |11\rangle &\quad \text{bra particle} &      & (0,+1,+1,-|q|)
 \end{alignedat}\right.
\end{aligned}
\end{equation}
The ket particle and hole both have $K_y>0$, and their bra counterparts have $K_y<0$. In the Fermi-sea picture the occupied ket (bra) particle modes have $K_y<0$ ($K_y>0$).

\section{Wilson loops and modular matrices}
\label{app:wilson-modular-data}

Wilson matrices describe how physical flux insertion mixes the ground states.
Wilson eigenbasis overlaps and rotation symmetry give the modular matrices, which encode mutual braiding and topological spin.
We first present the numerical Wilson loops and their finite-size convergence, then explain how their eigenbases and rotation symmetry determine the modular matrices.

\subsection{Finite-size convergence of the Wilson loops}
\label{app:wilson-numerics}

At each mesh point $j$ in physical twist space, the columns of $B_j$ are the three individually normalized projected kets $|\bar\Psi_{\boldsymbol\phi}\rangle$ at the chosen parton twists. Their Gram matrix $G_j=B_j^\dagger B_j$ gives the orthonormal frame $B_jG_j^{-1/2}$. For neighboring points $j$ and $j'$, the unitary part of the frame overlap defines the link
\begin{equation}
 U_{j,j'}=\operatorname{polar}
 \bigl(G_j^{-1/2}B_j^\dagger B_{j'}G_{j'}^{-1/2}\bigr),
 \qquad U_{j',j}=U_{j,j'}^\dagger .
 \label{eq:sm-wilson-link}
\end{equation}
For a singular-value decomposition $M=V\Sigma W^\dagger$, $\operatorname{polar}(M)=VW^\dagger$.

For Fig.~\ref{fig:mean-field}(e,f), the three parton twists are $(\varphi_1/3,\varphi_2/3+2\pi s/3)$ with $s=0,1,2$. Label the four corners of a mesh plaquette $1,2,3,4$ in counterclockwise order, starting along $+\varphi_1$. For a plaquette of area $\delta A$, the Berry curvature is $\mathcal F=\frac{\mathrm{i}}{\delta A}\log\!\left(U_{12}U_{23}U_{34}U_{41}\right)$,
where the matrix logarithm uses the principal branch.

For the Wilson loops, we use the three branches $(\varphi_1/3+2\pi/3,\varphi_2/3+2\pi s/3)$ and follow the flux cycles $l_{\mathbf T_1}(t)=(2\pi t,0)$ and $l_{\mathbf T_2}(t)=(0,2\pi t)$, $0\le t\le1$, keeping the boundary phases on the same bonds.
We compute the overlaps within and between successive frames by exact enumeration for $L\le3$ and with the estimator in Eq.~\ref{eq:sm-vmc-normalized-overlap} for larger sizes. For the Wilson-loop runs, we use 200 equilibration sweeps and 200 sweeps to estimate the norm fractions, then fix $c_s\propto1/\max(\hat f_s,10^{-8})$.

\begin{figure}[h]
 \centering
 \includegraphics[width=.76\textwidth]{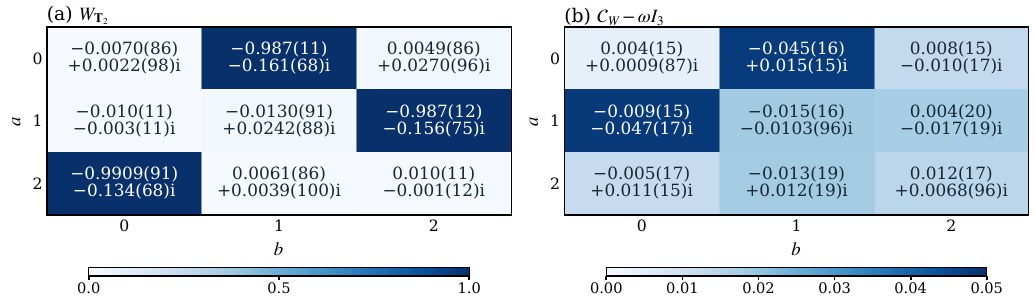}
 \caption{Wilson matrices at $L=16$ in the $W_{\mathbf T_1}$ eigenbasis.
 (a) $W_{\mathbf T_2}$. (b) $\mathcal C_W-\omega I_3$.
 Color gives the magnitude of each entry. The two lines give its real and imaginary parts, with one standard error in parentheses on the last digits.
 Each jackknife replicate uses the same branch ordering as Table~\ref{tab:app-wilson-eigenvalues} and fixes the eigenvector phases by taking $\langle a_{\mathbf T_1}|0_{\mathbf T_2}\rangle$ and $\langle 0_{\mathbf T_1}|b_{\mathbf T_2}\rangle$ positive.}
 \label{fig:sm-wilson-entries}
\end{figure}

The physical flux insertion $W_{\mathbf T_1,\mathbf T_2}$ can be interpreted as winding the Abelian anyon $a$ around the $l_{\mathbf T_1,\mathbf T_2}$.
Fig.~\ref{fig:sm-wilson-entries}(a) shows that $W_{\mathbf T_2}$ cycles the three $W_{\mathbf T_1}$ eigenstates, which suggests a $\mathbb Z_3$ fusion cycle through all three topological sectors, $0a\to 2a\to 1a\to 0a$. 
The commutator $\mathcal C_W=W_{\mathbf T_1}W_{\mathbf T_2}W_{\mathbf T_1}^{-1}W_{\mathbf T_2}^{-1}$ measures the mutual braiding of $a$, and Fig.~\ref{fig:sm-wilson-entries}(b) shows that $\mathcal C_W$ is close to $\omega I_3$ with $\omega=\mathrm{e}^{2\pi \mathrm{i}/3}$.

\begin{figure}[!htbp]
 \centering
 \includegraphics[width=.76\textwidth]{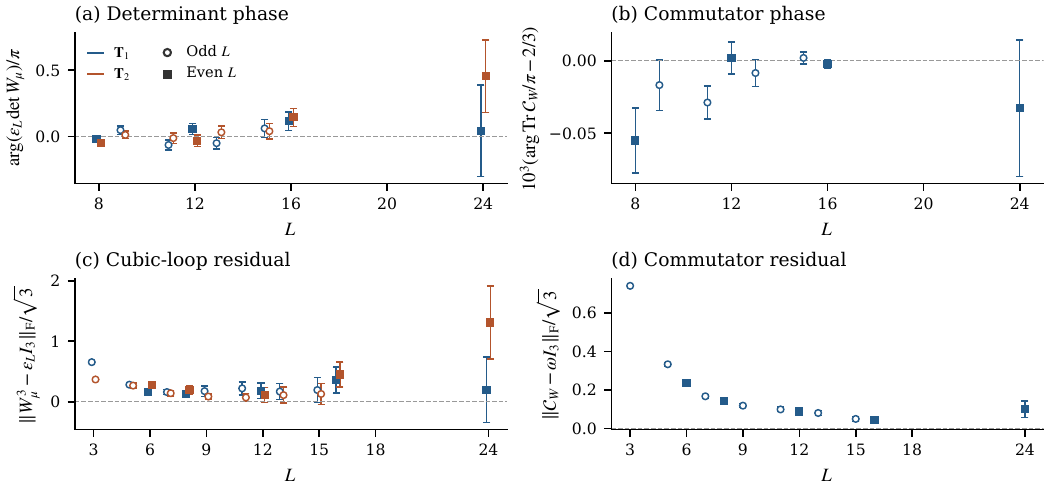}
 \caption{Wilson diagnostics with the original overall phases retained.
 (a,b) Determinant and commutator phases for $L\geq8$; zero denotes the corresponding prediction.
 (c,d) Deviations of $W_\mu^3$ from $\epsilon_LI_3$ and of $\mathcal C_W$ from $\omega I_3$ [Eq.~\ref{eq:app-wilson-residual}].
 Blue and orange distinguish $\mathbf T_1,\mathbf T_2$ in (a,c); open circles and filled squares distinguish odd and even sizes throughout.
 Error bars give one jackknife standard error.}
 \label{fig:app-wilson-convergence}
\end{figure}

We prove in Sec.~\ref{app:c2-wilson-determinants} and numerically confirm [Fig.~\ref{fig:app-wilson-convergence}(a)] that $\det W_{\mathbf T_1,\mathbf T_2}=\epsilon_L$ with $\epsilon_L=(-1)^{L+1}$.
Together with $\mathcal C_W=\omega I_3$, this predicts the eigenvalues of $W_{\mathbf T_1,\mathbf T_2}$ to approach $\epsilon_L\{1,\omega,\omega^2\}$, in agreement with Table~\ref{tab:app-wilson-eigenvalues}.

We further test $\mathcal C_W=\omega I_3$ through the residuals
\begin{equation}
\arg\Tr\,\mathcal{C}_W-\frac{2\pi}{3},    
\qquad
 \frac{\|W_{\mathbf T_1,\mathbf T_2}^3-\epsilon_L I_3\|_{\rm F}}{\sqrt3},
 \qquad
 \frac{\|\mathcal C_W-\omega I_3\|_{\rm F}}{\sqrt3}.
 \label{eq:app-wilson-residual}
\end{equation}
Figure~\ref{fig:app-wilson-convergence}(b,c,d) shows their size dependence.
For the sampled sizes from $L=5$ through $L=13$, we use the same protocol as at $L=16$: 12 steps per flux cycle and 28 independent chains, each with 1600 burn-in and 16000 measured sweeps. At $L=15,16,24$ we pool the raw Gram estimates over 24, 28, and 16 independent chains, respectively.
We compute jackknife errors by deleting one block at a time, using block lengths of 100–4000 sweeps and also treating each whole chain as a block. 
For each observable, we report the largest standard error across these block lengths.

\begin{table}[h]
 \centering\small
 \caption{Eigenvalues $\mathrm{e}^{\mathrm{i}\theta_{\mu,a}}$ of the two Wilson loops, including their overall phases. The errors in the exponents are one jackknife standard error in $\theta_{\mu,a}/\pi$. The $L=3$ row uses exact enumeration. Branches are ordered about $\epsilon_L\omega^{-a}$ for $W_{\mathbf T_1}$ and $\epsilon_L\omega^a$ for $W_{\mathbf T_2}$.}
 \label{tab:app-wilson-eigenvalues}
\begin{tabular}{rcccccc}
\hline\hline
 & \multicolumn{3}{c}{$W_{\mathbf T_1}$} & \multicolumn{3}{c}{$W_{\mathbf T_2}$}\\\cline{2-4}\cline{5-7}$L$ & $a=0$ & $a=1$ & $a=2$ & $a=0$ & $a=1$ & $a=2$\\\hline
3 & $\mathrm{e}^{\mathrm{i}\pi(0.000)}$ & $\mathrm{e}^{\mathrm{i}\pi(-0.579)}$ & $\mathrm{e}^{\mathrm{i}\pi(0.579)}$ & $\mathrm{e}^{\mathrm{i}\pi(-0.027)}$ & $\mathrm{e}^{\mathrm{i}\pi(0.607)}$ & $\mathrm{e}^{\mathrm{i}\pi(-0.684)}$ \\
5 & $\mathrm{e}^{\mathrm{i}\pi(0.001\pm0.005)}$ & $\mathrm{e}^{\mathrm{i}\pi(-0.629\pm0.005)}$ & $\mathrm{e}^{\mathrm{i}\pi(0.631\pm0.005)}$ & $\mathrm{e}^{\mathrm{i}\pi(-0.036\pm0.005)}$ & $\mathrm{e}^{\mathrm{i}\pi(0.638\pm0.005)}$ & $\mathrm{e}^{\mathrm{i}\pi(-0.683\pm0.005)}$ \\
6 & $\mathrm{e}^{\mathrm{i}\pi(1.003\pm0.007)}$ & $\mathrm{e}^{\mathrm{i}\pi(0.358\pm0.006)}$ & $\mathrm{e}^{\mathrm{i}\pi(1.648\pm0.007)}$ & $\mathrm{e}^{\mathrm{i}\pi(0.960\pm0.006)}$ & $\mathrm{e}^{\mathrm{i}\pi(1.637\pm0.006)}$ & $\mathrm{e}^{\mathrm{i}\pi(0.318\pm0.006)}$ \\
7 & $\mathrm{e}^{\mathrm{i}\pi(0.010\pm0.008)}$ & $\mathrm{e}^{\mathrm{i}\pi(-0.641\pm0.009)}$ & $\mathrm{e}^{\mathrm{i}\pi(0.657\pm0.008)}$ & $\mathrm{e}^{\mathrm{i}\pi(-0.024\pm0.007)}$ & $\mathrm{e}^{\mathrm{i}\pi(0.656\pm0.007)}$ & $\mathrm{e}^{\mathrm{i}\pi(-0.665\pm0.007)}$ \\
8 & $\mathrm{e}^{\mathrm{i}\pi(0.990\pm0.009)}$ & $\mathrm{e}^{\mathrm{i}\pi(0.341\pm0.008)}$ & $\mathrm{e}^{\mathrm{i}\pi(1.647\pm0.008)}$ & $\mathrm{e}^{\mathrm{i}\pi(0.968\pm0.009)}$ & $\mathrm{e}^{\mathrm{i}\pi(1.654\pm0.009)}$ & $\mathrm{e}^{\mathrm{i}\pi(0.329\pm0.010)}$ \\
9 & $\mathrm{e}^{\mathrm{i}\pi(0.018\pm0.011)}$ & $\mathrm{e}^{\mathrm{i}\pi(-0.640\pm0.011)}$ & $\mathrm{e}^{\mathrm{i}\pi(0.670\pm0.011)}$ & $\mathrm{e}^{\mathrm{i}\pi(-0.004\pm0.010)}$ & $\mathrm{e}^{\mathrm{i}\pi(0.667\pm0.010)}$ & $\mathrm{e}^{\mathrm{i}\pi(-0.651\pm0.010)}$ \\
11 & $\mathrm{e}^{\mathrm{i}\pi(-0.019\pm0.013)}$ & $\mathrm{e}^{\mathrm{i}\pi(-0.678\pm0.013)}$ & $\mathrm{e}^{\mathrm{i}\pi(0.633\pm0.012)}$ & $\mathrm{e}^{\mathrm{i}\pi(-0.012\pm0.013)}$ & $\mathrm{e}^{\mathrm{i}\pi(0.664\pm0.014)}$ & $\mathrm{e}^{\mathrm{i}\pi(-0.664\pm0.013)}$ \\
12 & $\mathrm{e}^{\mathrm{i}\pi(1.021\pm0.014)}$ & $\mathrm{e}^{\mathrm{i}\pi(0.357\pm0.014)}$ & $\mathrm{e}^{\mathrm{i}\pi(1.677\pm0.015)}$ & $\mathrm{e}^{\mathrm{i}\pi(0.982\pm0.015)}$ & $\mathrm{e}^{\mathrm{i}\pi(1.660\pm0.016)}$ & $\mathrm{e}^{\mathrm{i}\pi(0.325\pm0.014)}$ \\
13 & $\mathrm{e}^{\mathrm{i}\pi(-0.016\pm0.015)}$ & $\mathrm{e}^{\mathrm{i}\pi(-0.678\pm0.015)}$ & $\mathrm{e}^{\mathrm{i}\pi(0.643\pm0.015)}$ & $\mathrm{e}^{\mathrm{i}\pi(0.003\pm0.017)}$ & $\mathrm{e}^{\mathrm{i}\pi(0.680\pm0.017)}$ & $\mathrm{e}^{\mathrm{i}\pi(-0.652\pm0.015)}$ \\
15 & $\mathrm{e}^{\mathrm{i}\pi(0.024\pm0.022)}$ & $\mathrm{e}^{\mathrm{i}\pi(-0.646\pm0.022)}$ & $\mathrm{e}^{\mathrm{i}\pi(0.682\pm0.025)}$ & $\mathrm{e}^{\mathrm{i}\pi(0.008\pm0.020)}$ & $\mathrm{e}^{\mathrm{i}\pi(0.680\pm0.020)}$ & $\mathrm{e}^{\mathrm{i}\pi(-0.649\pm0.020)}$ \\
16 & $\mathrm{e}^{\mathrm{i}\pi(1.039\pm0.023)}$ & $\mathrm{e}^{\mathrm{i}\pi(0.374\pm0.023)}$ & $\mathrm{e}^{\mathrm{i}\pi(1.701\pm0.024)}$ & $\mathrm{e}^{\mathrm{i}\pi(1.042\pm0.022)}$ & $\mathrm{e}^{\mathrm{i}\pi(1.717\pm0.023)}$ & $\mathrm{e}^{\mathrm{i}\pi(0.385\pm0.022)}$ \\
24 & $\mathrm{e}^{\mathrm{i}\pi(1.012\pm0.113)}$ & $\mathrm{e}^{\mathrm{i}\pi(0.367\pm0.117)}$ & $\mathrm{e}^{\mathrm{i}\pi(1.662\pm0.117)}$ & $\mathrm{e}^{\mathrm{i}\pi(1.143\pm0.092)}$ & $\mathrm{e}^{\mathrm{i}\pi(1.830\pm0.090)}$ & $\mathrm{e}^{\mathrm{i}\pi(0.481\pm0.095)}$ \\
\hline\hline
\end{tabular}
\end{table}

These results strongly support a $\mathbb Z_3$ fusion group with $\omega$ as the full mutual-braiding phase of its generator, matching the charged-anyon fusion and braiding of $U(1)_3$.

\subsection{Modular matrices from Wilson loops and sixfold rotation}
\label{app:modular-extraction}

We extract $S$ from the Wilson eigenbasis overlaps and $T$ from sixfold rotation, after fixing the branch labels and eigenvector phases. Below $C_{6z}$ denotes the induced rotation on the same dual frame as the Wilson loops.
The general construction of standard Wilson eigenbases is given in Ref.~\cite{LiMong2022Modular}; here we describe the choices used in our numerical calculation.

Common geometric phases do not change the Wilson eigenstates.
Our flux-insertion loops represent an Abelian anyon winding around the torus, while local indistinguishability motivates treating the additional microscopic geometric phase as common to the three ground states:
\begin{equation}
 W_\mu\simeq \mathrm{e}^{\mathrm{i}\gamma_\mu}W_{l_\mu}^{\rm top}(1),\qquad \mu=\mathbf T_1,\mathbf T_2.
 \label{eq:sm-microscopic-topological-wilson}
\end{equation}
Here $W_l^{\rm top}(a)$ is the standard Wilson loop of anyon $a$ along the oriented cycle $l$; the approximation allows finite-size and numerical corrections.

\begin{figure}[!htbp]
 \centering
 \includegraphics[width=0.56\textwidth]{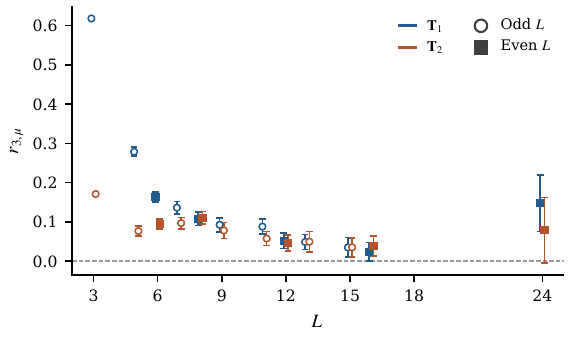}
 \caption{Deviation of $W_\mu^3$ from a scalar matrix, measured by $r_{3,\mu}$ in Eq.~\ref{eq:sm-wilson-cubic-spread}. The $L=3$ points use exact overlaps. Error bars at larger sizes give one jackknife standard error.}
 \label{fig:sm-wilson-cubic-spread}
\end{figure}

Twofold rotation fixes the reference branch used to remove this common phase.
The relation $C_{2z}W_\mu C_{2z}^\dagger=W_\mu^{-1}$, derived in Sec.~\ref{app:rotation-wilson-proof}, pairs inverse eigenvalues.
We label the unique $C_{2z}$-invariant branch by $a=0$, consistent with reversal of a standard loop sending $a\mapsto-a$ in the $\mathbb Z_3$ fusion group.
Its reference eigenvalue is $\epsilon_L=(-1)^{L+1}$ in the ideal limit, so $W_{l_\mu}^{\rm top}(1)\simeq\epsilon_L W_\mu$.
As a numerical check of a common phase and threefold fusion, we measure
\begin{equation}
 r_{3,\mu}
 =\frac{1}{\sqrt3}\left\|W_\mu^3-\frac{\operatorname{Tr}W_\mu^3}{3}I_3\right\|_{\rm F}
 =\sqrt{1-\frac{|\operatorname{Tr}W_\mu^3|^2}{9}}.
 \label{eq:sm-wilson-cubic-spread}
\end{equation}
The second equality uses unitarity.
This residual is independent of the common phase and vanishes when the three cubed eigenvalues coincide; it decreases with system size in Fig.~\ref{fig:sm-wilson-cubic-spread}.

\begin{figure}[htbp]
 \centering
 \includegraphics[width=0.98\textwidth]{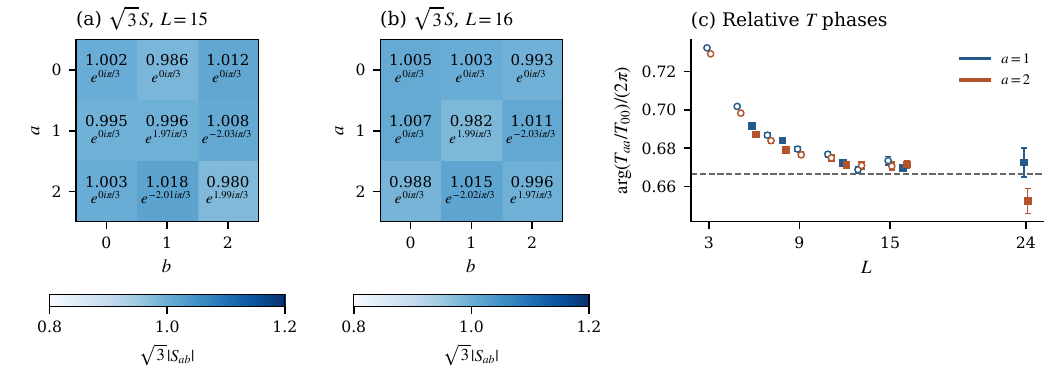}
 \caption{Modular matrices.
 (a,b) $\sqrt3S$ at $L=15,16$, respectively; $L=16$ is used instead of $L=24$ for its higher statistical precision. Entries show modulus and phase in units of $\pi/3$, with three significant digits for the phase.
 Each panel has its own modulus colorbar.
 (c) Relative $T$ phases; blue and orange denote $a=1,2$, open circles odd sizes, and filled squares even sizes. The dashed line is $2/3$.
 Matrix deviations and statistical errors are given in Table~\ref{tab:app-modular-data}.}
 \label{fig:wilson-loops}
\end{figure}

The modular $S$ transformation exchanges the two cycles as $(l_{\mathbf T_1},l_{\mathbf T_2})\mapsto(-l_{\mathbf T_2},l_{\mathbf T_1})$, and is obtained from their Wilson eigenbasis overlap~\cite{LiMong2022Modular}:
\begin{equation}
 S_{ab}=\langle a_{\mathbf T_1}|b_{\mathbf T_2}\rangle.
 \label{eq:sm-S-overlap}
\end{equation}
We order $|a_{\mathbf T_1}\rangle$ and $|b_{\mathbf T_2}\rangle$ by eigenvalues approaching $\epsilon_L\omega^{-a}$ and $\epsilon_L\omega^b$, respectively, and fix their phases by requiring $S_{0a},S_{a0}>0$.
The results in Fig.~\ref{fig:wilson-loops}(a,b) and Table~\ref{tab:app-modular-data} approach $S_{ab}=\omega^{ab}/\sqrt3$, as expected for $U(1)_3$.

Sixfold rotation provides the second modular matrix.
With $T:(l_{\mathbf T_1},l_{\mathbf T_2})\mapsto(l_{\mathbf T_1}-l_{\mathbf T_2},l_{\mathbf T_2})$, the combination $TS^{-1}$ has the same geometric action as $C_{6z}$: $(l_{\mathbf T_1},l_{\mathbf T_2})\mapsto(l_{\mathbf T_2},l_{\mathbf T_2}-l_{\mathbf T_1})$.
The universal wave-function overlap prescription~\cite{MeiWen2015} identifies their ground-state actions up to a common phase.
The Wilson loops also constrain any additional action of rotation on the anyon labels and phases.
Following Eq.~(113) of Ref.~\cite{LiMong2022Modular}, write
\begin{equation}
 C_{6z}W_l^{\rm top}(a)C_{6z}^\dagger
 =\eta_l(a)\,W_{C_6l}^{\rm top}\!\bigl(\rho_6(a)\bigr),
 \label{eq:sm-rotation-topological-loops}
\end{equation}
where $\rho_6$ permutes the anyon species and $\eta_l(a)$ is an additional phase.
The distinct fractional charges $q_a=ae/3\pmod e$ and charge conservation require $\rho_6(a)=a$.
Fusion constrains the phases by
\begin{equation}
 \eta_l(a+b)=\eta_l(a)\eta_l(b),\qquad \eta_l(a)^3=1.
 \label{eq:sm-rotation-fusion-phases}
\end{equation}
Applying $C_{2z}$ after Eq.~\ref{eq:sm-rotation-topological-loops} gives $\eta_l(a)W_{-C_6l}^{\rm top}(a)$, whereas applying it first gives $\eta_l(a)^{-1}W_{-C_6l}^{\rm top}(a)$.
The two rotations commute, so $\eta_l(a)^2=1$ and hence $\eta_l(a)=1$.
Thus $C_{6z}$ and $TS^{-1}$ act identically on the standard loops. Within this three-state space, a matrix commuting with both loops must be proportional to $I_3$, so the two operations differ only by a common phase.

We therefore use the already fixed $W_{\mathbf T_2}$ eigenbasis to calculate
\begin{equation}
 (T_{\rm num})_{ab}=\sum_{c=0}^2\langle a_{\mathbf T_2}|C_{6z}|c_{\mathbf T_2}\rangle S_{cb},
 \qquad
 \frac{T_{aa}}{T_{00}}=\frac{(T_{\rm num})_{aa}}{(T_{\rm num})_{00}}.
 \label{eq:sm-T-from-rotation}
\end{equation}
The ratio removes the common phase.
For fermionic $U(1)_3$ with periodic boundary conditions along both cycles (PP spin structure), the expected ratios are $T_{aa}/T_{00}=\exp[2\pi \mathrm{i}(a^2/6+a/2)]$, giving $T/T_{00}=\operatorname{diag}(1,\omega^2,\omega^2)$~\cite{Okuda2021SpinCS}.
The linear term $a/2$ accounts for the spin structure.
Figure~\ref{fig:wilson-loops}(c) and Table~\ref{tab:app-modular-data} show convergence to these values.

\begin{table}[htbp]
 \centering\small
 \caption{Relative modular phases $\arg(T_{aa}/T_{00})/(2\pi)$ modulo one.
 The $S$ residual is $\|S-S_{\rm ideal}\|_{\rm F}/\sqrt3$; the $T$ residual is $[\sum_{a\ne b}|(T_{\rm num})_{ab}|^2/3]^{1/2}$ before reference normalization.
 Eigenvector phases are fixed again after deleting the same sample block from all measured overlaps. A sweep consists of $L^2$ Metropolis proposals.}
 \label{tab:app-modular-data}
 \setlength{\tabcolsep}{5pt}
\begin{tabular}{rrrrrrr}
\hline\hline
$L$ & Steps & Sweeps/chain & $a=1$ & $a=2$ & $S$ residual & $T$ residual\\\hline
3 & 12 & Exact & $0.7323$ & $0.7292$ & $0.300$ & $0.356$ \\
5 & 12 & 16,000 & $0.7017\pm0.0009$ & $0.6983\pm0.0009$ & $0.146\pm0.004$ & $0.189\pm0.003$ \\
6 & 12 & 16,000 & $0.6914\pm0.0010$ & $0.6871\pm0.0012$ & $0.106\pm0.004$ & $0.134\pm0.004$ \\
7 & 12 & 16,000 & $0.6867\pm0.0012$ & $0.6839\pm0.0012$ & $0.071\pm0.004$ & $0.098\pm0.004$ \\
8 & 12 & 16,000 & $0.6840\pm0.0012$ & $0.6791\pm0.0014$ & $0.060\pm0.005$ & $0.082\pm0.005$ \\
9 & 12 & 16,000 & $0.6795\pm0.0014$ & $0.6766\pm0.0014$ & $0.048\pm0.005$ & $0.064\pm0.005$ \\
11 & 12 & 16,000 & $0.6768\pm0.0015$ & $0.6749\pm0.0015$ & $0.040\pm0.006$ & $0.052\pm0.005$ \\
12 & 12 & 16,000 & $0.6723\pm0.0016$ & $0.6712\pm0.0016$ & $0.038\pm0.006$ & $0.042\pm0.006$ \\
13 & 12 & 16,000 & $0.6687\pm0.0016$ & $0.6710\pm0.0018$ & $0.037\pm0.006$ & $0.040\pm0.006$ \\
15 & 12 & 16,000 & $0.6735\pm0.0022$ & $0.6706\pm0.0021$ & $0.020\pm0.007$ & $0.031\pm0.008$ \\
16 & 12 & 16,000 & $0.6697\pm0.0018$ & $0.6714\pm0.0021$ & $0.019\pm0.007$ & $0.026\pm0.007$ \\
24 & 12 & 4,000 & $0.6725\pm0.0076$ & $0.6524\pm0.0065$ & $0.008\pm0.010$ & $0.036\pm0.019$ \\
\hline\hline
\end{tabular}
\end{table}

\FloatBarrier

\section{Twofold rotation and Wilson-loop constraints}
\label{app:rotation-wilson-proof}

Twofold rotation reverses the Wilson loops and fixes their determinant signs through its eigenvalues at rotation-invariant twists.
We derive these constraints from the compact orbitals to establish the reference branch used in Sec.~\ref{app:wilson-modular-data}.
We implement the twists through phases on hoppings crossing the boundaries of a torus with $L_1=L_2=L$ unit cells along each lattice direction.

\subsection{Wilson-loop reversal under twofold rotation}
\label{app:c2-wilson-loops}
Twofold rotation sends $\mathbf k$ at parton twist $\boldsymbol\phi$ to $-\mathbf k$ at $-\boldsymbol\phi$. On the mesh $k_\mu=\mathbf k\cdot\mathbf T_\mu=(2\pi n_\mu+\phi_\mu)/L$, $n_\mu=0,\ldots,L-1$, we Fourier transform each parton creator at its position $\mathbf R$. For sublattice $a=1,2$, the Bloch creator and occupied orbital are
\begin{equation}
 \begin{aligned}
 f^\dagger_{\mathbf k,a,\alpha}
 &=L^{-1}\sum_{\mathbf R\in a}\mathrm{e}^{\mathrm{i}\mathbf k\cdot\mathbf R}f^\dagger_{\mathbf R,\alpha},
 &\mathcal S^\dagger_\alpha(\mathbf k)&=\sum_a s_a(\mathbf k)f^\dagger_{\mathbf k,a,\alpha},\\
 s_a(\mathbf k)&=\sum_{\mathbf R\in a\cap\mathcal N(\mathbf 0)}
 \mathrm{e}^{-\mathrm{i}\mathbf k\cdot\mathbf R}s_{\mathbf R,\mathbf 0},
 &S_{(a,\alpha),n}(\mathbf k)&=s_a(\mathbf k)\delta_{\alpha n}.
 \end{aligned}
 \label{eq:sm-site-bloch-convention}
\end{equation}
Here $\mathbf 0$ is the orbital center. Opposite bonds end on sublattices $a$ and $\bar a=3-a$ and have opposite amplitudes by Eq.~\ref{eq:app-tns-orbital}. On the finite torus, the representative of the rotated site is $\mathbf R'=L(\mathbf T_1+\mathbf T_2)-\mathbf R$, so $\mathrm{e}^{\mathrm{i}\mathbf k\cdot\mathbf R}=\mathrm{e}^{\mathrm{i}(\phi_1+\phi_2)}\mathrm{e}^{-\mathrm{i}\mathbf k\cdot\mathbf R'}$. Including the on-site rotation eigenvalue $-1$ gives
\begin{equation}
 C_{2z} f^\dagger_{\mathbf k,a,\alpha} C_{2z}^{-1}
 =-\mathrm{e}^{\mathrm{i}(\phi_1+\phi_2)} f^\dagger_{-\mathbf k,\bar a,\alpha},
 \qquad s_{\bar a}(-\mathbf k)=-s_a(\mathbf k).
 \label{eq:sm-bloch-c2}
\end{equation}
For $\mathbf k\ne\Gamma$ the two minus signs cancel in the occupied orbital. Filling the square mesh gives an even permutation under $\mathbf k\mapsto-\mathbf k$, and $C_{2z}$ commutes with $P_{\rm G}$. Normalizing the projected branches as $|\bar\Psi_{\boldsymbol\phi}\rangle=P_{\rm G}|\bar\Phi_{\boldsymbol\phi}\rangle/\|P_{\rm G}|\bar\Phi_{\boldsymbol\phi}\rangle\|$ gives
\begin{equation}
 \begin{aligned}
 C_{2z}\mathcal S^\dagger_\alpha(\mathbf k)C_{2z}^{-1}
 &=\mathrm{e}^{\mathrm{i}(\phi_1+\phi_2)}\mathcal S^\dagger_\alpha(-\mathbf k),\\
 C_{2z}|\bar\Psi_{\boldsymbol\phi}\rangle
 &=\mathrm{e}^{\mathrm{i}3L^2(\phi_1+\phi_2)}|\bar\Psi_{-\boldsymbol\phi}\rangle \qquad (\text{mesh avoiding }\Gamma).
 \end{aligned}
 \label{eq:app-projected-c2}
\end{equation}
At zero parton twist the mesh contains $\Gamma$, where $s_1(\Gamma)=s_2(\Gamma)=0$. We fill the limiting occupied orbital there. Indeed, $s_2(\mathbf k)=-s_1(-\mathbf k)$ makes the two linear terms equal, and $|s_1|^2+|s_2|^2\simeq 3|\mathbf k|^2/2$ [Eq.~\ref{eq:app-bulk-gaussian-kernel}] fixes the limiting sublattice line to $(1,1)^T$. Each on-site parton orbital transforms as $p_x-\mathrm{i} p_y$, which changes sign under a $\pi$ rotation. At $\Gamma$ the Bloch phase is unity, so Eq.~\ref{eq:sm-bloch-c2} gives
\begin{equation}
 \begin{aligned}
 f^\dagger_{\Gamma,+,\alpha}
 &=\frac{f^\dagger_{\Gamma,1,\alpha}+f^\dagger_{\Gamma,2,\alpha}}{\sqrt{2}},
 \qquad C_{2z}f^\dagger_{\Gamma,+,\alpha}C_{2z}^{-1}=-f^\dagger_{\Gamma,+,\alpha},\\
 C_{2z}|\bar\Psi_{\boldsymbol 0}\rangle
 &=(-1)^3|\bar\Psi_{\boldsymbol 0}\rangle .
 \end{aligned}
 \label{eq:sm-gamma-c2}
\end{equation}
The factor $(-1)^3$ comes from filling the $\Gamma$ orbital for all three colors. The remaining momenta give an even permutation since $\Gamma$ is fixed.

Together the two cases map the span of the twist branches at $\boldsymbol\varphi$ to that at $-\boldsymbol\varphi$. The overall phases drop out of the projector. The same holds for the conjugated states on which the Wilson loops act, since $C_{2z}$ is real in the occupation basis.
On a uniform grid around either flux cycle starting at zero twist, let $\Pi_j$ project onto this space at step $j$, with $\Pi_N=\Pi_0$. Then $C_{2z}\Pi_j C_{2z}^\dagger=\Pi_{N-j}$, and the rotated links traverse the Wilson loop in reverse, giving
\begin{equation}
 \begin{aligned}
 W_{\mathbf T_\mu}
 &=\prod_{j=0}^{N-1}\operatorname{polar}(\Pi_j\Pi_{j+1}),\\
 C_{2z}W_{\mathbf T_\mu}C_{2z}^\dagger
 &=\prod_{j=0}^{N-1}\operatorname{polar}(\Pi_{N-j}\Pi_{N-j-1})
 =W_{\mathbf T_\mu}^\dagger=W_{\mathbf T_\mu}^{-1}.
 \end{aligned}
 \label{eq:app-wilson-projector-product}
\end{equation}
The products are ordered from left to right in increasing $j$ and act on the ground-state space at zero twist.

\subsection{Wilson-loop determinants from rotation eigenvalues}
\label{app:c2-wilson-determinants}

The determinant of a Wilson loop is fixed by the $C_{2z}$ eigenvalues at its two invariant physical twists, $0$ and $\pi$~\cite{Alexandradinata2014Wilson}.
Consider the $\mathbf T_1$ cycle; the same steps apply to $\mathbf T_2$.
In the boundary convention used above, the projector obeys
\begin{equation}
 \Pi(\theta+2\pi,0)=\Pi(\theta,0),\qquad
 C_{2z}\Pi(\theta,0)C_{2z}^\dagger=\Pi(-\theta,0).
 \label{eq:sm-cycle-projector-symmetry}
\end{equation}
Thus $\Pi(0,0)$ and $\Pi(\pi,0)=\Pi(-\pi,0)$ are invariant under $C_{2z}$.
For $0=\theta_0<\cdots<\theta_N=\pi$, define the first half-cycle product by
\begin{equation}
 U_{\mathbf T_1}(0,\pi)=\prod_{j=0}^{N-1}
 \operatorname{polar}[\Pi(\theta_j,0)\Pi(\theta_{j+1},0)],
 \label{eq:sm-half-cycle-product}
\end{equation}
with the same left-to-right ordering as Eq.~\ref{eq:app-wilson-projector-product}.
It maps the subspace at $\pi$ to that at $0$.
Rotating each factor gives
\begin{equation}
 \begin{aligned}
 C_{2z}U_{\mathbf T_1}(0,\pi)C_{2z}^\dagger
 &=\prod_{j=0}^{N-1}\operatorname{polar}
 [\Pi(-\theta_j,0)\Pi(-\theta_{j+1},0)]\\
 &=\prod_{j=0}^{N-1}\operatorname{polar}
 [\Pi(2\pi-\theta_j,0)\Pi(2\pi-\theta_{j+1},0)]\\
 &=U_{\mathbf T_1}(2\pi,\pi)=U_{\mathbf T_1}(\pi,2\pi)^\dagger.
 \end{aligned}
 \label{eq:sm-rotated-half-cycle}
\end{equation}
The second equality uses $2\pi$ periodicity; the last reverses the path and the order of its factors.

Choose orthonormal bases $V_\mu(0)$ and $V_\mu(\pi)$ at the two invariant twists of each cycle, and define the rotation matrices in these bases by
\begin{equation}
 C_{2z}^{(\mu)}(\theta)=V_\mu(\theta)^\dagger C_{2z}V_\mu(\theta),\qquad
 C_{2z}^{(\mu)}(\theta)^2=I_3,\qquad \theta=0,\pi.
 \label{eq:sm-endpoint-rotation-matrices}
\end{equation}
At each endpoint, $-\theta$ and $\theta$ are identified by physical $2\pi$ periodicity. A change of orthonormal basis $V_\mu(\theta)\mapsto V_\mu(\theta)G$ with $G$ unitary conjugates $C_{2z}^{(\mu)}(\theta)$ by $G$, preserving its eigenvalues.
Writing the half-cycle products in the corresponding endpoint bases, Eq.~\ref{eq:sm-rotated-half-cycle} yields
\begin{equation}
 \begin{aligned}
 U_{\mathbf T_\mu}(\pi,2\pi)
 &=C_{2z}^{(\mu)}(\pi)U_{\mathbf T_\mu}(0,\pi)^\dagger C_{2z}^{(\mu)}(0)^\dagger,\\
 W_{\mathbf T_\mu}
 &=U_{\mathbf T_\mu}(0,\pi)U_{\mathbf T_\mu}(\pi,2\pi)\\
 &=U_{\mathbf T_\mu}(0,\pi)C_{2z}^{(\mu)}(\pi)U_{\mathbf T_\mu}(0,\pi)^\dagger C_{2z}^{(\mu)}(0)^\dagger.
 \end{aligned}
 \label{eq:sm-wilson-half-cycle-rotation}
\end{equation}
Let $\xi_{\mu,a}(\theta)=\pm1$ be the eigenvalues of $C_{2z}^{(\mu)}(\theta)$.
Since the half-cycle products are unitary between their endpoint spaces,
\begin{equation}
 \begin{aligned}
 \det W_{\mathbf T_\mu}
 &=\underbrace{\det U_{\mathbf T_\mu}(0,\pi)\,[\det U_{\mathbf T_\mu}(0,\pi)]^*}_{1}
   \det C_{2z}^{(\mu)}(\pi)\,[\det C_{2z}^{(\mu)}(0)]^*\\
 &=\prod_{a=1}^{3}\xi_{\mu,a}(\pi)\xi_{\mu,a}(0)\in\{+1,-1\}.
 \end{aligned}
 \label{eq:sm-wilson-fixed-point-eigenvalues}
\end{equation}
Assume the three branches $\phi_\mu=(\theta+2\pi a)/3$, $a=0,1,2$, with the other twist zero, span the ground-state space at each endpoint. Rotation sends $a\mapsto-a-\theta/\pi\pmod 3$. At $\theta=0$ the $a=0$ branch has zero parton twist and eigenvalue $(-1)^3$ by Eq.~\ref{eq:sm-gamma-c2}, while $a=1,2$ are exchanged. Since $C_{2z}^2=I$, we can choose the phases of each exchanged pair so that, in branch order $(0,1,2)$,
\begin{align}
 C_{2z}^{(\mu)}(0)&\sim
 \begin{pmatrix} (-1)^3&0&0\\0&0&1\\0&1&0\end{pmatrix},
 &C_{2z}^{(\mu)}(\pi)&\sim
 \begin{pmatrix}0&0&1\\0&\mathrm{e}^{\mathrm{i}3\pi L^2}&0\\1&0&0\end{pmatrix}.\label{eq:sm-invariant-twist-rotation-matrices}
\intertext{At $\theta=\pi$, the fixed branch is $a=1$ ($\phi_\mu=\pi$), and Eq.~\ref{eq:app-projected-c2} gives $\mathrm{e}^{\mathrm{i}3\pi L^2}=(-1)^L$. Identifying the $-\pi$ and $+\pi$ meshes permutes the occupied Bloch creators $\mathcal S^\dagger_\alpha(\mathbf k)$ with sign $+1$ per color and no sewing phase [Eq.~\ref{eq:sm-site-bloch-convention}]. Each exchanged pair contributes $-1$, so Eq.~\ref{eq:sm-wilson-fixed-point-eigenvalues} gives}
 \det W_{\mathbf T_\mu}
 &=\bigl[-(-1)^3\bigr]\bigl[-(-1)^L\bigr]
 =(-1)^{L+1}, &\mu&=1,2.
 \label{eq:sm-wilson-determinant-sign}
\end{align}

\FloatBarrier

\section{Bulk field theory of the projected norm}
\label{app:bulk-cft}

Inspired by the plasma analogy, one can view the projected wavefunction norm $\inn{\Psi|\Psi}$ as partition function of a $2$d classical statistical model, which can be further reformulated into a $(1+1)d$ path integral evolving under the transfer matrix of the TNS representation.
In this section, we first rewrite the norm as an integral of parton color rotation on the unprojected Slater determinants $\ket{\Phi}$, and divide it into a fermion sector and an $SU(3)$ color sector [Sec.~\ref{app:bulk-norm-construction}].
Then, the gradient expansion of the color sector gives the principal chiral model (PCM) action with a Wess--Zumino (WZ) phase, which flows to the $SU(3)_1$ Wess--Zumino--Witten (WZW) fixed point under coarse graining [Sec.~\ref{app:bulk-gradient-action}].
We then relate transfer eigenvalues to the energy, momentum, and color quantum numbers of the two CFT sectors [Sec.~\ref{app:bulk-transfer-energy-momentum}].
The final subsection establishes convergence and error bounds for the gradient expansion [Sec.~\ref{app:bulk-gradient-convergence}].

Throughout this section, $\Tr$ includes all matrix indices present, while $\tr$ denotes the fundamental color trace.
We label a parton site by its position $\mathbf R$ and use $\mathbf R_1,\mathbf R_2,\ldots$ for distinct site arguments. Continuum integrals use $d^2r=dx\,dy$.

\subsection{Fermion and color sectors of the projected norm}
\label{app:bulk-norm-construction}
The projected norm $\inn{\Psi|\Psi}=\inn{\Phi|P_{\rm G}|\Phi}$ separates into fermion and color contributions to a $(1+1)$-dimensional path integral. We derive this factorization and identify the uniform color configurations as saddles of maximal weight.  

The Gutzwiller projector $P_{\rm G}=\prod_{\mathbf R}P_{{\rm G},\mathbf R}$ can be written as an average over local color rotations $\hat g =\prod_{\mathbf R}\hat{g}_{\mathbf R}$.
The zero-, one-, two-, and three-parton states transform as $\mathbf1$, $\mathbf3$, $\overline{\mathbf3}$, and $\mathbf1$, respectively. Averaging over all $SU(3)$ rotations with equal group weight ($\int dg_{\mathbf R}=1$) retains only the two singlets, exactly as $P_{{\rm G},\mathbf R}$ does.
\begin{equation}
    \begin{aligned}
        P_{{\rm G},\mathbf R}=\int dg_{\mathbf R}\,\widehat g_{\mathbf R}
        =|000\rangle\langle000|+|111\rangle\langle111|\\
        \inn{\Psi|\Psi}=\inn{\Phi|P_{\rm G}|\Phi}=\int \prod_{\mathbf R}dg_{\mathbf R} \inn{\Phi|\hat g|\Phi}
    \end{aligned}
    \label{eq:app-bulk-haar}
\end{equation}

For a system with $N$ unit cells, the unnormalized $\ket{\Phi}$ occupies the $3N$ compact orbitals that form the columns of the $6N\times3N$ hopping matrix $S$ in Eq.~\ref{eq:app-tns-orbital}; boundary conditions exclude $\Gamma$ so these columns are linearly independent.
Let $\mathbf u$ be a set of $3N$ occupied site-color pairs $(\mathbf R,\alpha)$ in the ordering of Sec.~\ref{app:tns-column-transfer}. Its Slater amplitude is $\det S[\mathbf u]$, where $S[\mathbf u]$ selects the corresponding rows of $S$.
Writing $\hat{g}$ on the one-particle space as $g=\operatorname{diag}_{\mathbf R}(g_{\mathbf R})$, we have 
\begin{equation}
    \begin{aligned}
    \langle\Psi|\Psi\rangle
    &=\int\prod_{\mathbf R}dg_{\mathbf R}\,\inn{\Phi|\hat g|\Phi}
    =\int\prod_{\mathbf R}dg_{\mathbf R}\,
      \sum_{\mathbf u}(\det S[\mathbf u])^*\det\big((gS)[\mathbf u]\big)
    =\int\prod_{\mathbf R}dg_{\mathbf R}\,\det(S^\dagger gS).
    \end{aligned}
    \label{eq:app-bulk-slater-norm}
\end{equation}
Here $\mathbf u$ runs over all such configurations;
the last equality expands $\det(S^\dagger gS)$ as the sum of products of the corresponding $3N\times3N$ determinants.

Orthogonalizing the compact orbitals separates their normalization from the color rotations:
\begin{equation}
    D=S^\dagger S,\qquad U=SD^{-1/2},\qquad
    U^\dagger U=I_{3N},\qquad P=UU^\dagger=P_1\otimes I_3.
    \label{eq:app-bulk-orthogonalization}
\end{equation}
Because $S$ is identical for all colors, $P_1$ projects onto the occupied orbitals of one color and $I_3$ acts on the three colors.
The norm then factorizes into a Gaussian fermion determinant and an integral over the color field:
\begin{equation}
    \begin{aligned}
        \langle\Psi|\Psi\rangle&=\det D\,\mathcal Q,\qquad
        \mathcal Q=\int\prod_{\mathbf R} dg_{\mathbf R}\,\mathrm{e}^{-\mathcal S_G[g]},\\
        \mathcal S_G[g]&=-\log\det K[g],\qquad K[g]=U^\dagger gU.
    \end{aligned}
    \label{eq:app-bulk-statistical-action}
\end{equation}
Here $\mathcal S_G$ is defined modulo $2\pi \mathrm{i}$.
Since $U$ is color independent, $\mathcal{S}_G$ is invariant under two independent global color rotations: $g_{\mathbf R}\mapsto U_{\mathrm b}^\dagger g_{\mathbf R}U_{\mathrm k}$, with $U_{\mathrm k},U_{\mathrm b}\in SU(3)$.
For uniform $g_{\mathbf R}=g_0$, $K=I_{N}\otimes g_0$ and $\mathcal{S}_G[g_0]=0$. 
Below, $\det D$ gives the free-fermion sector, while $\mathcal{Q}$ describes the $SU(3)$ color field. Their independent contributions to the low-energy transfer spectrum are identified in Sec.~\ref{app:bulk-transfer-energy-momentum}.

The fermion factor is $\det D=\int\mathcal D\bar\chi\mathcal D\chi\,\mathrm{e}^{-\bar\chi D\chi}$, with independent Grassmann vectors $\chi,\bar\chi$ indexed by compact orbitals. Its kernel $D=S^\dagger S$ uses the same orbital coefficients as the parton Hamiltonian $H=-SS^\dagger$. 
\begin{equation}
    \begin{aligned}
    D(\mathbf k)=S^{\dagger}(\bf k)S(\bf{k})=d(\mathbf k)I_3,\qquad
    d(\mathbf k)&=6-2\cos k_1-2\cos k_2-2\cos(k_1+k_2)\\
    &\xlongequal{\bf{k}\rightarrow\Gamma}\frac32|\mathbf k|^2+O(|\mathbf k|^4).
    \end{aligned}
    \label{eq:app-bulk-gaussian-kernel}
\end{equation}
Here $k_i=\mathbf k\cdot\mathbf T_i$ with $\mathbf T_i$ the primitive translation vectors in Fig.~\ref{fig:mean-field}(a).
Since $\det D=\langle\Phi|\Phi\rangle$, Sec.~\ref{app:unprojected-transfer} identifies its $U(3)_1$ Dirac spectrum directly from the transfer matrix. At $\phi_2=\pi$, the lowest single-particle additions of Eq.~\ref{eq:app-unprojected-transfer-eigenvalues} give triplets at weights $(1/2,0)$ and $(0,1/2)$, consistent with the projected transfer levels in Table~\ref{tab:transfer-branches}.

The stationary configurations of $\mathcal{S}_{G}[g]$ follow from its equation of motion. Consider the variation $g_{\mathbf R}\mapsto g_{\mathbf R}\mathrm{e}^{\epsilon h_{\mathbf R}}$ with $h_{\mathbf R}^\dagger=-h_{\mathbf R}$ and $\tr h_{\mathbf R}=0$, writing $h=\operatorname{diag}_{\mathbf R}(h_{\mathbf R})$ and denote $\frac{d}{d{\epsilon}}$ by a dot. Then
\begin{equation}
    K=U^\dagger g U,\qquad
    \dot K=U^\dagger g hU,\qquad
    \ddot K=U^\dagger g h^2U.
    \label{eq:app-bulk-uniform-kernel-derivatives}
\end{equation}
The action varies as
\begin{equation}
    \delta_h\mathcal{S}_G[g]:=\left.\frac{d}{d\epsilon}\mathcal S_G[ge^{\epsilon h}]\right|_{\epsilon=0}
    =-\Tr(K^{-1}\dot K)
    =-\sum_{\mathbf R}\tr\!\left[h_{\mathbf R}
        (UK^{-1}U^\dagger g)_{\mathbf R,\mathbf R}\right].
    \label{eq:app-bulk-saddle-variation}
\end{equation}
Stationarity for any traceless $h$ gives the equation of motion
\begin{equation}
    (UK^{-1}U^\dagger g)_{\mathbf R,\mathbf R}
    =\frac13\tr\!\left[(UK^{-1}U^\dagger g)_{\mathbf R,\mathbf R}\right]I_3,
    \label{eq:app-bulk-saddle-equation}
\end{equation}
i.e., the traceless part of $(UK^{-1}U^\dagger g)_{\mathbf R,\mathbf R}$ vanishes.
A uniform $g_{\mathbf R}=g_0$ has $K=I_N\otimes g_0$ and $UK^{-1}U^\dagger g=P_1\otimes I_3$, which satisfies the equation of motion.
Also, $K=U^\dagger gU$ cannot increase a vector's norm: $U$ preserves it, $g$ is unitary, and $U^\dagger$ projects back onto the occupied orbitals. Thus every singular value of $K$ is at most one and $|\mathrm{e}^{-\mathcal S_G}|=|\det K|\leq1$. Uniform $g_0$ attains this maximum.

To test whether small changes increase the action near a uniform $g_0$, we calculate $\frac{d^2\mathcal S_G}{d\epsilon^2}$.
\begin{equation}
    \begin{aligned}
    \delta_h^2\mathcal{S}_G[g_0]:=\left.\frac{d^2}{d\epsilon^2}\mathcal S_G[g_0\mathrm{e}^{\epsilon h}]\right|_{\epsilon=0}
    &=\left.\Tr\!\left[K^{-1}\dot K K^{-1}\dot K-K^{-1}\ddot K\right]\right|_{\epsilon=0}\\
    &=\Tr\!\left[(U^\dagger hU)^2-U^\dagger h^2U\right]
      =\Tr(PhPh-Ph^2).
    \end{aligned}
    \label{eq:app-bulk-uniform-hessian-trace}
\end{equation}
Here we use $K\frac{d K^{-1}}{d\epsilon}+\dot K K^{-1}=0$; color independence gives $(I_{2N}\otimes g_0)U=U(I_N\otimes g_0)$ and cancels the $g_0$ factors. Using $P_1^\dagger=P_1$ and $P_1^2=P_1$, the second variation becomes
\begin{equation}
    \begin{aligned}
    \delta_h^2\mathcal{S}_G[g_0]
    &=-\sum_{\mathbf R}(P_1)_{\mathbf R,\mathbf R}\tr h_{\mathbf R}^2
      +\sum_{\mathbf R_1,\mathbf R_2}|(P_1)_{\mathbf R_1,\mathbf R_2}|^2
        \tr(h_{\mathbf R_1}h_{\mathbf R_2})\\
    &=-\frac12\sum_{\mathbf R_1,\mathbf R_2}|(P_1)_{\mathbf R_1,\mathbf R_2}|^2
        \tr\!\left[h_{\mathbf R_1}^2+h_{\mathbf R_2}^2-2h_{\mathbf R_1}h_{\mathbf R_2}\right]\\
    &=\frac12\sum_{\mathbf R_1,\mathbf R_2}|(P_1)_{\mathbf R_1,\mathbf R_2}|^2
        \tr\!\left[(h_{\mathbf R_1}-h_{\mathbf R_2})^\dagger(h_{\mathbf R_1}-h_{\mathbf R_2})\right]\geq0.
    \end{aligned}
    \label{eq:app-bulk-uniform-hessian}
\end{equation}
The quadratic cost is strictly positive for every nonuniform perturbation because the nonzero matrix elements of $P_1$ connect all physical sites. We show this first within each sublattice and then between them, using the real-space projector in the bulk limit:
\begin{equation}
    (P_1)_{\mathbf R_1,\mathbf R_2}
    =\left\langle \mathrm{e}^{\mathrm{i}\mathbf k\cdot(\mathbf R_1-\mathbf R_2)}
      [P_1(\mathbf k)]_{aa'}\right\rangle_{\rm BZ},\qquad
    a=a(\mathbf R_1),\quad a'=a(\mathbf R_2).
    \label{eq:app-bulk-projector-real-space}
\end{equation}
Here $a(\mathbf R)$ denotes the sublattice at $\mathbf R$, and the Brillouin-zone average is $\langle f\rangle_{\rm BZ}=A_{\rm cell}\int_{\rm BZ}\frac{d^2k}{(2\pi)^2}\,f$, with $A_{\rm cell}=\sqrt3/2$.
The occupied orbital in Eq.~\ref{eq:sm-site-bloch-convention} gives the diagonal entries
\begin{equation}
    [P_1(\mathbf k)]_{11}=1-[P_1(\mathbf k)]_{22}
    =\frac12-\frac{\sqrt3 F(\mathbf k)}{d(\mathbf k)},\qquad
    F(\mathbf k)=\sin(k_1+k_2)-\sin k_1-\sin k_2,
    \label{eq:app-bulk-projector-diagonal}
\end{equation}
where $k_i=\mathbf k\cdot\mathbf T_i$ and
$d=6-2\cos k_1-2\cos k_2-2\cos(k_1+k_2)$ as in Eq.~\ref{eq:app-bulk-gaussian-kernel}.
Since $F$ is odd under $\mathbf k\mapsto-\mathbf k$, only its sine transform contributes at nonzero separation.
The changes of variables $k_1\leftrightarrow k_2$ and $(k_1,k_2)\mapsto(k_1,-k_1-k_2)$ give
\begin{equation}
    \begin{aligned}
    \left\langle\frac{F\sin k_1}{d}\right\rangle_{\rm BZ}
    &=\left\langle\frac{F\sin k_2}{d}\right\rangle_{\rm BZ}
      =-\left\langle\frac{F\sin(k_1+k_2)}{d}\right\rangle_{\rm BZ}
      =-\frac13\left\langle\frac{F^2}{d}\right\rangle_{\rm BZ},\\
    |(P_1)_{\mathbf R_1,\mathbf R_2}|^2
    &=\frac13\left\langle\frac{F^2}{d}\right\rangle_{\rm BZ}^{\!2}
      \equiv J>0,\qquad \mathbf R_1-\mathbf R_2=\mathbf T_\mu,\quad\mu=1,2.
    \end{aligned}
    \label{eq:app-bulk-projector-connected-links}
\end{equation}
Each term in Eq.~\ref{eq:app-bulk-uniform-hessian} is nonnegative, and the links in Eq.~\ref{eq:app-bulk-projector-connected-links} therefore require $h_{\mathbf R_1}=h_{\mathbf R_2}$ on both sets of links for zero cost.
Since $\mathbf T_1,\mathbf T_2$ connect every site within each sublattice, a zero-cost field must be a constant matrix $h^{(1)}$ on sublattice 1 and a constant matrix $h^{(2)}$ on sublattice 2.
Any spatial variation within either sublattice has positive cost.

A relative rotation between the two sublattices also has positive cost.
Equation~\ref{eq:app-bulk-projector-gamma} gives $[P_1(\Gamma)]_{12}=1/2$ and continuity at $\Gamma$; hence $|[P_1(\mathbf k)]_{12}|>1/4$ throughout a sufficiently small disk around $\Gamma$, which has nonzero area.
Summing over sites on sublattice 2 and using the orthogonality of the Fourier components gives a positive total weight of cross-sublattice links from each site on sublattice 1:
\begin{equation}
    \sum_{\mathbf R_2:\,a(\mathbf R_2)=2}|(P_1)_{\mathbf R_1,\mathbf R_2}|^2
    =\left\langle|[P_1(\mathbf k)]_{12}|^2\right\rangle_{\rm BZ}>0,
    \qquad a(\mathbf R_1)=1.
    \label{eq:app-bulk-intersublattice-weight}
\end{equation}
For a field constant on each sublattice, $h_{\mathbf R}=h^{(a(\mathbf R))}$, the same-sublattice terms vanish, while the terms with $a(\mathbf R_1)\ne a(\mathbf R_2)$ in Eq.~\ref{eq:app-bulk-uniform-hessian} give the cost per unit cell
\begin{equation}
    \frac1N\left.\frac{d^2\mathcal S_G[g_0\mathrm{e}^{\epsilon h}]}{d\epsilon^2}\right|_{\epsilon=0}
    =\left\langle|[P_1]_{12}|^2\right\rangle_{\rm BZ}
      \tr\!\left[(h^{(1)}-h^{(2)})^\dagger(h^{(1)}-h^{(2)})\right].
    \label{eq:app-bulk-relative-rotation-cost}
\end{equation}
It vanishes only when $h^{(1)}=h^{(2)}$.
Thus zero cost requires a common spatially uniform color rotation.

All eight color directions have the same spatial cost.
Expand $h_{\mathbf R}=\mathrm{i}\sum_{A=1}^8h_{\mathbf R}^AT^A$, with real $h_{\mathbf R}^A$ and Hermitian generators normalized by $\tr(T^AT^B)=\delta^{AB}/2$.
Equation~\ref{eq:app-bulk-uniform-hessian} becomes
\begin{equation}
    \left.\frac{d^2\mathcal S_G[g_0\mathrm{e}^{\epsilon h}]}{d\epsilon^2}\right|_{\epsilon=0}
    =\frac14\sum_{A,\mathbf R_1,\mathbf R_2}
      |(P_1)_{\mathbf R_1,\mathbf R_2}|^2(h_{\mathbf R_1}^A-h_{\mathbf R_2}^A)^2.
    \label{eq:app-bulk-color-hessian-real-space}
\end{equation}
There are no mixed terms between different $A$, and the coefficient $|(P_1)_{\mathbf R_1,\mathbf R_2}|^2$ is the same for all eight components.
The quadratic cost is therefore positive for every nonuniform $h$, with only the eight common uniform rotations leaving the action unchanged.

\subsection{Color action and the WZW fixed point}
\label{app:bulk-gradient-action}
We expand $\mathcal S_G[g]$ in the characteristic wave number $|\mathbf q|\ll1$ of $g_{\mathbf R}$.
The field may accumulate a finite change across the system, $\max_{\mathbf R_1,\mathbf R_2}\|g_{\mathbf R_1}-g_{\mathbf R_2}\|=O(1)$, so this is more general than a small-amplitude perturbation about a uniform saddle, $g_0\mathrm{e}^{\epsilon h}$.
We first expand the variation $\frac{d\mathcal{S}_G[ge^{\epsilon h}]}{d\epsilon}$ for arbitrary smooth $g,h$, then integrate it to recover the action with $\mathcal S_G[I]=0$. This organizes the gradient expansion without restricting the field amplitude.
\begin{equation}
    \begin{aligned}
        \delta_h\mathcal{S}_G[g]
        &=-\Tr(K^{-1}U^\dagger ghU)\\
        &=-\Tr(Ph)-\Tr[K^{-1}U^\dagger g(1-P)hU]\\
        &=\Tr[K^{-1}U^\dagger[g,P][h,P]U].
    \end{aligned}
    \label{eq:app-bulk-exact-variation}
\end{equation}
We insert $P+(1-P)$ between $g$ and $h$ in the second equality, and the first term vanishes since $\Tr(Ph)=\sum_{\mathbf R}(P_1)_{\mathbf R,\mathbf R}\tr h_{\mathbf R}=0$.
In the last equality, we use $U^\dagger P=U^\dagger$ and $PU=U$:
\begin{equation}
    \begin{aligned}
        -U^\dagger g(1-P)=U^\dagger gP-U^\dagger g=U^\dagger gP-U^\dagger Pg=U^\dagger[g,P]\\
        (1-P)hU=hU-PhU=hPU-PhU=[h,P]U.
    \end{aligned}
    \label{eq:app-bulk-projected-commutators}
\end{equation}
The commutators then supply the small gradient factors of order $|\mathbf q|$:
\begin{equation}
    [f,P]_{\mathbf R_1,\mathbf R_2}=(P_1)_{\mathbf R_1,\mathbf R_2}[f_{\mathbf R_1}-f_{\mathbf R_2}],
    \qquad f=g,h.
    \label{eq:app-bulk-field-differences}
\end{equation}

Define the positive matrix
\begin{equation}
    E=I-K^\dagger K
    =U^\dagger g^\dagger(1-P)gU
    =U^\dagger[P,g]^\dagger[P,g]U\geq0.
    \label{eq:app-bulk-leakage}
\end{equation}
It measures how much $g$ takes occupied orbitals outside the occupied subspace. We use the operator norm $\|A\|_{\rm op}=\sup_{\|v\|_2=1}\|Av\|_2$, the largest singular value of $A$.
The two commutators in $E$ give the bound $\|E\|_{\rm op}=O(|\mathbf q|^2)$, proved in Sec.~\ref{app:bulk-gradient-convergence}. For sufficiently slow variation, the following series therefore converge:
\begin{equation}
    \begin{aligned}
        K^{-1}&=(I-E)^{-1}K^\dagger
        =\sum_{n=0}^{\infty}E^nK^\dagger,\\
        \delta_h\mathcal S_G
        &=\sum_{n=0}^{\infty}
        \Tr\!\left[E^nK^\dagger U^\dagger[g,P][h,P]U\right].
    \end{aligned}
    \label{eq:app-bulk-variation-series}
\end{equation}
The leading term with $n=0$ gives
\begin{equation}
    \begin{aligned}
        \delta_h\mathcal S_G
        &=\Tr\!\left(Pg^{-1}[g,P][h,P]\right)+O(N|\mathbf q|^4)\\
        &=\sum_{\mathbf R_1,\mathbf R_2,\mathbf R_3}(P_1)_{\mathbf R_1,\mathbf R_2}(P_1)_{\mathbf R_2,\mathbf R_3}(P_1)_{\mathbf R_3,\mathbf R_1}
        \tr[g_{\mathbf R_2}^{-1}(g_{\mathbf R_2}-g_{\mathbf R_3})(h_{\mathbf R_3}-h_{\mathbf R_1})]+O(N|\mathbf q|^4),
    \end{aligned}
    \label{eq:app-bulk-leading-trace}
\end{equation}
where $N$ is the number of unit cells and the remainder $O(N|\mathbf q|^4)$ is estimated in Sec.~\ref{app:bulk-gradient-convergence}.

Expanding each field difference to first order in the displacement expresses the leading variation through momentum derivatives of the projector: 
\begin{equation}
    \begin{aligned}
    \delta_h\mathcal S_G[g]
    &=\sum_{\mathbf R_1,\mathbf R_2,\mathbf R_3}
       (P_1)_{\mathbf R_1,\mathbf R_2}(P_1)_{\mathbf R_2,\mathbf R_3}(P_1)_{\mathbf R_3,\mathbf R_1}
       (\mathbf R_2-\mathbf R_3)_\mu(\mathbf R_3-\mathbf R_1)_\nu \tr\!\left[g_{\mathbf R_2}^{-1}(\partial_\mu g)_{\mathbf R_3}(\partial_\nu h)_{\mathbf R_3}\right]
       +O\!\left(N|\mathbf q|^4\ln\frac1{|\mathbf q|}\right)\\
    &=-\int d^2r\int_{\rm BZ}\frac{d^2k}{(2\pi)^2}
       \tr_{\rm orb}\!\left[P_1(\bf k)\partial_{k_\mu}\!P_1(\bf k)\partial_{k_\nu}\!P_1(\bf k)\right]
       \tr\!\left[g^{-1}(\mathbf r)\partial_\mu g(\mathbf r)\partial_\nu h(\mathbf r)\right]
       +O\!\left(N|\mathbf q|^4\ln\frac1{|\mathbf q|}\right)\\
    &=-\int d^2r\int_{\rm BZ}\frac{d^2k}{(2\pi)^2}
       \Tr\!\left[P(\partial_{k_\mu}P)(\partial_{k_\nu}P)
       g^{-1}(\partial_\mu g)(\partial_\nu h)\right]
       +O\!\left(N|\mathbf q|^4\ln\frac1{|\mathbf q|}\right).
    \end{aligned}
    \label{eq:app-bulk-geometric-variation}
\end{equation}
Repeated $\mu,\nu=x,y$ are summed over the orthogonal Cartesian directions; $\tr_{\rm orb}$ traces over the two sublattices.
We use the Fourier convention
\[
    (P_1)_{\mathbf R_1,\mathbf R_2}
    =A_{\rm cell}\int_{\rm BZ}\frac{d^2k}{(2\pi)^2}
      \mathrm{e}^{\mathrm{i}\mathbf k\cdot(\mathbf R_1-\mathbf R_2)}[P_1(\mathbf k)]_{aa'},
\]
where $a=a(\mathbf R_1)$ and $a'=a(\mathbf R_2)$, and $A_{\rm cell}=\sqrt3/2$.
The remainder of order $O(N|\mathbf q|^4\ln\frac1{|\mathbf q|})$ is estimated in Sec.~\ref{app:bulk-gradient-convergence}.

Using the total quantum metric and Berry curvature of the dispersive bands,
\begin{equation}
    \begin{aligned}
        &\mathfrak g_{\mu\nu}
        =\frac12\Tr[(\partial_{k_\mu}P)(\partial_{k_\nu}P)],\qquad
        \Omega=\mathrm{i}\Tr\!\left(P[\partial_{k_x}P,\partial_{k_y}P]\right),\\
        &\Tr[P(\partial_{k_\mu}P)(\partial_{k_\nu}P)]
        =\mathfrak g_{\mu\nu}-\frac{\mathrm{i}}{2}\Omega\epsilon_{\mu\nu},
        \qquad \epsilon_{xy}=-\epsilon_{yx}=1,
    \end{aligned}
    \label{eq:app-bulk-total-geometry}
\end{equation}
Eq.~\ref{eq:app-bulk-geometric-variation} becomes
\begin{equation}
    \delta_h\mathcal S_G[g]
    \approx-\frac13\int d^2r\left[\int_{\rm BZ}\frac{d^2k}{(2\pi)^2}
    \left(\mathfrak g_{\mu\nu}-\frac{\mathrm{i}}{2}\Omega\epsilon_{\mu\nu}\right)\right]
    \tr(J_\mu\partial_\nu h),
    \qquad J_\mu=g^{-1}\partial_\mu g.
    \label{eq:app-bulk-split-variation}
\end{equation}
The factor $1/3$ removes the color multiplicity already included in the total band geometry.
The matrix $J_\mu=g^{-1}\partial_\mu g$ measures the spatial change of $g$ and is anti-Hermitian.
For smooth $g$, we have the identity $\partial_xJ_y-\partial_yJ_x=-[J_x,J_y]$, and integration by parts gives
\begin{equation}
    \begin{aligned}
        \delta_h\mathcal S_G[g]
        &\approx-2\kappa_{\mu\nu}\int d^2r\,\tr(J_\mu\partial_\nu h)
        -\frac{\mathrm{i}k}{4\pi}\int d^2r\,\tr(h[J_x,J_y]),\\
        \kappa_{\mu\nu}&=\frac16\int_{\rm BZ}\frac{d^2k}{(2\pi)^2}\,\mathfrak g_{\mu\nu},\qquad
        k=\frac{C}3=1.
    \end{aligned}
    \label{eq:app-bulk-pcm-wz-variation}
\end{equation}
The first term is the variation of the PCM action, which penalizes spatial changes of $g$; the second is the variation of the WZ phase~\cite{Witten1984Bosonization}. The two functionals are
\begin{equation}
    \begin{aligned}
        \mathcal S_{\rm PCM}[g]&=\kappa_{\mu\nu}\int_\Sigma d^2r\,
        \tr(\partial_\mu g^\dagger\partial_\nu g)=-\kappa_{\mu\nu}\int_\Sigma d^2r\,
        \tr(J_\mu J_\nu),\\
        \mathcal S_{\rm WZ}[g]&=\frac1{12\pi}\int_B
        \tr[(\widetilde g^{-1} d\widetilde g)^{\wedge3}]=\frac1{12\pi}\int_B
        \tr(\widetilde J^{\wedge3}),\\
        &=\frac1{12\pi}\int_B d^3u\sum_{a,b,c=1}^3\epsilon^{abc}
        \tr\!\left[(\widetilde g^{-1}\partial_a\widetilde g)
        (\widetilde g^{-1}\partial_b\widetilde g)
        (\widetilde g^{-1}\partial_c\widetilde g)\right].
    \end{aligned}
    \label{eq:app-bulk-pcm-wz-definitions}
\end{equation}
The second expression for the PCM action uses $J_\mu^\dagger=(\partial_\mu g^\dagger)g=-g^\dagger\partial_\mu g=-J_\mu$.
In $\mathcal S_{\rm WZ}$, $\widetilde g$ extends $g$ into a three-dimensional region $B$ with boundary $\Sigma$, and $\widetilde J=\widetilde g^{-1}d\widetilde g$. 
Different extensions change $\mathcal S_{\rm WZ}$ by an integer multiple of $2\pi$, so integer $k$ makes the weight well defined.
$\wedge$ indicates the wedge product, $du^a\wedge du^b=-du^b\wedge du^a$ where $u=(u^1,u^2,u^3)$ is the local coordinate on $B$; its component sum is written above where $\epsilon^{abc}$ is the totally antisymmetric tensor with $\epsilon^{123}=1$.

To verify the variation of $\mathcal S_{\rm PCM}$ and $\mathcal S_{\rm WZ}$, we first derive the variation of $J_\mu$.
Since $\delta_hg=gh$ and $\delta_hg^{-1}=-g^{-1}(\delta_hg)g^{-1}=-hg^{-1}$,
\begin{equation}
    \begin{aligned}
    \delta_hJ_\mu
    &=(\delta_hg^{-1})\partial_\mu g+g^{-1}\partial_\mu(\delta_hg)
    =-hJ_\mu+g^{-1}\partial_\mu(gh)\\
    &=-hJ_\mu+J_\mu h+\partial_\mu h
    =\partial_\mu h+[J_\mu,h].
    \end{aligned}
    \label{eq:app-bulk-current-variation}
\end{equation}
For $\delta_h\mathcal S_{\rm PCM}$, $\tr(ABC)=\tr(BCA)$ cancels the commutator terms. Using $\kappa_{\mu\nu}=\kappa_{\nu\mu}$ gives
\begin{equation}
    \delta_h\mathcal S_{\rm PCM}
    =-\kappa_{\mu\nu}\int_\Sigma d^2r\,\tr[\delta_h(J_\mu J_\nu)]
    =-2\kappa_{\mu\nu}\int_\Sigma d^2r\,\tr(J_\mu\partial_\nu h),
    \label{eq:app-bulk-pcm-variation}
\end{equation}
which is the first term in Eq.~\ref{eq:app-bulk-pcm-wz-variation}.
The antisymmetric second term in Eq.~\ref{eq:app-bulk-pcm-wz-variation} is supplied by the WZ action.

The WZ variation reduces to a boundary integral.
Let $\widetilde h$ be a smooth extension of $h$ into $B$, with $\widetilde h|_\Sigma=h$ and $\delta_{\widetilde h}\widetilde g=\widetilde g\widetilde h$. The same calculation as for $J_\mu$ gives $\delta_{\widetilde h}\widetilde J=d\widetilde h+[\widetilde J,\widetilde h]$.
Starting from the action in Eq.~\ref{eq:app-bulk-pcm-wz-definitions}, we obtain
\begin{equation}
    \begin{aligned}
    \delta_{\widetilde h}\mathcal S_{\rm WZ}
    &=\frac1{12\pi}\int_B\delta_{\widetilde h}\tr(\widetilde J^{\wedge3})\\
    &=\frac1{12\pi}\int_B\tr\!\left[(\delta_{\widetilde h}\widetilde J)\wedge\widetilde J\wedge\widetilde J
      +\widetilde J\wedge(\delta_{\widetilde h}\widetilde J)\wedge\widetilde J
      +\widetilde J\wedge\widetilde J\wedge(\delta_{\widetilde h}\widetilde J)\right]\\
    &=\frac1{4\pi}\int_B\tr\!\left[(d\widetilde h+[\widetilde J,\widetilde h])\wedge\widetilde J^{\wedge2}\right]
    =\frac1{4\pi}\int_B\tr(d\widetilde h\wedge\widetilde J^{\wedge2}).
    \end{aligned}
    \label{eq:app-bulk-wz-cubic-variation}
\end{equation}
The commutator contribution cancels because $\tr(\widetilde J\widetilde h\wedge\widetilde J^{\wedge2})=\tr(\widetilde h\widetilde J^{\wedge3})$.
Moreover, applying the product rule to $\widetilde J=\widetilde g^{-1}d\widetilde g$ and using $d^2\widetilde g=0$ gives
\begin{equation}
    \begin{aligned}
    d\widetilde J
    &=d\widetilde g^{-1}\wedge d\widetilde g
      +\widetilde g^{-1}d^2\widetilde g\\
    &=-\widetilde g^{-1}d\widetilde g\,\widetilde g^{-1}\wedge d\widetilde g
      =-\widetilde J\wedge\widetilde J,\\
    d(\widetilde J^{\wedge2})
    &=d\widetilde J\wedge\widetilde J-\widetilde J\wedge d\widetilde J
      =-\widetilde J^{\wedge3}+\widetilde J^{\wedge3}=0.
    \end{aligned}
    \label{eq:app-bulk-wz-current-square-closed}
\end{equation}
Thus, $\tr(d\widetilde h\wedge\widetilde J^{\wedge2})=d\,\tr(\widetilde h\widetilde J^{\wedge2})$ and
\begin{equation}
    \begin{aligned}
    \delta_{\widetilde h}\mathcal S_{\rm WZ}
    &=\frac1{4\pi}\int_B d\,\tr(\widetilde h\widetilde J^{\wedge2})
      =\frac1{4\pi}\int_\Sigma\tr[h(g^{-1}dg)^{\wedge2}]\\
    &=\frac1{4\pi}\int_\Sigma d^2r\,\tr(h[J_x,J_y]).
    \end{aligned}
    \label{eq:app-bulk-wz-boundary-variation}
\end{equation}
Using $\widetilde h|_\Sigma=h$ in the boundary integral, $\delta_h\mathcal S_{\rm PCM}-\mathrm{i}k\delta_{\widetilde h}\mathcal S_{\rm WZ}$ reproduces Eq.~\ref{eq:app-bulk-pcm-wz-variation}.
Consider a smooth path $\{g(s)|g(0)=I, g(1)=g,s\in[0,1]\}$ whose tangent $h(s)=g(s)^{-1}\partial_s g(s)$ is also slowly varying.
Integrating $\delta_{h} \mathcal S_G$ along this path, with $\mathcal S_G[I]=0$, recovers $\mathcal{S}_G[g]$ with the same remainder estimate:
\begin{equation}
    \mathcal S_G[g]=\mathcal S_{\rm PCM}[g]
    -\mathrm{i}\frac{C}{3}\mathcal S_{\rm WZ}[g]
    +O\!\left(N|\mathbf q|^4\ln\frac1{|\mathbf q|}\right)\pmod{2\pi \mathrm{i}}.
    \label{eq:app-bulk-leading-action}
\end{equation}
The parton mean-field Hamiltonian enters the PCM--WZ action through the stiffness $\kappa_{\mu\nu}$ and the signed coefficient $-\mathrm{i}C/3$ of $\mathcal S_{\rm WZ}$. The affine level is $|C|/3=1$. 
$C_{6z}$ symmetry enforces $\kappa_{\mu\nu}=\kappa\delta_{\mu\nu}$.
The pointwise bound $\mathfrak g_{xx}+\mathfrak g_{yy}\geq|\Omega|$~\cite{Roy2014Geometry} implies
\begin{equation}
    \kappa=\frac1{12}\int_{\rm BZ}\frac{d^2k}{(2\pi)^2}\,(\mathfrak g_{xx}+\mathfrak g_{yy})
    \geq\frac1{12}\int_{\rm BZ}\frac{d^2k}{(2\pi)^2}\,|\Omega|
    \geq\frac{|C|}{24\pi}
    =\frac1{8\pi}\equiv\kappa_*.
    \label{eq:app-bulk-stiffness-bound}
\end{equation}
Equality holds when the ideal trace condition $\mathfrak g_{xx}+\mathfrak g_{yy}=|\Omega|$ is saturated throughout the Brillouin zone and the Berry curvature does not change sign~\cite{Wang2021IdealFlatband}.
Hence $\kappa-\kappa_*$ measures the integrated departure from ideal band geometry.
In the present lattice units, the full Bloch column also gives $\mathfrak g_{xx}+\mathfrak g_{yy}+\Omega=1$.
With the Brillouin-zone area $|{\rm BZ}|=8\pi^2/\sqrt3$, this fixes $\kappa=1/(6\sqrt3)-1/(8\pi)\simeq1.41840\kappa_*$.
For the PCM--WZ action with level $1$, the one-loop flow is~\cite{Demulder2018WZ}
\begin{equation}
    \frac{d\kappa}{d\ell}
    =-\frac3{8\pi}\left[1-\left(\frac{\kappa_*}{\kappa}\right)^2\right]
    +\text{higher-loop terms},\qquad \ell=\ln(\text{length}).
    \label{eq:app-bulk-rg}
\end{equation}
The inequality Eq.~\ref{eq:app-bulk-stiffness-bound} enforces $\kappa\geq\kappa_*$, where the one-loop flow approaches $\kappa_*$ from above.
The fixed-point stiffness $\kappa=\kappa_*$ is precisely the value that saturates the ideal trace condition. At this point the theory is the $SU(3)_1$ WZW CFT with central charge $c=2$~\cite{Witten1984Bosonization}.
Beyond the one-loop expansion, the massless thermodynamic Bethe ansatz proposed for the integrable PCM with a WZ term gives consistent results~\cite{Ravanini1992TBA}:
specializing its central-charge formulas to $SU(3)$ at level $1$ gives $c_{\rm UV}=8$ and $c_{\rm IR}=2$.
We numerically test the corresponding transfer spectrum in Sec.~\ref{app:transfer-spectrum-data}.

The effective PCM-WZ action $\mathcal{S}_G[g]$ is invariant under global bra and ket $SU(3)$ rotations $g\mapsto U_{\mathrm b}^\dagger gU_{\mathrm k}$.
At the WZW fixed point, the corresponding Noether currents of the bra and ket $SU(3)$ rotations have definite chiralities~\cite{Witten1984Bosonization}.
With the WZ sign in Eq.~\ref{eq:app-bulk-leading-action} ($C/3=+1$), the level-one WZW Noether current is antiholomorphic for ket rotation and holomorphic for bra rotation.
Also, since each rotation acts on site along the cut [Eqs.~\ref{eq:app-ket-cut-rotation} and \ref{eq:app-bra-cut-rotation}], anomaly cancellation requires equal total holomorphic and antiholomorphic current levels for the bra and ket rotations~\cite{Wen2013GaugeAnomalies,Bastianelli1997AnomalyMatching}.
Such a cancellation must be realized by another low-energy $U(3)_1$ Dirac sector whose three complex fermions supply a single $SU(3)_1$ color current algebra per chirality~\cite{Witten1984Bosonization}.
To balance the WZW levels for ket and bra separately, the holomorphic Dirac branch must therefore carry ket color and the antiholomorphic branch bra color.

For the WZW fixed point stability, the $C_{6z}$ symmetry requires the conformal spin $h-\bar h\in6\mathbb Z$, so that any perturbation with scaling dimension $\Delta=h+\bar h\leq4$ is a scalar ($h=\bar h$).
According to Eqs.~\ref{eq:app-color-primary-pairing} and \ref{eq:app-affine-current-modes}, the possible relevant and marginal perturbations with $\Delta\leq 2$ are $g$, $g^\dagger$, and $\mathcal J_g^A\bar{\mathcal J}_g^B$ transforming as $(\mathbf3,\overline{\mathbf3})$, $(\overline{\mathbf3},\mathbf3)$, and $(\mathbf8,\mathbf8)$, respectively. 
Independent ket and bra $SU(3)$ rotations thus exclude them and protect the WZW fixed point against relevant and marginal perturbations.
The holomorphic WZW stress tensor $T_g=\tfrac14\sum_{A=1}^8:\!\mathcal J_g^A\mathcal J_g^A\!:$ and its antiholomorphic counterpart $\bar T_g$ are the lowest chiral singlets above the identity, so the leading symmetry-allowed perturbation is $T_g\bar T_g$ with $(h,\bar h)=(2,2)$.
Its leading correction to the dimensionless transfer gap $\Delta(L_2)=-\frac{L_2}{\pi\sqrt3}\ln|\lambda/\lambda_0|$ should scale as $L_2^{2-4}=L_2^{-2}$, consistent with Fig.~\ref{fig:app-transfer-levels}.

\subsection{CFT levels and their transfer eigenvalues}
\label{app:bulk-transfer-energy-momentum}

The low-energy transfer spectrum combines the fermion and color spectra. For a segment of $L_1$ columns, let $\lambda_j$ be the eigenvalues of the transfer matrix in Sec.~\ref{app:tns-column-transfer} and $w_j$ the weight depending on the boundary condition. 
The norm is
\begin{equation}
    \langle\Psi|\Psi\rangle
    =\sum_j w_j\lambda_j^{L_1}.
    \label{eq:app-transfer-norm-spectrum}
\end{equation}
In the factorized norm [Eq.~\ref{eq:app-bulk-statistical-action}], both the quadratic fermion action $\bar\chi D\chi$ and the long-wavelength PCM--WZ action [Eq.~\ref{eq:app-bulk-leading-action}] are local. 
Therefore, the identical columns give a length-independent transfer matrix for each sector. Denoting their eigenvalues by $\lambda_a^f,\lambda_b^g$ and their boundary weights by $w_a^f,w_b^g$, we obtain
\begin{equation}
    \begin{aligned}
        \det D&=\sum_a w_a^f(\lambda_a^f)^{L_1},\qquad
        \mathcal Q\simeq\sum_b w_b^g(\lambda_b^g)^{L_1},\\
        \langle\Psi|\Psi\rangle
        &\simeq\sum_{a,b}w_a^fw_b^g
          (\lambda_a^f\lambda_b^g)^{L_1}.
    \end{aligned}
    \label{eq:app-transfer-factorized-norm}
\end{equation}
The approximation indicates the infrared. Each pair with $w_a^fw_b^g\ne0$ contributes a product $\lambda_a^f\lambda_b^g$ to Eq.~\ref{eq:app-transfer-norm-spectrum} and the relative eigenvalue factorizes:
\begin{equation}
    \frac{\lambda_{a,b}}{\lambda_0}
    \simeq
    \frac{\lambda_a^f}{\lambda_0^f}
    \frac{\lambda_b^g}{\lambda_0^g}.
    \label{eq:app-transfer-sector-product}
\end{equation}
In Sec.~\ref{app:bulk-gradient-action}, we derive the effective field theory with the Cartesian axes, so the transfer geometry is an infinite right cylinder with imaginary time $x$ and periodic space $y\sim y+L_2$ in a right-handed $(x,y)$ frame. 
In these Cartesian axes the unit lattice translations are
\begin{equation}
    \mathbf T_1=\left(\frac{\sqrt3}{2},-\frac12\right),
    \qquad \mathbf T_2=(0,1).
    \label{eq:app-transfer-oblique-metric}
\end{equation}
A step along $+\mathbf T_1$ advances the imaginary time by $\sqrt3/2$ and lowers the coordinate $y$ of a fixed boundary index $\ell_2$ by $1/2$, so $y_{+\mathbf T_1}(\ell_2)=y_0(\ell_2)-1/2$. Evolution at fixed $y$ aligns the input index $\ell_2$ with the output index $\ell_2+1/2$. With $H$ generating time evolution and $\mathsf t_d=\mathrm{e}^{-\mathrm{i}P_{\mathbf T_2}}$ the active cut translation in Eq.~\ref{eq:app-twisted-transverse-momentum}, the low-energy propagator is
\begin{equation}
    \mathcal T \propto \mathrm{e}^{-\sqrt3H/2}\mathrm{e}^{-\mathrm{i}P_{\mathbf T_2}/2}.
    \label{eq:app-transfer-energy-momentum}
\end{equation}
Taking logarithms in Eq.~\ref{eq:app-transfer-sector-product} gives additive energy gaps. The momenta add because the two sectors translate together.

At the conformal fixed point, we can label the transfer eigenstate by the conformal weights $h$ and $\bar h$. 
Consider the map from the cylinder coordinate $w=x+\mathrm{i}y$ to the plane~\cite{Cardy2008Lectures}
\begin{equation}
    \begin{aligned}
        z(w)&=\mathrm{e}^{2\pi w/L_2},
        \qquad w\sim w+\mathrm{i}L_2.
    \end{aligned}
    \label{eq:app-transfer-cylinder-plane-map}
\end{equation}
A slice at fixed $x$ maps to a circle. 
Due to the state-operator correspondence in CFT, each transfer eigenstate can be generated by inserting a scaling field $\mathcal O_j$ at the origin ($x=-\infty$), i.e., $\mathcal O_j(0)|0\rangle$ with $\ket{0}$ the CFT vacuum. 
The holomorphic and antiholomorphic weights $h_j,\bar h_j$ are the exponents in the field transformation
\begin{equation}
    \begin{aligned}
        z'&=\mathrm{e}^{s+\mathrm{i}\theta}z,\qquad \bar z'=\mathrm{e}^{s-\mathrm{i}\theta}\bar z,\\
        \mathcal O'_j(z',\bar z')
        &=\mathrm{e}^{-h_j(s+\mathrm{i}\theta)}\mathrm{e}^{-\bar h_j(s-\mathrm{i}\theta)}
          \mathcal O_j(z,\bar z)\\
        &=\mathrm{e}^{-s(h_j+\bar h_j)}\mathrm{e}^{-\mathrm{i}\theta(h_j-\bar h_j)}
          \mathcal O_j(z,\bar z).
    \end{aligned}
    \label{eq:app-transfer-conformal-weights}
\end{equation}
The common dilation of $z$ and $\bar z$ gives the weight sum $\Delta_j=h_j+\bar h_j$, while their opposite rotation phases give the difference $h_j-\bar h_j$.
For the $U(3)_1\times SU(3)_1$ theory, the dominate transfer eigenstate at $\phi_2=\pi$ corresponds to the CFT vacuum with $h_0=\bar h_0=0$. Subtracting its energy, we can reformulate Eq.~\ref{eq:app-transfer-energy-momentum} as 
\begin{equation}
    \begin{aligned}
        E_j-E_0&=\frac{2\pi}{L_2}(h_j+\bar h_j)
        =\frac{2\pi}{L_2}\Delta_j,\\
        k_{2,j}-k_{2,0}&=-\frac{2\pi}{L_2}(h_j-\bar h_j)
        =-\frac{2\pi}{L_2}K_{y,j},\\
        \frac{\lambda_j}{\lambda_0}
        &= \exp\!\left[-\frac{\pi\sqrt3}{L_2}\Delta_j
                       +\frac{\mathrm{i}\pi}{L_2}K_{y,j}\right].
    \end{aligned}
    \label{eq:app-transfer-cft-levels}
\end{equation}

The modulus gives the scaled transfer gap in Eq.~\ref{eq:bulk-scaled-transfer-gap}. The phase gives an independent measurement of $K_y$ [Sec.~\ref{app:transfer-spectrum-data}]. The same weights fix the plane two-point function at nonzero separation,
\begin{equation}
    \langle\mathcal O_j(z,\bar z)\mathcal O_j^\dagger(0)\rangle_{\rm plane}
    \propto\frac{1}{z^{2h_j}\bar z^{2\bar h_j}},
    \qquad
    \left|\langle\mathcal O_j(z,\bar z)\mathcal O_j^\dagger(0)\rangle_{\rm plane}\right|
    \propto |z|^{-2\Delta_j}.
    \label{eq:app-transfer-plane-power-law}
\end{equation}
The scaled transfer gap thus gives half the plane decay exponent.

We then summarize the quantum numbers $(R_{\mathrm{k}},R_{\mathrm{b}},\delta Q,K_y)$ and scaling dimensions $\Delta$ of the $U(3)_1\!\times\!SU(3)_1$ CFT.
\begingroup
\setlength{\abovedisplayskip}{6pt plus 1pt minus 1pt}
\setlength{\belowdisplayskip}{6pt plus 1pt minus 1pt}
\setlength{\abovedisplayshortskip}{3pt plus 1pt}
\setlength{\belowdisplayshortskip}{4pt plus 1pt minus 1pt}

\paragraph{Auxiliary Dirac fields, $U(3)_1$.}
In the holomorphic ket sector, the primary fields $F_n$ are the local operators of smallest $\Delta$ built from $n$ factors of $\psi^\dagger_{\mathrm k}$ and their derivatives, carrying charge $\delta Q=n>0$. 
Each $\psi^\dagger_{\mathrm k\alpha}$ contributes $1/2$ to $\Delta$, and each $\partial_z$ adds one. Fermi statistics allow each pair $(j,\alpha)$ in $\partial_z^j\psi^\dagger_{\mathrm k\alpha}(z)$ only once. Thus for $n=3m+r$, $r=0,1,2$, we use all three colors for $j<m$ and any $r$ colors $\alpha_1<\cdots<\alpha_r$ for $j=m$,
\begin{equation}
 (F_{3m+r})_{\alpha_1\ldots\alpha_r}(z)
 \propto
 :\!\left(\prod_{j=0}^{m-1}\prod_{\alpha=1}^{3}
       \partial_z^j\psi^\dagger_{\mathrm k\alpha}(z)\right)
    \left(\prod_{a=1}^{r}\partial_z^m\psi^\dagger_{\mathrm k\alpha_a}(z)\right)\!: .
\end{equation}
Colons denote normal ordering with respect to the vacuum.
Replacing $\psi^\dagger_{\mathrm k}$ by $\psi_{\mathrm k}$ gives $F_{-n}$.
The currents form a density singlet $\rho_f=:\!\psi^\dagger_{\mathrm k}\psi_{\mathrm k}\!:$ (summed over colors) and a color octet $j_f$ with components $j_f^A=:\!\psi^\dagger_{\mathrm k}t^A\psi_{\mathrm k}\!:$, where $t^A$ are the $SU(3)$ generators~\cite{Ginsparg1988CFT}.
Their conformal weights and quantum numbers are summarized below.
\begin{equation}
    \begin{array}{l|c|c|c}
        \text{Holomorphic field} &(R_{\mathrm{k}},R_{\mathrm{b}})&\delta Q&K_y=\Delta\\ \hline
        F_n,\quad n=0\bmod 3 &(\mathbf1,\mathbf1)&n&n^2/6\\
        F_n,\quad n=1\bmod 3 &(\mathbf3,\mathbf1)&n&n^2/6+1/3\\
        F_n,\quad n=2\bmod 3 &(\overline{\mathbf3},\mathbf1)&n&n^2/6+1/3\\
        \rho_f &(\mathbf1,\mathbf1)&0&1\\
        j_f &(\mathbf8,\mathbf1)&0&1
    \end{array}
    \label{eq:app-fermion-primary-weights}
\end{equation}
Starting from the ket particle $F_n(z)$ with $n>0$, replacing $\psi_{\rm k}^\dagger$ by $\psi_{\rm k}$ gives the ket hole $F_{-n}(z)$, with $\delta Q=-n$, conjugate ket representation, and the same $K_y=\Delta$.
Applying $\mathrm k\to\mathrm b$ and $z\to\bar z$ (hence $\partial_z\to\partial_{\bar z}$) to these two fields gives the bra particle $\bar F_n(\bar z)$ built from $\psi_{\rm b}^\dagger$ and the bra hole $\bar F_{-n}(\bar z)$ built from $\psi_{\rm b}$, respectively.
Each bra field inherits $\delta Q$ and $\Delta$ from its ket partner, exchanges $R_{\rm k}$ and $R_{\rm b}$, and has $K_y=-\Delta\leq0$. The same layer and coordinate replacements give the currents $\bar\rho_f$ and $\bar j_f^A$.

\paragraph{Collective color field, $SU(3)_1$.}
Apart from the vacuum $I$, the primary fields $g$ and $g^\dagger$ in this sector involve both the holomorphic and antiholomorphic components.
The holomorphic currents $\mathcal J_g^A=-\tr[t^A(\partial_zg)g^{-1}]$ form a bra color octet, and the antiholomorphic currents $\bar{\mathcal J}_g^A=\tr[t^A g^{-1}\partial_{\bar z}g]$ form a ket color octet~\cite{Witten1984Bosonization,Gawedzki2004WZWQuantization,Tu2014SUNWZW}.
The table lists the primary fields and the currents that generate their descendants.
\begin{equation}
    \begin{array}{l|c|c|c|c}
        \text{Color field}&(R_{\mathrm{k}},R_{\mathrm{b}})&\delta Q&K_y&\Delta\\ \hline
        I                 &(\mathbf1,\mathbf1)&0&0&0\\
        g                 &(\mathbf3,\overline{\mathbf3})&0&0&2/3\\
        g^\dagger         &(\overline{\mathbf3},\mathbf3)&0&0&2/3\\
        \mathcal J_g        &(\mathbf1,\mathbf8)&0&1&1\\
        \bar{\mathcal J}_g  &(\mathbf8,\mathbf1)&0&-1&1
    \end{array}
    \label{eq:app-color-primary-pairing}
\end{equation}

\paragraph{General states of the $U(3)_1 \times SU(3)_1$ CFT.}
Let $J$ denote any holomorphic current $\rho_f$, $j_f^A$, or $\mathcal J_g^A$. Each mode carries the same color representation as $J$. Its mode expansion and the changes induced by $J_m$ are
\begin{equation}
    J(z)=\sum_{m\in\mathbb Z}J_m z^{-m-1},\qquad
    (\delta Q,\delta K_y,\delta\Delta)=(0,-m,-m),\qquad m\in\mathbb Z .
    \label{eq:app-affine-current-modes}
\end{equation}
For the corresponding antiholomorphic currents $\bar\rho_f$, $\bar j_f^A$, and $\bar{\mathcal J}_g^A$, replace $z,J_m$ by $\bar z,\bar J_m$ and reverse the sign of $\delta K_y$.
The total CFT Hilbert space is spanned by primary states and their descendants generated by products of current modes with $m<0$.
Their $\delta Q$, $K_y$, and $\Delta$ add across the constituent fields and modes. The allowed color multiplets occur in the tensor products taken separately for ket and bra. Table~\ref{tab:transfer-branches} compares these predictions with the transfer spectrum.
\endgroup

\subsection{Error bounds for the gradient expansion}
\label{app:bulk-gradient-convergence}
\begingroup
\setlength{\abovedisplayskip}{4pt plus 1pt minus 1pt}
\setlength{\belowdisplayskip}{4pt plus 1pt minus 1pt}
\setlength{\abovedisplayshortskip}{2pt plus 1pt}
\setlength{\belowdisplayshortskip}{3pt plus 1pt minus 1pt}
\setlength{\jot}{2pt}
\paragraph{Lemma (bounds on the orbital projector).}
\begin{equation}
 [P_1(\mathbf k)]_{aa'}=\tfrac12+O(|\mathbf k|),\qquad
 \|\nabla_{\mathbf k}^s P_1(\mathbf k)\|_{\rm op}
 \le C_s|\mathbf k|^{1-s},\qquad \mathbf k\rightarrow0 .
 \label{eq:app-bulk-projector-gamma}
\end{equation}
\begin{equation}
 \sup_{\mathbf k\in{\rm BZ}}\|P_1(\mathbf k)-P_1(\mathbf k-\mathbf p)\|_{\rm op}
 \le C|\mathbf p|.
 \label{eq:app-global-projector-difference}
\end{equation}
\emph{Proof.} The finite bond sums $s_a(\mathbf k)$ vanish at $\Gamma$ and satisfy $s_2(\mathbf k)=-s_1(-\mathbf k)$ [Eqs.~\ref{eq:sm-site-bloch-convention} and \ref{eq:sm-bloch-c2}], so $s_2(\mathbf k)-s_1(\mathbf k)=O(|\mathbf k|^2)$. With $d(\mathbf k)=|s_1(\mathbf k)|^2+|s_2(\mathbf k)|^2$ from Eq.~\ref{eq:app-bulk-gaussian-kernel}, the common quadratic term cancels in $[P_1]_{aa'}-\tfrac12=(s_a(\mathbf k)s_{a'}(\mathbf k)^*-d(\mathbf k)/2)/d(\mathbf k)=O(|\mathbf k|^3)/(\tfrac32|\mathbf k|^2+O(|\mathbf k|^4))$. The numerator and denominator are real analytic, so each derivative costs at most one inverse power of $|\mathbf k|$, giving Eq.~\ref{eq:app-bulk-projector-gamma}. For $s=1$, Eq.~\ref{eq:app-bulk-projector-gamma} gives $\|\nabla_{\mathbf k}P_1\|_{\rm op}\le C_1$ near $\Gamma$. Outside this neighborhood, $d(\mathbf k)$ has a positive minimum, so the derivative of $s_a(\mathbf k)s_{a'}(\mathbf k)^*/d(\mathbf k)$ is bounded there too. The mean value theorem then gives Eq.~\ref{eq:app-global-projector-difference}.

\paragraph{Truncating the matrix series.}
For $f_{\mathbf R}=\sum_{\mathbf p}\mathrm{e}^{\mathrm{i}\mathbf p\cdot\mathbf R}\widehat f_{\mathbf p}$ ($f=g,h$), $P_1$ acts on sublattices and $\widehat f_{\mathbf p}$ on colors. Hence
\begin{equation}
 ([P,f]v)(\mathbf k)=\sum_{\mathbf p}
 \bigl[P_1(\mathbf k)-P_1(\mathbf k-\mathbf p)\bigr]\otimes\widehat f_{\mathbf p}\,
 v(\mathbf k-\mathbf p).
 \label{eq:app-commutator-bloch-action}
\end{equation}
Applying Eq.~\ref{eq:app-global-projector-difference} and using that momentum shifts preserve $\|v\|_2$ gives
\begin{equation}
 \|[P,f]\|_{\rm op}\le C\sum_{\mathbf p}|\mathbf p|\,\|\widehat f_{\mathbf p}\|_{\rm op}
 =O(|\mathbf q|),\qquad f=g,h.
 \label{eq:app-commutator-norm-bound}
\end{equation}
Equation~\ref{eq:app-bulk-leakage} then gives $\|E\|_{\rm op}\le\|[P,g]\|_{\rm op}^2=O(|\mathbf q|^2)$. With $\|K\|_{\rm op}\le1$ and $3N$ occupied orbitals, the tail of Eq.~\ref{eq:app-bulk-variation-series} satisfies
\begin{equation}
 \left|\sum_{n=m}^{\infty}\Tr\!\left[E^nK^\dagger U^\dagger[g,P][h,P]U\right]\right|
 \le 3N\,\frac{\|E\|_{\rm op}^m}{1-\|E\|_{\rm op}}\,
       \|[g,P]\|_{\rm op}\,\|[h,P]\|_{\rm op}
 = O(N|\mathbf q|^{2m+2}).
 \label{eq:app-bulk-complex-error}
\end{equation}
Retaining $n=0$ gives $m=1$ and an $O(N|\mathbf q|^4)$ error.

\paragraph{Replacing field differences by gradients.}
At Fourier momenta $\mathbf q_1,\mathbf q_2,\mathbf q_3$ of $g^{-1},g,h$, the orbital factor $F$ multiplies the color trace,
\begin{equation}
 \Tr(Pg^{-1}[g,P][h,P])
 =\sum_{\mathbf q_1+\mathbf q_2+\mathbf q_3=0}
 \tr\!\left[\widehat{g^{-1}}_{\mathbf q_1}
 \widehat g_{\mathbf q_2}\widehat h_{\mathbf q_3}\right]
 \sum_{\mathbf k} F(\mathbf k;\mathbf q_1,\mathbf q_2,\mathbf q_3).
 \label{eq:app-orbital-factor-trace}
\end{equation}
Each commutator $[f,P]$ gives a finite difference, linearizing the difference gives $F_2$:
\begin{equation}
 \begin{aligned}
 F&=\tr_{\rm orb}\!\left[P_1(\mathbf k)(P_1(\mathbf k+\mathbf q_3)-P_1(\mathbf k-\mathbf q_1))(P_1(\mathbf k)-P_1(\mathbf k+\mathbf q_3))\right],\\
 F_2&=\tr_{\rm orb}\!\left[P_1(\mathbf k)
   (\mathbf q_2\cdot\nabla_{\mathbf k}P_1(\mathbf k))
   (\mathbf q_3\cdot\nabla_{\mathbf k}P_1(\mathbf k))\right].
 \end{aligned}
 \label{eq:app-bulk-leading-trace-momentum}
\end{equation}
Take $|\mathbf q|=\max_i|\mathbf q_i|$. Inside $|\mathbf k|<2|\mathbf q|$, Eq.~\ref{eq:app-global-projector-difference} bounds each difference in $F$ by $O(|\mathbf q|)$, and Eq.~\ref{eq:app-bulk-projector-gamma} at $s=1$ gives the same bound for each gradient factor in $F_2$. Thus $|F-F_2|=O(|\mathbf q|^2)$.

Inversion exchanges the two sublattices [Eq.~\ref{eq:sm-bloch-c2}], so the orbital trace in Eq.~\ref{eq:app-bulk-leading-trace-momentum} gives
\begin{equation}
 F(-\mathbf k;\{-\mathbf q_i\})=F(\mathbf k;\{\mathbf q_i\}),\qquad
 \langle F-F_2\rangle_{\rm BZ}
 =\left\langle\tfrac12\sum_{\pm}(F-F_2)(\mathbf k;\{\pm\mathbf q_i\})\right\rangle_{\rm BZ}.
 \label{eq:app-inversion-average}
\end{equation}
For $2|\mathbf q|\le|\mathbf k|<1$, the shifted momenta stay at least $|\mathbf k|/2$ from $\Gamma$. The $s=2,3$ bounds in Eq.~\ref{eq:app-bulk-projector-gamma} control the quadratic terms and Taylor remainders when we expand the two differences in Eq.~\ref{eq:app-bulk-leading-trace-momentum}, using $\mathbf q_1+\mathbf q_2+\mathbf q_3=0$,
\begin{equation}
\begin{aligned}
 P_1(\mathbf k+\mathbf q_3)-P_1(\mathbf k-\mathbf q_1)
 &=-(\mathbf q_2\cdot\nabla_{\mathbf k})P_1(\mathbf k)+\tfrac12((\mathbf q_3\cdot\nabla_{\mathbf k})^2-(\mathbf q_1\cdot\nabla_{\mathbf k})^2)P_1(\mathbf k)
    +O(|\mathbf q|^3/|\mathbf k|^2),\\
 P_1(\mathbf k)-P_1(\mathbf k+\mathbf q_3)
 &=-(\mathbf q_3\cdot\nabla_{\mathbf k})P_1(\mathbf k)-\tfrac12(\mathbf q_3\cdot\nabla_{\mathbf k})^2P_1(\mathbf k)
    +O(|\mathbf q|^3/|\mathbf k|^2).
\end{aligned}
 \label{eq:app-projector-difference-expansion}
\end{equation}
Subtracting the product of the linear terms leaves
\begin{equation}
\begin{aligned}
 F-F_2
 &=\tfrac12\tr_{\rm orb}\!\left\{P_1(\mathbf k)
   [(\mathbf q_2\cdot\nabla_{\mathbf k})P_1(\mathbf k)][(\mathbf q_3\cdot\nabla_{\mathbf k})^2P_1(\mathbf k)]\right\}\\
 &\quad-\tfrac12\tr_{\rm orb}\!\left\{P_1(\mathbf k)
   [((\mathbf q_3\cdot\nabla_{\mathbf k})^2-(\mathbf q_1\cdot\nabla_{\mathbf k})^2)P_1(\mathbf k)][(\mathbf q_3\cdot\nabla_{\mathbf k})P_1(\mathbf k)]\right\}
   +O(|\mathbf q|^4/|\mathbf k|^2).
\end{aligned}
 \label{eq:app-cubic-orbital-remainder}
\end{equation}
Each displayed term in Eq.~\ref{eq:app-cubic-orbital-remainder} contains three factors of external momentum, so the two signs in Eq.~\ref{eq:app-inversion-average} cancel its contribution. For $|\mathbf k|\ge1$ and $|\mathbf q|\ll1$, the same expansion applies with uniformly bounded projector derivatives. After the sign average the remainder is therefore $O(|\mathbf q|^4)$ throughout this finite-area region.
The three regions thus give
\begin{equation}
 \left|\tfrac12\sum_{\pm}(F-F_2)(\mathbf k;\{\pm\mathbf q_i\})\right|
 \le C
 \begin{cases}
 |\mathbf q|^2, & |\mathbf k|<2|\mathbf q|,\\
 |\mathbf q|^4/|\mathbf k|^2, & 2|\mathbf q|\le|\mathbf k|<1,\\
 |\mathbf q|^4, & |\mathbf k|\ge1 .
 \end{cases}
 \label{eq:app-piecewise-orbital-bound}
\end{equation}
Integrating over the three regions gives
\begin{equation}
 \left|\langle F-F_2\rangle_{\rm BZ}\right|
 \le C |\mathbf q|^4\left(1+\int_{2|\mathbf q|<|\mathbf k|<1}\frac{d^2\mathbf k}{|\mathbf k|^2}\right)
 = O\!\left(|\mathbf q|^4\ln\frac1{|\mathbf q|}\right).
 \label{eq:app-bulk-triangle-integrated-bound}
\end{equation}
\endgroup

\FloatBarrier

\begingroup
\setlength{\intextsep}{8pt plus 2pt minus 2pt}
\setlength{\textfloatsep}{12pt plus 2pt minus 2pt}
\setlength{\floatsep}{8pt plus 2pt minus 2pt}

\clearpage
\section{Finite-size scaling of the transfer spectrum}
\label{app:transfer-spectrum-data}
\label{app:transfer-scaling-data}
\label{app:transfer-twist-data}
\label{app:transfer-free-energy-data}

Table~\ref{tab:transfer-branches} and Fig.~\ref{fig:app-transfer-levels} compare the extrapolated transfer gaps at three twists with the bulk CFT predictions.
A field of charge $\delta Q$ acting on the dominant state reaches the cut sector $Q=\delta Q$. Braces group candidate fields with identical $(R_{\rm k},R_{\rm b},Q,K_y)$ and the same predicted gap.

\makeatletter
\begin{figure}[!b]
    \centering
    \begin{minipage}[t]{0.36\linewidth}
    \vspace{0pt}\centering
    \includegraphics[width=0.97\linewidth]{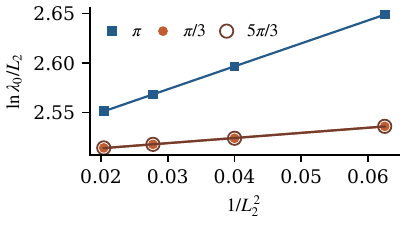}
    \caption{Cylinder free energy and linear fits for $L_2=4$--$7$. Filled and open circles mark the coincident outer twists.}
    \label{fig:app-casimir-linear}
    \end{minipage}\hfill
    \begin{minipage}[t]{0.60\linewidth}
    \vspace{0pt}\centering
    \def\@captype{table}
    \caption{Fits of $\ln\lambda_0/L_2$ versus $L_2^{-2}$ with equal weights, and the vacuum shift relative to $\phi_2=\pi$. The outer twists agree to the precision shown.}
    \label{tab:app-casimir-windows}
    \small
    \setlength{\tabcolsep}{3.5pt}
    \renewcommand{\arraystretch}{1.12}
    \begin{tabular}{lcrr}
        \hline\hline
        Quantity & $L_2$ & $\phi_2=\pi$ & $\phi_2=\pi/3,\,5\pi/3$\\
        \hline
        $c_{\rm fit}$, linear fit & $3$--$7$ & 5.2035 & 1.2267\\
                                 & $4$--$7$ & 5.1039 & 1.1421\\
                                 & $5$--$7$ & 5.0785 & 1.0943\\
        $R^2$, linear fit         & $4$--$7$ & 0.999994 & 0.999570\\
        $c_{\rm fit}$, including $L_2^{-4}$ & $4$--$7$ & 5.0107 & 0.9695\\
        \hline
        $\frac{L_2}{\pi\sqrt3}\ln[\lambda_0(\pi)/\lambda_0(\phi_2)]$ & $7$ & 0 & 0.33304\\
        \hline\hline
    \end{tabular}
    \end{minipage}
\end{figure}
\makeatother
At each twist $\phi_2$, the ratio to the leading eigenvalue $\lambda_0(\phi_2)$ gives the transfer gap above that twist's vacuum and the relative transverse momentum [Eq.~\ref{eq:app-transfer-cft-levels}],
\begin{equation}
 \Delta_{\rm eff}(L_2,\phi_2)=-\frac{L_2}{\pi\sqrt3}\ln\left|\frac{\lambda_j}{\lambda_0}\right|
 \ \longrightarrow\ \Delta_j(\phi_2)-\Delta_0(\phi_2),
 \qquad
 K_y\simeq\frac{L_2}{\pi}\arg\!\left(\frac{\lambda_j}{\lambda_0}\right)\pmod{L_2}.
\end{equation}

The bra and ket twist phases cancel, so the WZW field obeys $g(y+L_2)=g(y)$ for every $\phi_2$.
The auxiliary fermion sector $\det D$ feels $\phi_2$ and Sec.~\ref{app:unprojected-transfer} shows that the dominate transfer mode at $\phi_2=\pi$ corresponds to the identity vacuum, and we use $\delta=(\pi-\phi_2)/(2\pi)$ and $|\delta|<1/2$ to parameterize the twist.
Applying the free-fermion twist-field formula~\cite{Ginsparg1988CFT,Dixon1987Orbifolds} to $U(3)_1$ gives the cylinder dimension and momentum of $F_n$, including the twisted vacuum at $n=0$,
\begin{subequations}\label{eq:app-twist-level-shifts}
\begin{equation}
 \begin{aligned}
 \Delta_{F_n}(\phi_2)&=\Delta_{F_n}(\pi)+n\delta+3\delta^2,&
 K_{y,F_n}(\phi_2)&=\Delta_{F_n}(\pi)+n\delta.
 \end{aligned}
\end{equation}
For any state $j$ with holomorphic and antiholomorphic net fermion numbers $n,\bar n$, including current descendants and products with WZW fields, the two observables above are
\begin{equation}
 \Delta_j(\phi_2)-\Delta_0(\phi_2)=\Delta_j(\pi)+(n-\bar n)\delta,\qquad
 K_{y,j}(\phi_2)=K_{y,j}(\pi)+(n+\bar n)\delta .
\end{equation}

The above argument for $U(3)_1$ agrees with explict twisting the transfer matrix of $\det D$ in Sec.~\ref{app:unprojected-transfer}. 
Let $q=(2\pi m-\pi)/L_2\in(-\pi,\pi]$, $m\in\mathbb Z$, run over the transverse momenta at $\phi_2=\pi$. At twist $\phi_2$ the momenta are $q+\epsilon$ with $\epsilon=2\pi\delta/L_2$. Expanding $\lambda_0^\Phi(\phi_2)=\prod_q\lambda_+(q+\epsilon)^3$ gives
\begin{equation}
 \ln\frac{\lambda_0^\Phi(\phi_2)}{\lambda_0^\Phi(\pi)}
 =\frac{3\epsilon^2}{2}\sum_q\frac{\partial^2}{\partial q^2}\ln\lambda_+(q)+o(L_2^{-1})
 =-\frac{\pi\sqrt3}{L_2}\,3\delta^2+o(L_2^{-1}).
\end{equation}
\end{subequations}
This yields $\Delta_0(\phi_2)=3\delta^2$. For an excitation, Eqs.~\ref{eq:app-unprojected-transfer-eigenvalues} and \ref{eq:app-unprojected-scaled-gap} give a gap $|K_y|$ per occupied low-energy mode. Particle and hole momenta shift by $+\delta$ and $-\delta$, so the holomorphic and antiholomorphic gaps shift by $n\delta$ and $-\bar n\delta$, respectively. At $\phi_2=0\bmod 2\pi$ the $q=0$ modes give degenerate vacua.

At $\phi_2=\pi$, the color field $g$ and the particle--hole field $\psi_{\rm k}^\dagger\psi_{\rm b}$ share a sector but have distinct gaps. The gap of the mixed field $\psi_{\rm k}^\dagger g$ approaches the sum of its constituents' gaps. At the outer twists $\phi_2=\pi/3,5\pi/3$ the chiral $n$-fermion branches have predicted gap and momentum $n[1-\phi_2/(2\pi)]$ for $n=1,2,3$. 
The lowest neutral particle--hole field is $\psi_{\rm k}^\dagger\psi_{\rm b}$ at $5\pi/3$ and its conjugate at $\pi/3$.

The leading eigenvalue $\lambda_0(\phi_2)$ also determines the cylinder free energy. With a common tensor normalization, we extract the effective central charge $c_{\rm fit}(\phi_2)$ from its width dependence~\cite{Blote1986CentralCharge,Affleck1986ConformalAnomaly},
\begin{equation}
    \frac{\ln\lambda_0(\phi_2)}{L_2}
    =A+\frac{\pi\sqrt3}{12L_2^2}c_{\rm fit}(\phi_2)+O(L_2^{-4}),
    \qquad c_{\rm fit}(\phi_2)\longrightarrow c-12\Delta_0(\phi_2).
    \label{eq:app-casimir-fit}
\end{equation}
For $c=5$, the coefficient tends to $5$ at $\pi$ and $1$ at the outer twists, which is consistent with the expected $\Delta_0(\pi/3)=\Delta_0(5\pi/3)=1/3$. 
Figure~\ref{fig:app-casimir-linear} shows the nearly linear scaling. Table~\ref{tab:app-casimir-windows} quantifies its sensitivity to the fit window and the $L_2^{-4}$ correction.

\begin{table}[!htbp]
\centering
\caption{Transfer gaps at three parton twists. Unweighted linear fits in $L_2^{-2}$ use at least three widths and a free intercept. Parentheses give one regression standard error. A dash denotes an unavailable level or fit.}
\label{tab:transfer-branches}
\label{tab:app-transfer-phases}
\small
\setlength{\tabcolsep}{3.5pt}
\renewcommand{\arraystretch}{1.03}
\newcommand{\mergedrowrule}{%
  \noalign{\vskip .45pt{\color[gray]{.65}\hrule height .35pt}\vskip .45pt}}
\begin{tabular}{@{}llrrrrrcrc@{}}
\hline\hline
Field(s) & $(R_{\rm k},R_{\rm b},Q,K_y)$ & \multicolumn{5}{c}{$\Delta_{\rm eff}(L_2)$} & Window & Extrapolated $\Delta_{\rm eff}$ & CFT\\
 & & $3$ & $4$ & $5$ & $6$ & $7$ & & & \\
\hline \multicolumn{10}{l}{$\phi_2=\pi$}\\
$\psi_{\rm k}^\dagger$ & $(\mathbf{3},\mathbf{1},1,1/2)$ & 0.4777 & 0.4920 & 0.4965 & 0.4982 & 0.4990 & 4--7 & 0.50277(59) & $1/2$ \\
$j_f$ & $(\mathbf{8},\mathbf{1},0,1)$ & 1.0295 & 1.0104 & 1.0043 & 1.0019 & 1.0010 & 4--7 & 0.99576(84) & $1$ \\
$(\psi_{\rm k}^\dagger)^3$ & $(\mathbf{1},\mathbf{1},3,3/2)$ & 1.2589 & 1.4057 & 1.4574 & 1.4785 & 1.4880 & 4--7 & 1.5322(59) & $3/2$ \\
$\psi_{\rm b}^\dagger$ & $(\mathbf{1},\mathbf{3},1,-1/2)$ & 0.4777 & 0.4920 & 0.4965 & 0.4982 & 0.4990 & 4--7 & 0.50277(59) & $1/2$ \\
$\psi_{\rm k}^\dagger\psi_{\rm b}$ & $(\mathbf{3},\overline{\mathbf{3}},0,0)$ & 1.0453 & 1.0117 & 1.0049 & 1.0025 & 1.0014 & 4--7 & 0.99578(97) & $1$ \\
$g$ & $(\mathbf{3},\overline{\mathbf{3}},0,0)$ & 0.5807 & 0.6123 & 0.6313 & 0.6419 & 0.6485 & 4--7 & 0.66576(31) & $2/3$ \\
$\bar{\mathcal J}_g$ & $(\mathbf{8},\mathbf{1},0,-1)$ & 1.2106 & 1.0821 & 1.0486 & 1.0329 & 1.0240 & 4--7 & 0.9946(18) & $1$ \\
$\mathcal J_g$ & $(\mathbf{1},\mathbf{8},0,1)$ & 1.2106 & 1.0821 & 1.0486 & 1.0329 & 1.0240 & 4--7 & 0.9946(18) & $1$ \\
$\psi_{\rm k}^\dagger g$ & $(\overline{\mathbf{3}},\overline{\mathbf{3}},1,1/2)$ & 1.0116 & 1.0822 & 1.1153 & 1.1326 & 1.1428 & 4--7 & 1.17253(54) & $7/6$ \\
$\psi_{\rm k}^\dagger\psi_{\rm b}g$ & $(\mathbf{6},\overline{\mathbf{6}},0,0)$ & 1.4620 & 1.5507 & 1.6031 & 1.6289 & 1.6421 & 4--7 & 1.6887(29) & $5/3$ \\
$(\psi_{\rm k}^\dagger)^2$ & $(\overline{\mathbf{3}},\mathbf{1},2,1)$ & 0.8837 & 0.9581 & 0.9818 & 0.9909 & 0.9949 & 4--7 & 1.0149(31) & $1$ \\
$\bar\rho_f$ & $(\mathbf{1},\mathbf{1},0,-1)$ & 0.8089 & 0.9320 & 0.9705 & 0.9853 & 0.9919 & 4--7 & 1.0245(50) & $1$ \\
$\psi_{\rm b}^\dagger\psi_{\rm k}$ & $(\overline{\mathbf{3}},\mathbf{3},0,0)$ & 1.0453 & 1.0117 & 1.0049 & 1.0025 & 1.0014 & 4--7 & 0.99578(97) & $1$ \\
$g^\dagger$ & $(\overline{\mathbf{3}},\mathbf{3},0,0)$ & 0.5807 & 0.6123 & 0.6313 & 0.6419 & 0.6485 & 4--7 & 0.66576(31) & $2/3$ \\
\mergedrowrule
$\{\psi_{\rm k}^\dagger\psi_{\rm b}\,g^\dagger,\,\psi_{\rm b}^\dagger\psi_{\rm k}\,g\}$ & $(\mathbf{1},\mathbf{1},0,0)$ & \begin{tabular}[c]{@{}r@{}}1.1816\\1.5399\end{tabular} & \begin{tabular}[c]{@{}r@{}}1.4009\\1.8461\end{tabular} & \begin{tabular}[c]{@{}r@{}}1.4787\\1.7607\end{tabular} & \begin{tabular}[c]{@{}r@{}}1.5300\\1.7170\end{tabular} & \begin{tabular}[c]{@{}r@{}}1.5653\\1.6952\end{tabular} & \begin{tabular}[c]{@{}r@{}}4--7\\4--7\end{tabular} & \begin{tabular}[c]{@{}r@{}}1.6394(78)\\1.6185(41)\end{tabular} & $5/3$ \\
\mergedrowrule
$\{\psi_{\rm k}^\dagger\psi_{\rm b}\,g^\dagger,\,\psi_{\rm b}^\dagger\psi_{\rm k}\,g\}$ & $(\mathbf{8},\mathbf{8},0,0)$ & \begin{tabular}[c]{@{}r@{}}1.3214\\1.7817\end{tabular} & \begin{tabular}[c]{@{}r@{}}1.4341\\1.6678\end{tabular} & \begin{tabular}[c]{@{}r@{}}1.5223\\1.6546\end{tabular} & \begin{tabular}[c]{@{}r@{}}1.5769\\1.6535\end{tabular} & -- & \begin{tabular}[c]{@{}r@{}}4--6\\4--6\end{tabular} & \begin{tabular}[c]{@{}r@{}}1.6887(63)\\1.6399(59)\end{tabular} & $5/3$ \\
\hline \multicolumn{10}{l}{$\phi_2=\pi/3$}\\
$\psi_{\rm k}^\dagger$ & $(\mathbf{3},\mathbf{1},1,5/6)$ & 0.8279 & 0.8338 & 0.8343 & 0.8342 & 0.8340 & 4--7 & 0.83426(29) & $5/6$ \\
$(\psi_{\rm k}^\dagger)^2$ & $(\overline{\mathbf{3}},\mathbf{1},2,5/3)$ & 1.2504 & 1.4815 & 1.5864 & 1.6270 & 1.6451 & 4--7 & 1.734(13) & $5/3$ \\
$(\psi_{\rm k}^\dagger)^3$ & $(\mathbf{1},\mathbf{1},3,5/2)$ & 1.2900 & 1.8118 & 2.2545 & 2.3886 & 2.4368 & 4--7 & 2.795(82) & $5/2$ \\
$\psi_{\rm b}^\dagger\psi_{\rm k}$ & $(\overline{\mathbf{3}},\mathbf{3},0,0)$ & 0.3227 & 0.3294 & 0.3315 & 0.3324 & 0.3328 & 4--7 & 0.33460(26) & $1/3$ \\
\mergedrowrule
$\{g,\,(\psi_{\rm b}^\dagger\psi_{\rm k})^2\}$ & $(\mathbf{3},\overline{\mathbf{3}},0,0)$ & \begin{tabular}[c]{@{}r@{}}0.5934\\0.6391\end{tabular} & \begin{tabular}[c]{@{}r@{}}0.6200\\0.6562\end{tabular} & \begin{tabular}[c]{@{}r@{}}0.6346\\0.6618\end{tabular} & \begin{tabular}[c]{@{}r@{}}0.6439\\0.6641\end{tabular} & \begin{tabular}[c]{@{}r@{}}0.6497\\0.6652\end{tabular} & \begin{tabular}[c]{@{}r@{}}4--7\\4--7\end{tabular} & \begin{tabular}[c]{@{}r@{}}0.66353(99)\\0.66999(66)\end{tabular} & $2/3$ \\
\mergedrowrule
$g^\dagger$ & $(\overline{\mathbf{3}},\mathbf{3},0,0)$ & 0.6020 & 0.6247 & 0.6376 & 0.6457 & 0.6509 & 4--7 & 0.66309(80) & $2/3$ \\
$\bar\rho_f$ & $(\mathbf{1},\mathbf{1},0,-1)$ & 0.9424 & 0.9827 & 0.9933 & 0.9971 & 0.9986 & 4--7 & 1.0075(16) & $1$ \\
\mergedrowrule
$\{\psi_{\rm b}^\dagger\psi_{\rm k}\,g,\,(\psi_{\rm b}^\dagger\psi_{\rm k})^3\}$ & $(\mathbf{1},\mathbf{1},0,0)$ & \begin{tabular}[c]{@{}r@{}}0.9186\\0.9566\end{tabular} & \begin{tabular}[c]{@{}r@{}}0.9489\\0.9816\end{tabular} & \begin{tabular}[c]{@{}r@{}}0.9653\\0.9913\end{tabular} & \begin{tabular}[c]{@{}r@{}}0.9754\\0.9953\end{tabular} & \begin{tabular}[c]{@{}r@{}}0.9818\\0.9972\end{tabular} & \begin{tabular}[c]{@{}r@{}}4--7\\4--7\end{tabular} & \begin{tabular}[c]{@{}r@{}}0.99718(85)\\1.00548(99)\end{tabular} & $1$ \\
\mergedrowrule
$\psi_{\rm b}^\dagger\psi_{\rm k}\,g$ & $(\mathbf{8},\mathbf{8},0,0)$ & 0.9224 & 0.9516 & 0.9670 & 0.9767 & -- & 4--6 & 0.9963(13) & $1$ \\
$(\psi_{\rm b}^\dagger\psi_{\rm k})^2g^\dagger$ & $(\mathbf{8},\mathbf{8},0,0)$ & 1.2418 & 1.2814 & 1.3001 & 1.3104 & -- & 4--6 & 1.333558(25) & $4/3$ \\
\hline \multicolumn{10}{l}{$\phi_2=5\pi/3$}\\
$\psi_{\rm k}^\dagger$ & $(\mathbf{3},\mathbf{1},1,1/6)$ & 0.1628 & 0.1653 & 0.1660 & 0.1663 & 0.1665 & 4--7 & 0.167126(96) & $1/6$ \\
$(\psi_{\rm k}^\dagger)^2$ & $(\overline{\mathbf{3}},\mathbf{1},2,1/3)$ & 0.3244 & 0.3302 & 0.3319 & 0.3326 & 0.3329 & 4--7 & 0.33439(22) & $1/3$ \\
$(\psi_{\rm k}^\dagger)^3$ & $(\mathbf{1},\mathbf{1},3,1/2)$ & 0.4850 & 0.4947 & 0.4977 & 0.4988 & 0.4993 & 4--7 & 0.50179(38) & $1/2$ \\
$\psi_{\rm k}^\dagger\psi_{\rm b}$ & $(\mathbf{3},\overline{\mathbf{3}},0,0)$ & 0.3227 & 0.3294 & 0.3315 & 0.3324 & 0.3328 & 4--7 & 0.33460(26) & $1/3$ \\
\mergedrowrule
$\{g^\dagger,\,(\psi_{\rm k}^\dagger\psi_{\rm b})^2\}$ & $(\overline{\mathbf{3}},\mathbf{3},0,0)$ & \begin{tabular}[c]{@{}r@{}}0.5934\\0.6391\end{tabular} & \begin{tabular}[c]{@{}r@{}}0.6200\\0.6562\end{tabular} & \begin{tabular}[c]{@{}r@{}}0.6346\\0.6618\end{tabular} & \begin{tabular}[c]{@{}r@{}}0.6439\\0.6641\end{tabular} & \begin{tabular}[c]{@{}r@{}}0.6497\\0.6652\end{tabular} & \begin{tabular}[c]{@{}r@{}}4--7\\4--7\end{tabular} & \begin{tabular}[c]{@{}r@{}}0.66353(99)\\0.66999(66)\end{tabular} & $2/3$ \\
\mergedrowrule
$g$ & $(\mathbf{3},\overline{\mathbf{3}},0,0)$ & 0.6020 & 0.6247 & 0.6376 & 0.6457 & 0.6509 & 4--7 & 0.66309(80) & $2/3$ \\
$\bar\rho_f$ & $(\mathbf{1},\mathbf{1},0,-1)$ & 0.9424 & 0.9827 & 0.9933 & 0.9971 & 0.9986 & 4--7 & 1.0075(16) & $1$ \\
\mergedrowrule
$\{\psi_{\rm k}^\dagger\psi_{\rm b}\,g^\dagger,\,(\psi_{\rm k}^\dagger\psi_{\rm b})^3\}$ & $(\mathbf{1},\mathbf{1},0,0)$ & \begin{tabular}[c]{@{}r@{}}0.9186\\0.9566\end{tabular} & \begin{tabular}[c]{@{}r@{}}0.9489\\0.9816\end{tabular} & \begin{tabular}[c]{@{}r@{}}0.9653\\0.9913\end{tabular} & \begin{tabular}[c]{@{}r@{}}0.9754\\0.9953\end{tabular} & \begin{tabular}[c]{@{}r@{}}0.9818\\0.9972\end{tabular} & \begin{tabular}[c]{@{}r@{}}4--7\\4--7\end{tabular} & \begin{tabular}[c]{@{}r@{}}0.99718(85)\\1.00548(99)\end{tabular} & $1$ \\
\mergedrowrule
$\psi_{\rm k}^\dagger\psi_{\rm b}\,g^\dagger$ & $(\mathbf{8},\mathbf{8},0,0)$ & 0.9224 & 0.9516 & 0.9670 & 0.9767 & -- & 4--6 & 0.9963(13) & $1$ \\
$(\psi_{\rm k}^\dagger\psi_{\rm b})^2g$ & $(\mathbf{8},\mathbf{8},0,0)$ & 1.2418 & 1.2814 & 1.3001 & 1.3104 & -- & 4--6 & 1.333558(25) & $4/3$ \\
\hline \multicolumn{10}{l}{$\phi_2=\pi$, singlet descendants}\\
\mergedrowrule
$\left\{\begin{smallmatrix}\bar\partial\bar\rho_f,\ {:}\bar\rho_f^2{:},\\ {:}\bar j_f^2{:},\ {:}\bar{\mathcal J}_g^2{:}\end{smallmatrix}\right\}$ & $(\mathbf{1},\mathbf{1},0,-2)$ & \begin{tabular}[c]{@{}r@{}}1.0300\\1.0706\\1.3580\\1.6391\end{tabular} & \begin{tabular}[c]{@{}r@{}}1.3558\\1.5067\\1.7522\\2.0690\end{tabular} & \begin{tabular}[c]{@{}r@{}}1.6357\\1.7771\\1.8924\\2.0425\end{tabular} & \begin{tabular}[c]{@{}r@{}}1.7940\\1.8889\\1.9514\\2.0229\end{tabular} & \begin{tabular}[c]{@{}r@{}}1.8788\\1.9414\\1.9784\\2.0136\end{tabular} & \begin{tabular}[c]{@{}r@{}}4--7\\4--7\\4--7\\4--7\end{tabular} & \begin{tabular}[c]{@{}r@{}}2.1366(49)\\2.173(29)\\2.099(15)\\1.9870(25)\end{tabular} & $2$ \\
\hline\hline
\end{tabular}
\par\smallskip
\begin{minipage}{0.96\linewidth}\small\raggedright
Stacked gaps are ordered by increasing value at each width and fitted by rank. Singlet gaps exclude the vacuum. The direct particle--hole fit in Table~\ref{tab:transfer-dimensions} uses $L_2=4$--$7$. The conjugate block is listed separately.
\end{minipage}
\end{table}

\begin{figure}[!htbp]
    \centering
    \includegraphics[width=0.96\linewidth,height=0.88\textheight,keepaspectratio]{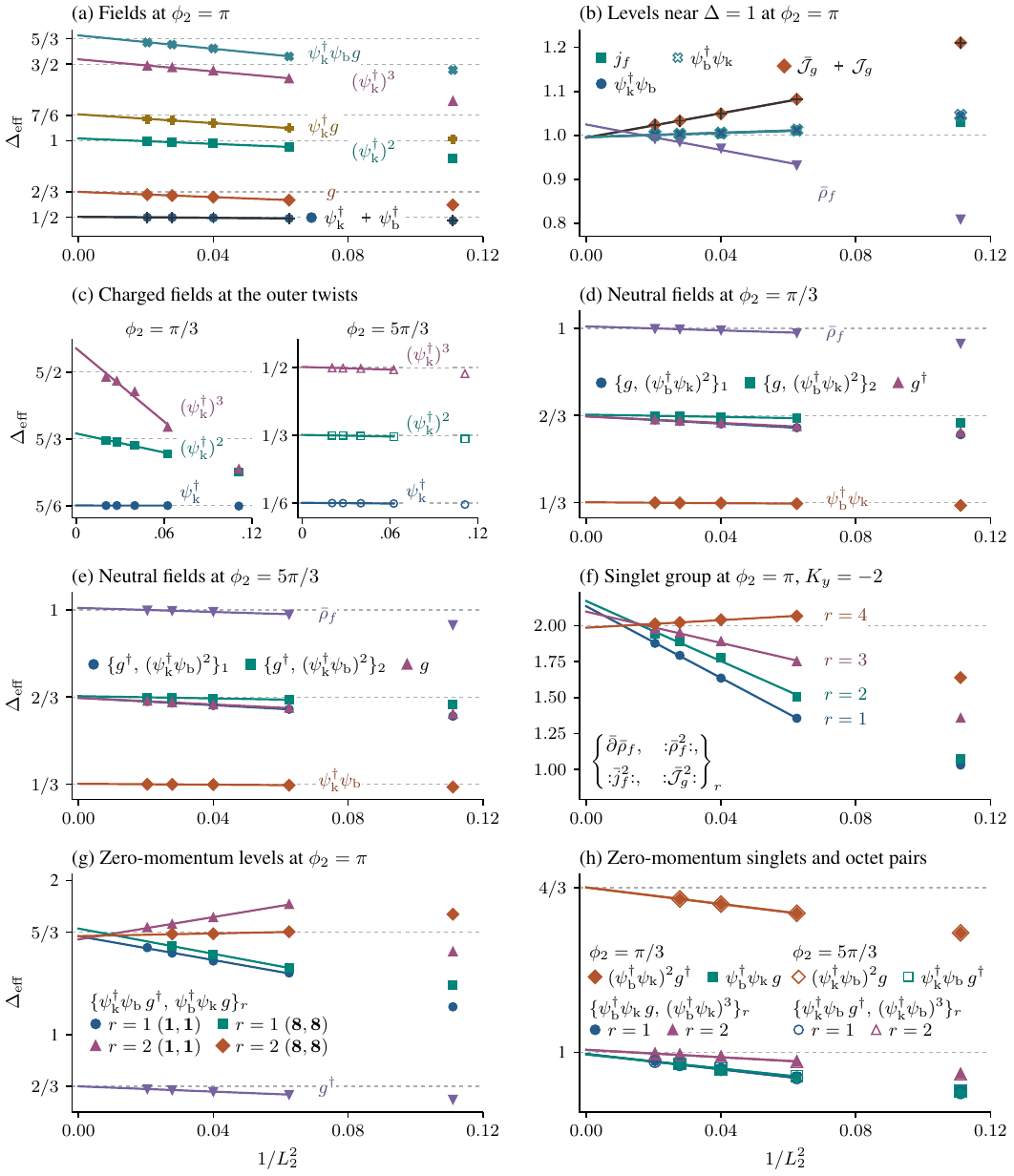}
    \caption{Gaps from Table~\ref{tab:transfer-branches} versus $1/L_2^2$. Solid lines give fits and dashed lines CFT predictions. Colored field labels identify individual curves. For a candidate set $\{\cdots\}_r$, $r$ ranks the levels by increasing gap within the same symmetry block at each width. (a,b) Central twist and detail near one. Black plus signs mark overlapping bra branches. (c--e) Charged and neutral fields at the outer twists. (f) Four $K_y=-2$ singlets share one set of normal-ordered descendants. (g,h) Zero-momentum fields. In (g), $(R_{\rm k},R_{\rm b})$ distinguishes two projections of the same set. Open markers in (c,h) denote $5\pi/3$.}
    \label{fig:app-transfer-levels}
\end{figure}

\clearpage
\section{Chiral towers in the entanglement spectrum}
\label{app:entanglement-spectrum}
\label{app:numerical-details}

Each eigenstate $|j\rangle$ of the half-cylinder density matrix $\rho_0$ has eigenvalue $p_j$ and entanglement energy $\zeta_j=-\ln p_j$.
At all three twists the dominant fixed points have $Q=0$, hence $|\mathbf n_{\rm k}|+|\mathbf n_{\rm b}|=3L_2$. Thus $\rho_0$ separates into blocks of momentum and occupation $N_\partial=|\mathbf n_{\rm k}^c|=|\mathbf n_{\rm b}|$. The plotted edge charge counts this occupation in electron units, $Q_{\rm e}=N_\partial/3$. Each ket hole carries $\overline{\mathbf3}$ [Eq.~\ref{eq:app-ket-cut-rotation}], so the $SU(3)$ center element $\mathrm{e}^{2\pi \mathrm{i}/3}\mathbb I$ acts by $\mathrm{e}^{-2\pi \mathrm{i}N_\partial/3}$. On a singlet this phase is one, hence $Q_{\rm e}\in\mathbb Z$.

We use the momentum convention of Sec.~\ref{app:transfer-quantum-numbers} for $K_y$. For each $Q_{\rm e}$, let $K_{y,0}$ be the $K_y$ of the tower bottom. The plotted difference $m=(K_{y,0}-K_y)\bmod L_2$ is then an integer.

At $L_2=7$ the tower bottoms in Fig.~\ref{fig:app-entanglement-spectrum}(a--c) have $(Q_{\rm e},K_{y,0})=(3,13/2),(3,5/2),(4,6)$, respectively. The $Q_{\rm e}=3,4$ bottoms at $\phi_2=\pi$ are degenerate, and Fig.~\ref{fig:transfer-entanglement}(a) shows $Q_{\rm e}=3$. The low levels in each tower have multiplicities $1,1,2,3,5,7$ at $m=0,\ldots,5$, the numbers of partitions of $m$ for one chiral boson.

For a chiral $U(1)_K$ edge, a low-lying descendant at level $m$ has entanglement energy~\cite{Wen1990ChiralLuttinger,Qi2012EntanglementEdge},
\begin{equation}
 \zeta_j(\phi_2,Q_{\rm e})\simeq
 \zeta^{\rm off}_{\phi_2}+
 \frac{\varepsilon_{\phi_2}K}{2}
 \left(Q_{\rm e}-Q_*-\frac{a(\phi_2)}{K}\right)^2+\varepsilon_{\phi_2}m .
 \label{eq:app-entanglement-tower-bottom}
\end{equation}
Here $Q_*$ is the charge center at $\pi$, $a=(-1,0,+1)$ in twist order, and $\zeta^{\rm off}_{\phi_2}$ is an additive offset.
We fit the $m=0$ tower bottoms to a free parabola at each twist using $Q_{\rm e}\in\{2,3,4\}$, $\{2,3,4,5\}$ and $\{3,4,5\}$, respectively.
The oscillator spacing $\varepsilon_{\phi_2}$ from the tower bottom to the lowest $m=1$ level is $(0.32936,0.37296,0.32936)$.
Dividing the fitted curvature by this spacing gives $K=(3.0749,3.0567,3.0749)$. The fitted centers advance by $0.32962$ electron charges per $2\pi/3$ increase in twist. Identifying this shift with $1/K$ gives $K\simeq3.0338$, also close to the Laughlin value $3$.

\begin{figure}[!htbp]
    \centering
    \includegraphics[width=0.84\textwidth]{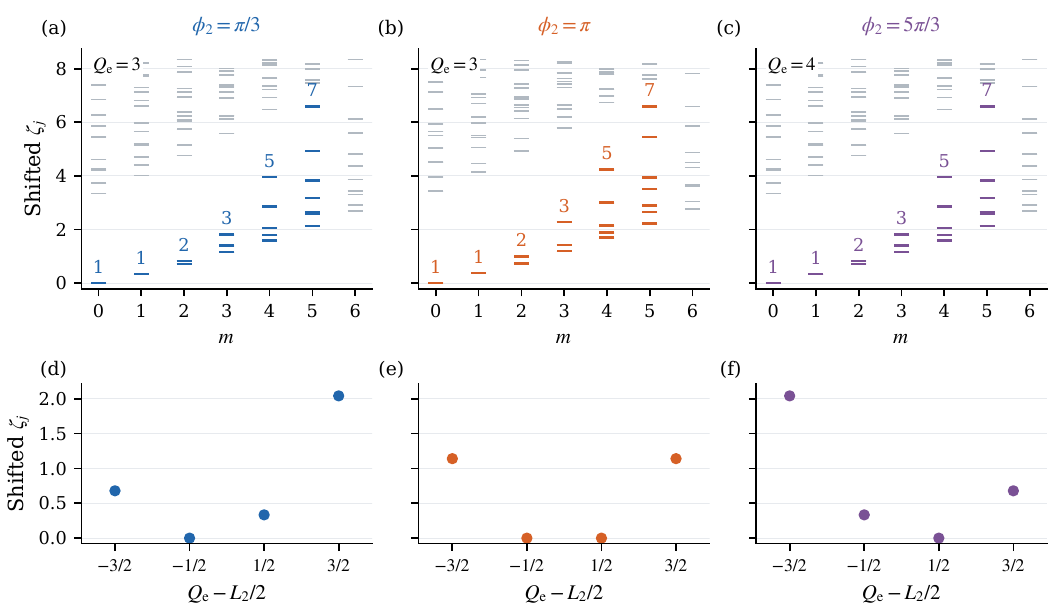}
    \caption{Half-cylinder entanglement spectra at $L_2=7$, with the minimum energy at each twist subtracted.
    (a--c) Levels below $8.35$ in the indicated charge sector, plotted against $m=(K_{y,0}-K_y)\bmod L_2$. Colors mark the chiral-boson counts.
    (d--f) Lowest levels at $Q_{\rm e}=2,3,4,5$. The horizontal coordinate $Q_{\rm e}-L_2/2$ is half-integer at this odd width. Reflection about $Q_{\rm e}=7/2$ exchanges the outer twists.}
    \label{fig:app-entanglement-spectrum}
\end{figure}
\FloatBarrier
\endgroup

\FloatBarrier